\documentclass[onecolumn,letterpaper,journal,11pt]{IEEEtran}

\usepackage[utf8]{inputenc}
\usepackage[T1]{fontenc}
\usepackage{amsmath,amstext,amsfonts,amssymb,url,psfrag,epsfig,dsfont}
\usepackage{hyperref,overpic,verbatim,enumerate}
\usepackage[usenames]{color}
\usepackage{mathrsfs}

\newtheorem{lemma}{Lemma} 
\newtheorem{theorem}{Theorem} 
\newtheorem{corollary}{Corollary}   
\newtheorem{definition}{Definition}

\newcommand{\const}[1]   {{\color{black}\mathit{#1}}}   
\newcommand{\varia}[1]   {{\color{black}\mathsf{#1}}}   
\newcommand{\set}[1]     {{\color{black}\mathcal{#1}}} 
\newcommand{\dva}[1]     {{\color{black}\mathrm{#1}}} 
\newcommand{\PDS}[1]     {{\color{black}\mathscr{P}({#1})}} 
\newcommand{\PWS}[1]     {{\color{black}\wp({#1})}} 
\newcommand{\SC}[0]      {{\color{black}\mathbb{Q}}}

\newcommand{\TVO}[0]     {{\color{black} \Delta}} 
\newcommand{\mtp}[0] {{{\color{black} \lambda }}}
\newcommand{\TV}[2]        {\TVO\!\left({#1}, {#2}\right)} %Total Varition 
\newcommand{\emp}[1]  {\dva{Q}_{\{#1\}}}

\newcommand{\PX}[1]      {{\bf P}\!\left[{#1}\right]} %Probability  
\newcommand{\EX}[1]      {{\bf E}\!\left[{#1}\right]} %Expectation  
\newcommand{\PXS}[2]     {{\bf P}\!\thinspace^{(#1)}\left[{#2}\right] }  %Probability
\newcommand{\EXS}[2]     {{\bf E}\!\thinspace^{(#1)}\left[{#2}\right] }  %Probability

\newcommand{\PCX}[2]     {\PX{\left.\! {#1} \right| {#2}}} %Conditional Probability  
\newcommand{\ECX}[2]     {\EX{\left.\! {#1} \right| {#2}}} %Conditional Expectation  
\newcommand{\PXD}[0]     {\mathtt{P}}
\newcommand{\PXAD}[1]    {\mathtt{P}_{\{\!{#1}\!\}}}   
\newcommand{\PXA}[2]     {{\bf P}_{\!\{\!{#1}\!\}\!}\left[{#2}\right]} %Probability  
\newcommand{\Era}[0]      {{\bf x}}
\newcommand{\Ero}[0] 	    {{\bf e}}

\newcommand{\Pe}[0]     {{\it P_{\Ero}}}         %Probability of error   
\newcommand{\Per}[0]    {{\it P_{\Era}}}         %Probability of erasure    
\newcommand{\Pem}[1]  	{{\it P}_{\Ero|{#1}}}         %Probability of error   
\newcommand{\Perm}[1] 	{{\it P}_{\Era|{#1}}}         %Probability of erasure    
\newcommand{\Pemb}[2]     {{\it P_{\Ero|{#1}}}({#2}) }         %Bit Probability of error   
\newcommand{\Peb}[1]         {{\it P_{\Ero}}({#1}) }         %Bit Probability of error   
\newcommand{\Pec}[1]          {{\it P_{\Ero}}\{{#1}\}}         %Message subSet Probability of error   

\newcommand{\mes}[0]     {\varia{M}}
\newcommand{\est}[0]     {\varia{\widehat{M}}}      

\newcommand{\tmes}[0]    {\!~\!_{\mathtt{t}}\!\varia{M}}
\newcommand{\test}[0]    {\!~\!_{\mathtt{t}}\!\varia{\widehat{M}}}

\newcommand{\dmes}[0]   {\varia{m}}

\newcommand{\inp}[0]    {\varia{X}} 
\newcommand{\dinp}[0]   {\varia{x}}

\newcommand{\out}[0]    {\varia{Y}} 
\newcommand{\dout}[0]   {\varia{y}}
 
\newcommand{\dt}[0]     {\varia{T}}
\newcommand{\wdt}[0]    {\widetilde{\varia{T}}}
\newcommand{\ddt}[0]    {\varia{t}}
\newcommand{\dwdt}[0]   {\tilde{\varia{t}}}

\newcommand{\dum}[0]    {\varia{Z}} 
\newcommand{\ddum}[0]   {\varia{z}}

\newcommand{\tvs}[0]    {\varia{S}}

\newcommand{\rep}[0]    {\varia{L}} 
\newcommand{\drep}[0]   {\varia{l}}

\newcommand{\mar}[1]    {\varia{{#1}}}

\newcommand{\CMIX}[3]   {\varia{I} \left( {#1}; {#2} \left\vert {#3} \right. \right)}
\newcommand{\HX}[0]     {\varia{H}}

\newcommand{\dumS}[0]   {\set{Z}}
\newcommand{\outS}[0]   {\set{Y}}
\newcommand{\inpS}[0]   {\set{X}} 
\newcommand{\mesS}[0]   {\set{M}}
\newcommand{\estS}[0]   {\set{\widehat{M}}}
\newcommand{\tmesS}[0]  {\!~\!_{\mathtt{t}}\!\set{M}}

\newcommand{\event}[0]  {\Gamma }
\newcommand{\ldsf}[0]   {\set{A}}
\newcommand{\lds}[1]    {\ldsf({#1})}
\newcommand{\ldsn}[2]   {\ldsf_{#1}({#2})}

\newcommand{\dtr}[0]    {\const{\gamma}}

\newcommand{\decsx}[0]   {\set{G}}
\newcommand{\decpx}[0]   {\set{B}}

\newcommand{\decs}[1]   {\decsx[{#1}]}
\newcommand{\decp}[1]   {\decpx[{#1}]}

\newcommand{\decsg}[1]  {\decsx_{\dtr}[{#1}]}
\newcommand{\decpg}[1]  {\decpx_{\dtr}[{#1}]}

\newcommand{\inx}[0]    {\dva{\kappa}}
\newcommand{\blx}[0]    {\dva{n}}
\newcommand{\tin}[0]    {\dva{\tau}}

\newcommand{\ENC}[0]          {\dva{\Phi}}
\newcommand{\DEC}[0 ]         {\dva{\Psi}}

\newcommand{\tsc}[0]            {\dva{\alpha}}

\newcommand{\fr}[0]              {\dva{\eta}}
\newcommand{\afr}[0]            {\dva{\nu}}

\newcommand{\ex}[0]             {\dva{E}}
\newcommand{\rate}[0]          {\dva{R}} 
\newcommand{\arate}[1]        {\dva{r}_{{#1}}}

\newcommand{\Emd}[0]          {{\ex_{\textrm{md}}}}  %Miss Detection Exponent

\newcommand{\idis}[1]           {\dva{\mu}_{#1}}
\newcommand{\odis}[1]          {\bar{\dva{\mu}}_{#1}}

\newcommand{\MI}[2]                {\dva{I}\left({#1} , {#2} \right) }   % Mutual Information
\newcommand{\ENT}[1]  	   {\dva{H} \left( {#1} \right)}

\newcommand{\KLD}[2]             {\dva{D}\left( \left.\! {#1} \right\|{#2} \right)}

\newcommand{\CTM}[0]      {\const{W}}  
\newcommand{\CT}[2]         {{\CTM}_{\!{#1}}({#2})} 

\newcommand{\rexp}[0]     {\const{E}_{r}}
\newcommand{\CX}[0]        {\const{C}}%Capacity  
\newcommand{\TEX}[0]      {\const{E}}%Capacity  
\newcommand{\DX}[0]        {\const{D}}%Exponent
  
\newcommand{\xa}[0]         {\const{{\bf a}}}%Exponent  
\newcommand{\xr}[0]         {\const{{\bf r}}}%Exponent  

\newcommand{\JX}[1]         {\const{J}\!\left({#1}\right) }
\newcommand{\jx}[1]         {\const{j}\!\left({#1}\right) }
\newcommand{\bent}[1]       {\const{h}\!\left({#1}\right) }

\newcommand{\nlay}[0]      {{\ell}}

\newcommand{\real}[0] {{\bf R}}

\newcommand{\pkld}[2]{{\xi}_{{#1},{#2}}}
\newcommand{\epsq}[1] {{\varepsilon}_{{#1}}}
\newcommand{\eps}[2]  {{\varepsilon}_{{#1}}{\scriptstyle{ ({#2})}}}
\newcommand{\epst}[1] {\tilde{{\epsilon}}_{{#1}}}
\newcommand{\DEF}[0]    {{\triangleq}}  
\newcommand{\ald}[0] {{\it ALD}}
\newcommand{\nald}[0] {{\it NALD}}
\newcommand{\uep}[0] {{\it UEP}}
\newcommand{\IND}[1]    {{\mathds{1}}_{\left\{ {#1} \right\} }}  %Indicator function
\newcommand{\ord}[1]  {{{\it o}}({#1})}

\newcommand{\revs}[0] {rates-exponents vectors}
\newcommand{\rev}[0] {rates-exponents vector}
\newcommand{\hv}[0]     {\varia{H}} 

\newcommand{\hvS}[0]     {\set{H}} 
\newcommand{\gv}[0]     {\varia{G}} 
\newcommand{\dgv}[0]   {\varia{g}}
\newcommand{\gvS}[0]   {\set{G}}
\newcommand{\zv}[0]     {\varia{Z}} 
\newcommand{\dzv}[0]   {\varia{z}}

\title{Bit-wise Unequal Error Protection\\ for Variable Length Block Codes with Feedback}
\author{
Bar\i\c{s} Nakibo\u{g}lu
\quad \and
Siva K. Gorantla
\quad \and
Lizhong Zheng
\quad \and
Todd P. Coleman
\thanks{Authors acknowledge  the support of the NSF through  the NSF Cyberphysical Systems Program (NSF0932410) and  the NSF Science \& Technology Center (CCF-0939370) and support of  DARPA through the ITMANET program (W911NF-07-1-0029)}
\footnote{This paper was presented in part at ISIT 2010, \cite{uep-2010}.}
} 

\begin{document}
\maketitle
\begin{abstract}
The \emph{bit-wise} unequal error protection problem, for the case when the number of groups of 
bits \( \nlay\) is  fixed, is considered for variable length block codes with feedback. 
An  encoding scheme based on fixed length block codes with erasures is used to establish inner bounds to the 
achievable performance for finite expected decoding time. 
A new technique for bounding the  performance of variable length block codes is used to establish outer bounds   to the performance for a given expected decoding time.
The inner and the outer bounds match one another asymptotically and  characterize the achievable 
region of  \revs, completely.  The single message \emph{message-wise} unequal error protection problem
for variable length block codes with feedback
 is also solved as a necessary step on  the way.
 \end{abstract}

% !TeX root = uep.tex

\section{Introduction}
In the conventional  formulation of digital  communication problem, the primary concern
is the correct transmission of the message; hence there is no distinction between different 
error events. In other words, there is a tacit assumption that all error  events are equally
 undesirable; incorrectly decoding  to a message \(\bar{\dmes}\)    when  a message \(\tilde{\dmes}\)  
  is transmitted,  is as   undesirable as incorrectly decoding   to a message 
\(\bar{\bar{\dmes}}\)    when  a message \(\tilde{\tilde{\dmes}}\)   
is transmitted,  for any \(\bar{\dmes}\) other than \(\tilde{\dmes}\)   and
 \(\bar{\bar{\dmes}}\) other than  \(\tilde{\tilde{\dmes}}\). 
Therefore   the performance criteria  used in the conventional formulation  
(minimum distance  between codewords, maximum conditional error
 probability among messages,  average error probability, etc.) are  
 oblivious to any precedence order  that might exist among the error events.

 In many applications, however,  there is a clear order of precedence   among the error events. 
 For example  in Internet communication, packet headers are more important than the actual 
 payload data. Hence, a code used for Internet communication, can enhance  the protection 
 against   the  erroneous  transmission of the 
 packet headers  at the expanse of  the  protection against the erroneous  
 transmission of payload data.  In order to appreciate such a coding scheme, one may 
  analyze  error probability of the packet headers and error probability of
  payload data separately, instead of analyzing the  error probability of the 
  overall message composed of  packet header and payload data. Such 
  a formulation for Internet communication is   an unequal error protection (\uep) 
  problem, because of the separate calculation of the error probabilities of the
  parts of the messages.

Problems capturing the disparity of undesirability among various classes 
of   error events, by assigning  and analyzing  distinct   performance criteria for 
different  classes of error events, are called
 unequal error protection (\uep) problems. \uep~problems have already been studied 
 widely  by researchers in communication theory, coding theory, and computer networks 
 from the perspectives of their respective fields. In this paper we  enhance the information
 theoretic perspective on \uep~problems \cite{csiszar1}, \cite{bnz}  for  variable length block 
 codes by generalizing the results of \cite{bnz} to the rates below capacity.

In information theoretic \uep,  error   events are grouped into different classes and  the 
probabilities associated with these   different classes of error events are  analyzed 
separately. In order to prioritize protection against one or the other class of error events,
 corresponding error exponent is increased at  the expense of the other error exponents.
   There are various ways to choose the error event classes but two specific choices of error 
   event classes stand out   because of their intuitive familiarity and practical relevance; they 
   correspond to the  \emph{message-wise} \uep~and  the \emph{bit-wise} \uep. Below, we 
 first describe these two types of \uep~then specify the  \uep~problems we are 
 interested in this manuscript.

   In the \emph{message-wise} \uep, the message set \(\mesS\) is assumed to be the union of
   \( \nlay\)   disjoint sets for some fixed   \( \nlay\), i.e.,  
\(\mesS=\cup_{j=1}^{\nlay}\mesS_j\) where \( \mesS_{i} \cap \mesS_{j} =\emptyset\) for all \( i\neq  j\).
%\begin{align*}
%\mesS&=\bigcup_{j=1}^{ \nlay} \mesS_j 
%&\mbox{~where~}&      \mesS_{i} \cap \mesS_{j} =\emptyset~~   \forall i\neq  j.
%\end{align*}
 For each set    \(\mesS_j\),  the
 maximum error   probability\footnote{This formulation is called the missed detection
  formulation of     the  \emph{message-wise} \uep~problem in \cite{bnz}. 
   If  \( \PCX{\est\!\neq \dmes}{\mes\!=\dmes}\)  is replaced 
   with    \( \PCX{\est\!= \dmes}{\mes\! \neq\dmes}\)   we get the   false alarm formulation of 
   the \emph{message-wise} \uep~problem.   In this paper  we restrict our   discussion to the
    missed detection problem and  use  \emph{message-wise} \uep~without any qualifications
     to refer to the missed 
        detection formulation of the  \emph{message-wise} \uep~problem.}  \(\Pec{j}\),    
        the rate \(\rate_{\{j\}}\)  and  the error exponent  \(\ex_{\{ j\}} \)  are defined as the
        corresponding quantities  defined in the conventional problem,  i.e., 
\(\Pec{j}=\max\nolimits_{\dmes\in \mesS_j} \PCX{\est\!\neq \dmes}{\mes\!=\dmes}\), 
\(\rate_{\{ j\}}=\tfrac{\ln |\mesS_j|}{\blx}\),  
\(\ex_{\{ j\}}= \tfrac{- \ln \Pec{j}}{\blx}\),  
for all \( j\) in \(\left\{1,2,\ldots, \nlay\right\}\) 
%\begin{align*}
%\Pec{j}&=\max \limits_{\dmes\in \mesS_j} \PCX{\est\!\neq \dmes}{\mes\!=\dmes}
%&
%\rate_{\{ j\}}&=\frac{|\mesS_j|}{\blx}
%&
%\ex_{\{ j\}}&= \frac{- \ln \Pec{j}}{\blx}
%&\forall j\in \left\{1,2,\ldots, \nlay\right\}
%\end{align*}  
where \(\blx\) is the length of the code.       
 The ultimate aim is  calculating the achievable region of    rate vector   error exponent vector pairs, 
 \((\rate_{\{ \cdot\}},\ex_{\{ \cdot\}})\)'s    where\footnote{ Here  \(  \nlay \) is assumed to be a fixed
 integer. All \revs, achievable or not, are in  the region of  
 \(\real^{2\nlay}\)  in  which \(\rate_{\{ j\}}\!\geq\!0\) and \(\ex_{\{ j\}}\!\geq\!0\) for all 
 \(1\leq j \leq  \nlay\). 
  \(\real^{2\nlay}\) is the \(2\nlay \) dimensional real vector space  with the norm
  \(\lVert \vec{X} \rVert =\sup_{j} |x_j|\)}
 \(\rate_{\{ \cdot\}}=(\rate_{\{1\}},\rate_{\{ 2\}},\ldots,\rate_{\{ \nlay\}})\)  and 
 \(\ex_{\{ \cdot\}}=(\ex_{\{ 1\}},\ex_{\{ 2\}},\ldots,\ex_{\{  \nlay\}})\). The \emph{message-wise} 
 \uep~problem was the first information theoretic \uep~problem to be considered;    
 it was considered by  Csisz\'ar  in his work on joint source   channel coding \cite{csiszar1}. 
 Csisz\'ar showed that for any  integer \( \nlay\), block length  \(\blx\) and \( \nlay\)-dimensional 
 rate vector   \( \rate_{\{ \cdot\}}\)  such that  \(0\leq \rate_{\{ j\}} \leq \CX \) for \(j=1,2,\ldots, \nlay\), 
 there exists a  length \( \blx \) block code with message set \(\mesS=\cup_{j=1}^{ \nlay} \mesS_j\)
where \(|\mesS_j|=e^{\blx (\rate_{\{ j\}}-\epsq{\blx} )}\)  such that the  conditional error probability of 
each message in each  \( \mesS_j \)  is less then \(e^{-\blx (\rexp(\rate_{\{ j\}})-\epsq{\blx})}\) where 
\(\rexp(\cdot)\) is the random coding exponent and \( \epsq{\blx}\) converges to zero as \( \blx\) 
diverges.\footnote{Csisz\'ar proved the above result not only for the case when \( \nlay\) is  constant 
for all \( \blx \) but also for the case when \(  \nlay_{\blx}\)  is a sequence such 
that \( \lim_{\blx \to \infty} \tfrac{\ln  \nlay_{\blx}}{\blx}=0\). See \cite[Theorem 5]{csiszar1}.}

The \emph{bit-wise} \uep~problem is the other canonical form of the information theoretic
\uep~problems. In the \emph{bit-wise} \uep~problem the message set \(\mesS\) is assumed to be 
the Cartesian product of \(\mesS_1\), \(\mesS_2\),  \(\ldots\), \(\mesS_ \nlay\) for some fixed \( \nlay\), 
i.e., \(\mesS=\mesS_1 \times \mesS_2\times \ldots \times  \mesS_ \nlay.\)
% \begin{align*}
%   \mesS=\mesS_1 \times \mesS_2\times \ldots \times  \mesS_ \nlay.
%\end{align*}
Thus the transmitted message \(\mes\) and the decoded message \(\est\)  are given by 
\(\mes=(\mes_1, \mes_2, \ldots,  \mes_ \nlay)\) and \(\est=(\est_1, \est_2, \ldots,  \est_ \nlay)\),  receptively.
Furthermore,   \(\mes_j\)'s and \(\est_j\)'s are called the transmitted and decoded sub-messages, respectively.
%Thus the transmitted message \(\mes\) and the decoded message \(\est\)  are composed of sub-messages  
%\begin{align*}
%\mes&=(\mes_1, \mes_2, \ldots,  \mes_ \nlay)
%&
%\est&=(\est_1, \est_2, \ldots,  \est_ \nlay).
%\end{align*}
  The error  events  of interest in the  \emph{bit-wise}    
  \uep~problem are the ones  corresponding to  the erroneous  transmission of the 
  sub-messages. The error probability \(\Peb{j}\), rate \(\rate_{j}\) and the error
  exponent \(\ex_{j}\) of sub-messages  are   given by \(\Peb{j}=\PX{\est_j\neq \mes_j}\),
 \(\rate_j=\frac{\ln |\mesS_j|}{\blx}\),
\(\ex_j=\frac{-\ln \Peb{j}}{\blx}\) 
for all \( j\) in \(\left\{1,2,\ldots, \nlay\right\}\)
 % \begin{align*}
%\Peb{j}&=\PX{\est_j\neq \mes_j}
%&
%\rate_j&=\frac{\ln |\mesS_j|}{\blx}
%&
%\ex_j&=\frac{-\ln \Peb{j}}{\blx}
%&\forall j\in \left\{1,2,\ldots, \nlay\right\}
%\end{align*}
where  \(\blx\) is the block length. As  was the case in the \emph{message-wise} \uep~problem,
 the ultimate aim in the \emph{bit-wise}   \uep~problem is determining the achievable region of 
 the  rate  vector error exponent vector    pairs\footnote{Similar to the \emph{message-wise} 
 \uep~problem discussed above, in the current formulation of    \emph{bit-wise}    \uep~problem
 we assume \(  \nlay \) to be fixed. Thus   all \revs, achievable or not, are in  region  of \(\real^{2 \nlay}\)  
in which   \(\rate_{j}\!\geq\!0\) and   \(\ex_j\!\geq\!0\) for all \(1\leq j \leq  \nlay\), by definition.}  
 \((\vec{\rate},\vec{\ex})\).   The formulation of  Internet   communication  problem we have       
 considered above, with packet header and payload data, is a \emph{bit-wise} \uep~problem 
 with two sub-messages, i.e., with \( \nlay=2\).

There is some resemblance in the definitions of  \emph{message-wise}  and  \emph{bit-wise}
 \uep~problems, but they  have very different behavior in many problems. For example,
consider  the  \emph{message-wise} \uep~problem and the  \emph{bit-wise} \uep~problem
  with  \(\nlay=2\), \( \mesS_1=\{1,2\} \) and \(  \mesS_2=\{3,4,\ldots,e^{\blx (\CX-\epsq{\blx})}\} \)
  for some \( \epsq{\blx}\) that goes to zero as  \( \blx\) diverges.
   It is shown in  \cite[Theorem 1]{bnz} that  if  \( \mesS=\mesS_1 \times \mesS_2 \) 
   and \( \PX{\mes_2\neq \est_2} \leq \widetilde{\epsq{\blx}} \) for some \(\widetilde{\epsq{\blx}}\) 
   that  goes to zero as  \( \blx\) diverges then\footnote{The channel  is assumed to have no zero  
   probability transition.}  \( \ex_1=0 \). Thus in the \emph{bit-wise} \uep~problem even 
     a bit can not have a positive error exponent. As result of \cite[Theorem 5]{csiszar1},  on the other hand,  
     if \(\mesS=\mesS_1 \cup \mesS_2\) we know   that  \(\mesS_1\) can have an error  exponent
       \(\ex_{\{1\}}\)  as high as  $\rexp(0)>0$ while having a small error probability for    \(\mesS_2\),
       i.e.,        \(\max\limits_{\dmes\in \mesS_2} \PCX{\est\!\neq \dmes}{\mes\!=\dmes}\leq \widetilde{\epsq{\blx}} \)
 for some  \(\widetilde{\epsq{\blx}}\)   that  goes to zero as  \( \blx\) diverges.  Thus in  
 the  \emph{message-wise} \uep~problem it is possible to give an error exponent  as 
 high as  \( \rexp(0) \) to \(\mesS_1\).

The   \emph{message-wise} and the  \emph{bit-wise} \uep~problems cover a wide range of problems 
of practical interest. Yet,  as noted in \cite{bnz}, there are  many  \uep~problems of practical importance
that are neither  \emph{message-wise}  nor \emph{bit-wise} \uep~problems.  One of our aims in studying
the \emph{message-wise}  and the  \emph{bit-wise} \uep~problems  is gaining insights  and devising tools 
for the analysis of those more complicated problems.

In the above discussion the \uep~problems  are described for fixed length block codes for the sake of   simplicity.  
One can, however, easily define the corresponding problems for various families of codes:with or without 
feedback, fixed or variable length, by  modifying the definitions of the error probability, the rate and the error 
exponent appropriately. Furthermore parameter \( \nlay\) representing the number of groups of bits or messages 
is assumed to be fixed in the above discussion for simplicity.  However, both the   \emph{message-wise} and the  
\emph{bit-wise} \uep~problems can be defined for \( \nlay\)'s that are increasing with block length \( \blx\) in fixed 
length block codes and for \(  \nlay\)'s that are increasing with expected block length \( \EX{\dt}\) in variable length 
block codes. In fact Csisz\'ar's  result discussed above, \cite[Theorem 5]{csiszar1}, is proved not only for constant 
\(  \nlay\) but also for any \(  \nlay_{\blx} \) sequence satisfying 
\( \lim_{\blx \to \infty} \tfrac{\ln  \nlay_{\blx}}{\blx}=0 \).

In this manuscript we consider two closely related \uep~problems for variable length block codes over a discrete memoryless channels with noiseless feedback: the \emph{bit-wise} \uep~problem and the single message \emph{message-wise} \uep~problem.

\begin{itemize}
\item In the \emph{bit-wise} \uep~problem there are \( \nlay\) sub-messages each with different priority and rate.
For all fixed values of \(  \nlay\)  we  characterize the trade-off between the rates and the error exponents of these 
sub-messages by revealing the region of achievable rate vector, exponent vector pairs. For fixed \(  \nlay\) this
 problem is simply the  variable length code version  of the above described \emph{bit-wise} \uep~problem. 

\item  In the single message  \emph{message-wise} \uep~problem,
 we characterize the trade-off between the exponents of the minimum and the average 
 conditional error probability.
Thus this problem is similar to the above described \emph{message-wise} \uep~problem 
for the case \( \nlay=2\) and \(\mesS_1=\{1\}\). But unlike that problem we work with variable length 
codes and average conditional error probability rather than fixed length codes and the maximum 
error probability.  
\end{itemize}
The \emph{bit-wise} \uep~problem  for fixed number of groups of bits, i.e., fixed \( \nlay\), and the 
single message  \emph{message-wise} \uep~problem  were first considered in  \cite{bnz}, for the 
case when the rate is  (very close to) the channel  capacity; we solve  both of these problems for
 all achievable  rates.

In fact,  in  \cite{bnz} single message \emph{message-wise}~\uep~problem is solved not only at capacity,
 but  also for all the rates below capacity both for fixed length block codes without feedback and
  for variable length block codes with feedback, but only for case when overall error exponent is zero 
  (see  \cite[Appendix D]{bnz}).   Recently Wang, Chandar, Chung and Wornell  \cite{wang12} 
put forward  a new proof based on method of types for the same problem.\footnote{In addition
to their new proof in missed-detection  problem \cite[Theorem 1]{wang12}
Wang, Chandar, Chung and Wornell  present a completely new result
on the false-alarm formulation of the problem \cite[Theorem 5]{wang12}.}
Nazer,  Shkel and Draper \cite{NSD}, on the other hand, investigated the problem for 
 fixed length block  codes on additive white   Gaussian noise channels at zero  error exponent
and derived the exact analytical expression in terms of rate and power constraints.

Before starting our presentation, let us give a brief outline of the paper.  
In Section \ref{sec:model}, we specify the  channel model and make a brief overview of stopping times 
and variable length block codes. In Section \ref{sec:psres}, we first present the single message  
\emph{message-wise} \uep~problem and  fixed \( \nlay\) version of  the \emph{bit-wise}  \uep~problem 
for variable length  block codes; then we state the solutions of these two \uep~problems.
 In Section \ref{sec:ach} we present inner bounds  for both  the single message  \emph{message-wise}
 \uep~problem and    the \emph{bit-wise}  \uep~problem. 
 In Section \ref{sec:con} we introduce a new  technique,  Lemma \ref{lem:con}, for deriving  outer  bounds
  for  variable length block codes and apply it to  the two \uep~problems we are interested in.
Finally in Section \ref{sec:conc} we  discuss the qualitative ramifications of our results in terms  the design
 of communication systems with \uep~and the limitations of our analysis. 
 The proofs of the propositions in Sections \ref{sec:psres}, \ref{sec:ach}, \ref{sec:con} are  deferred to 
  the Appendices.

\section{Preliminaries}\label{sec:model}

As it is customary we  use upper case letters, e.g., \(\mes\), \(\inp\), \(\out\), \(\dt\) for random variables    and lower 
case letters, e.g., \(\dmes\), \(\dinp\), \(\dout\), \(\ddt\) for their sample values.

We denote discrete sets by capital letters with calligraphic fonts,  e.g., \(\mesS\), \(\inpS\), \(\outS\) and power sets of discrete sets by \(\PWS{\cdot}\), e.g.,    \(\PWS{\mesS}\), \(\PWS{\inpS}\), \(\PWS{\outS}\). In order to denote  the set of all probability distributions on a discrete set we use \(\PDS{\cdot}\), e.g., \(\PDS{\mesS}\), \(\PDS{\inpS}\), \(\PDS{\outS}\).
\begin{definition}[Total Variation]
For any discrete set \(\dumS\) and for any \(\idis{1},\idis{2} \in \PDS{\dumS}\) the total variation 
\(\TV{\idis{1}}{\idis{2}}\) is defined as,
\begin{equation}
\label{eq:deftv}
\TV{\idis{1}}{\idis{2}}=\tfrac{1}{2}\sum\nolimits_{\ddum\in \dumS} |\idis{1}(\ddum) - \idis{2}(\ddum)  |.
\end{equation} 
\end{definition}

We denote the indicator function by  \(\IND{\cdot}\), i.e., \(\IND{\event}=1\) when event \(\event\) happens \(\IND{\event}=0\) otherwise.

We denote the binary entropy function by \(\bent{\cdot}\), i.e.,
\begin{align}
\bent{s}\DEF -s \ln s - (1-s)\ln(1-s) && \forall s\in [0,1]. 
\end{align} 
\subsection{Channel Model}\label{sec:cm}
We consider a discrete memoryless channel (DMC) with input alphabet $\inpS$, output
 alphabet  $\outS$ and  $|\inpS|-\mbox{by}-|\outS|$  transition probability matrix $\CTM$. 
Each row of \(\CTM\) corresponds to a probability distribution on \(\outS\), i.e.,
 \(\CTM_{\dinp} \in \PDS{\outS}\) for all   \(\dinp \in \inpS\). For the reasons   that will 
become clear shortly, in Section \ref{sec:relsec},
 we  assume that  $\CT{\dinp}{\dout}>0$  for all $\dinp \in \inpS$ and
 $\dout \in \outS$ and    denote the smallest transitions  probability by $\mtp$:
 \begin{equation}
\label{eq:lambda}
\mtp \DEF \min_{\dinp,\dout}\CT{\dinp}{\dout}>0.
\end{equation}
    The input and output letters at time $\tin$, up to time $\tin$ and between time $\tin_1$ and
   $\tin_2$ are denoted by  $\inp_{\tin}$, $\out_{\tin}$, $\inp^{\tin}$, $\out^{\tin}$,
    $\inp_{\tin_1}^{\tin_2}$ and $\out_{\tin_1}^{\tin_2}$   respectively.  DMCs are 
    both memoryless and stationary, hence  the    conditional probability of $\out_{\tin}=\dout$ given 
     $(\inp^{\tin},\out^{\tin-1})$ is given by
\begin{equation*}
  \PCX{\out_{\tin}=\dout}{\inp^{\tin},\out^{\tin-1}}= \CT{\inp_{\tin}}{\dout}.
\end{equation*}
\begin{definition}[Empirical Distribution]
For any  \(\tin_2\geq \tin_1\) and any sequence \(\ddum_{\tin_1}^{\tin_2}\) such that \(\ddum_{j} \in \dumS\) for all \(j\in [\tin_1 ,\tin_2]\),   the empirical distribution \(\emp{\ddum_{\tin_1}^{\tin_2}}\) is given by 
\begin{equation}
\label{eq:defemp}
\emp{\ddum_{\tin_1}^{\tin_2}}(\ddum) = \frac{1}{\tin_2-\tin_1+1} \sum\nolimits_{\tin=\tin_1}^{\tin_2} \IND{\ddum_\tin=\ddum} \qquad \forall \ddum\in\dumS.
\end{equation} 
\end{definition}
Note that if we replace  \(\ddum_{\tin_1}^{\tin_2}\)  by  \(\dum_{\tin_1}^{\tin_2}\) when the empirical distribution \(\emp{\dum_{\tin_1}^{\tin_2}}(\ddum) \)	becomes a random variable for each \(\ddum \in \dumS\). 

 \subsection{Stopping Times}\label{sec:stop}   
Stopping times are central in the  formal  treatment of variable length codes;  it is not possible to 
define or comprehend variable length codes without a solid understanding of stopping times.
For those readers who are not already familiar with the concept of  the stopping times, we present 
a brief overview  in this section.

  In order to make  our presentation more accessible,  we use the concept of power sets,  rather than sigma-
  fields in the definitions. We can do that only because the random variables we use to define stopping times 
are discrete random variables. In the general case, when  the underlying variables are not necessarily discrete, 
one needs to use the concept of sigma fields instead of power set.

Let us start with introducing the concept of  Markov times.  For an infinite sequence of random variables 
\(\dum_1,\dum_2,\ldots\), a positive, \emph{integer\(^{*}\)} 
valued\footnote{\emph{Integer\(^{*}\)} is the set of all  integers together with  two infinities, i.e., 
\(\{-\infty,\ldots,-1,0,1,\ldots,\infty\}\).}  
function  \(\dt\) defined on  \(\dumS^{\infty}\) is  a Markov time,  if for all positive integers  \(\tin\) it is 
possible   determine whether \(\dt=\tin\) or not  by considering \(\dum^{\tin}\) only, i.e., if  \(\IND{\dt=\tin}\)
 is not only a function of \(\dum^{\infty}\) but also  a function of \(\dum^{\tin}\)  for all positive integers
 \(\tin\). The formal definition  is given  below.
\begin{definition}[Markov Time]
Let \(\dum_{1}^{\infty}\) be an infinite sequence of \(\dumS\) valued random variables \(\dum_{\tin}\) for \(\tin \in  \{1,2,\ldots\}\) and  \(\dt\) be a function of \(\dum^{\infty}\) which takes values from 
 the set \(\{1,2,\ldots,\infty \}\).
  Then the random variable \(\dt\) is a Markov time  with respect to \(\dum^{\tin}\) if 
\begin{equation}
\label{eq:def-markov-time}
\{\ddum^{\infty}: \dt =\tin \mbox{~if~} \dum^{\infty}=\ddum^{\infty}\} \in \PWS{\dumS^{\tin}} \times \{\dumS_{\tin+1}^{\infty}\}
\qquad~\quad\forall \tin \in  \{1,2,\ldots\}.
%\{\dt =\tin\} \in \PWS{\dumS^{\tin}}  
\end{equation}
where \(\PWS{\dumS^{\tin}} \times \{\dumS_{\tin+1}^{\infty}\}\) is the Cartesian  product of the power 
set of \(\dumS^{\tin}\) and the one element set \(\{\dumS_{\tin+1}^{\infty}\}\).
\end{definition}
 We denote  \(\dum_{\tin}\)'s from \(\tin=1\) to \(\tin=\dt\) by \(\dum^{\dt}\) and their sample values by 
\(\ddum^{\ddt}\). The set of all sample values of \(\dum^{\dt}\) such that \(\dt=\tin\), on the other hand,  is 
denoted by  \( \dumS^{\tin}_{\{\dt=\tin\}}\). We denote union of all \( \dumS^{\tin}_{\{\dt=\tin\}}\)'s for finite 
\(\tin\)'s by  \(\dumS^{\dt*}\) and the union of all  \( \dumS^{\tin}_{\{\dt=\tin\}}\)'s   by  \(\dumS^{\dt}\),  i.e.,
\begin{subequations}
\label{eq:def-stop-dom}
\begin{align}
 \dumS^{\tin}_{\{\dt=\tin\}}
\label{eq:def-stop-dom-a}
 &\DEF \{\ddum^{\tin}:  \dt=\tin \mbox{~if~} \dum^{\tin}=\ddum^{\tin}  \}&
 &\tin\in \{1,2,\ldots,\infty\}\\
 \dumS^{\dt*}
 \label{eq:def-stop-dom-b}
&\DEF \bigcup_{1\leq \tin < \infty}\dumS^{\tin}_{\{\dt=\tin\}}&
& \\
\dumS^{\dt}
\label{eq:def-stop-dom-c}
 &\DEF \dumS^{\dt*} \bigcup\dumS^{\infty}_{\{\dt=\infty\}}.&
&
 \end{align}
\end{subequations}
For an arbitrary,  positive, \emph{integer\(^{*}\)} valued  function \(\dt\) of \(\dum^{\infty}\), however,  one 
can  not talk about  \(\dum^{\dt}\),  because the value of \(\dt\) can in principle depend on 
\(\dum_{\dt+1}^{\infty}\). For a  Markov time   \(\dt\), however,  the value of \(\dt\) does not  depend on 
\(\dum_{\dt+1}^{\infty}\). That is  why we can define   \(\dum^{\dt}\), \( \dumS^{\tin}_{\{\dt=\tin\}}\), 
\(\dumS^{\dt*}\)  and  \(\dumS^{\dt}\) for any Markov  time  \(\dt\).

Given an infinite sequence of \(\ddum_{\tin}\)'s, i.e., \(\ddum^{\infty}\),
 either \(\ddum^{\infty} \in  \dumS^{\infty}_{\{\dt=\infty\}}\)
or  \(\ddum^{\infty}\)  has a unique  subsequence  \(\ddum^{\tin}\) that is in \(\dumS^{\dt*}\).

In most practical situations, one is interested in Markov times that are guaranteed to have a finite value; those Markov times   are called Stopping times.
\begin{definition}[Stopping Time]~
A Markov time \(\dt\) with respect to \(\dum^{\tin}\) is a Stopping Time iff  \(\PX{\dt<\infty}\). 
\end{definition}

Note that if \(\dt\) is a stopping time then \(\PX{\dum^{\dt} \in \dumS^{\dt*}}=1\).  Furthermore 
unlike \( \dumS^{\dt}\), \(\dumS^{\dt*}\) is  a countable set for all  stopping times \(\dt\) because \(|\dumS|\) is finite.\footnote{\(\dumS^{\dt*}\) is  a countable set even when   \(|\dumS|\) is countably infinite.}

\subsection{Variable Length Block Codes}\label{sec:vlbc}
A variable length block code on a  DMC is given by a random decoding time $\dt$, an encoding scheme $\ENC$ and a decoding rule $\DEC$ satisfying \(\PX{\dt<\infty}=1\).  
\begin{itemize}
\item \emph{Decoding time} $\dt$ is a Markov  time with respect to the receiver's
 observation \(\out^{\tin}\), i.e., given $\out^\tin$ receiver knows whether \(\dt=\tin\) or
 not. Hence \(\dt\) is a random quantity rather than a constant, thus neither the decoder
nor the receiver  knows the  value  of \(\dt\) \emph{a priori}.  But as time passes,  both 
the decoder and the encoder (because of feedback link) will be able to decide 
whether \(\dt\)  has been reached or not, just by considering the current and past 
channel outputs. 
\item  \emph{Encoding scheme} \(\ENC\) is  a collection of mappings which determines 
the input letter at time $(\tin+1)$ for each message in the finite  message set  $\mesS$, for each
 $\dout^{\tin}\in \outS^{\tin}$ such that $\dt>\tin$,
\begin{equation*}
  \ENC(\cdot,\dout^\tin): \mesS \rightarrow \inpS \qquad \forall \dout^{\tin}: \dt>\tin.
\end{equation*}
\item  \emph{Decoding Rule} is a mapping from the set of output sequences \(\dout^{\tin}\)
 such that \(\dt=\tin\) to the finite  message set  $\mesS$ which determines the decoded message,
  \(\est\). With a slight abuse  of notation we denote 
  the set of all, possibly infinite,  output sequences \(\dout^{\tin}\)  
  such that      \(\{\dt=\tin \mbox{~if~}\out^{\tin}=\dout^{\tin} \}\) by\footnote{See equation (\ref{eq:def-stop-dom}).} 
  \(\outS^{\dt}\)  and write the decoding rule \(\DEC\) as,
\begin{equation*}
  \DEC(\cdot): \outS^{\dt}\rightarrow \mesS.
\end{equation*}
\item Note that because of the condition  \(\PX{\dt<\infty}=1\), decoding time is not only a Markov time, but also a Stopping time.\footnote{Having  a finite decoding time  with probability one, i.e., \(\PX{\dt<\infty}=1\), does not imply having a finite expected value for the decoding time, i.e., \(\EX{\dt}<\infty\). Thus a variable length code can, in principle, have an infinite expected decoding time.}
  \end{itemize}

 At time zero the message $\mes$ chosen uniformly at random from $\mesS$ is given to the transmitter; 
the  transmitter uses the codeword associated  \(\mes\), i.e., $\ENC(\mes,\cdot)$, to convey the message
  $\mes$ until the decoding  time $\dt$. Then the receiver chooses  the decoded message \(\est\)  using its 
  observation    \(\out^{\dt}\)  and the decoding rule \(\DEC\), i.e.,  $\hat{\mes}=\DEC(\out^{\dt})$. The error 
  probability,  the rate  and the error exponent of a variable length block code are given by
 \begin{align}
\Pe&=  \PX{\est \neq \mes} & 
\rate&=\frac{\ln |\mesS|}{\EX{\dt}} & 
\ex&=\frac{-\ln  \Pe}{\EX{\dt}} .
\end{align}
Indeed one can interpret the  variable length block codes on DMCs as trees, for a more detailed discussion of this interpretation readers may go over  \cite[Section II]{bnrt}.

\subsection{Reliable Sequences for Variable Length Block Codes}\label{sec:relsec}
In order to  suppress  the secondary terms while discussing the main results,
 we use the concept of reliable sequences.
In a sequence of codes we denote the error probability and the message set of the \(\inx^{th}\) code 
of the  sequence by  \(\Pe^{(\inx)}\) and   \(\mesS^{(\inx)}\), respectively.
\begin{definition}[Reliable Sequence]
 A sequence of variable length block codes $\SC$ is reliable if  the  error probabilities of the codes  
 vanish and the size of the message sets of the codes diverge:\footnote{Recall  that the  decoding time 
 of a variable length block code is finite with probability one. Thus
 \(\PXS{\inx}{\dt^{(\inx)}<\infty }=1\)  for all \(\inx\) for a reliable sequence.}
\begin{align*}
 \lim_{\inx\rightarrow \infty} \left(\Pe^{(\inx)}+\tfrac{1}{| \mesS^{(\inx)}|}\right)
 &=0.
\end{align*}
where  \(\Pe^{(\inx)}\) and \(\mesS^{(\inx)}\) are the error probability  and the message set for the \(\inx^{th}\) code of the reliable sequence, respectively. 
\end{definition} 

Note that in a sequence of codes, each code has an associated probability space.  We denote 
the random variables in these probability spaces together with a  superscript corresponding 
to the code. For example the decoding time of the 	\(\inx^{th}\) code in the sequence is denoted 
by \(\dt^{(\inx)}\). The expected value of random variables in the probability space associated 
with the \(\inx^{th}\) code  in the sequence is denoted\footnote{Evidently it is possible to come up with
a probability space that includes all of the codes in a reliable sequence and invoke independence between
random quantities associated with different codes. We choose the current convention to emphasize
 independence explicitly in the notation we use.}  
by   \(\EXS{\inx}{\cdot}\). 
\begin{definition}[Rate of a Reliable Sequence]
  The rate of a reliable sequence $\SC$ is the limit infimum of the rates of the individual codes,
\begin{align*}
\rate_{\SC}&\DEF\liminf_{\inx\rightarrow \infty} \frac{\ln| \mesS^{(\inx)}|}{\EXS{\inx}{\dt^{(\inx)}}}.
\end{align*}
 \end{definition}
\begin{definition}[Capacity]
The capacity of a channel for variable length block codes is the supremum of the rates
 of the all reliable sequences.
\begin{align*}
\CX&\DEF \sup_{\SC} \rate_{\SC}.
\end{align*}
\end{definition} 
The capacity of a DMC for  variable length block codes is identical  to the usual channel capacity,     \cite{bur}.  
Hence,
\begin{equation}
\label{eq:CX}
\CX=\max_{\idis{} \in \PDS{\inpS}} \sum\nolimits_{\dinp,\dout} \idis{}(\dinp) \CT{\dinp}{\dout} \ln \frac{\CT{\dinp}{\dout}}{ \odis{}(\dout)} 
\end{equation}
where \(\odis{}(\dout)=\sum_{\dinp} \idis{}(\dinp) \CT{\dinp}{\dout} \).

\begin{definition}[Error Exponent of a Reliable Sequence]
  The  error exponent of a reliable sequence $\SC$ is the limit infimum  of the error exponents of the individual codes,
\begin{align*}
\ex_{\SC}&\DEF\liminf_{\inx\rightarrow \infty} \frac{-\ln \Pe^{(\inx)}}{\EXS{\inx}{\dt^{(\inx)}}}.
\end{align*}
 \end{definition}

\begin{definition}[Reliability  Function]
 The reliability function of a channel for variable length block codes at rate 
 \(\rate \in [0,\CX]\) is the  supremum of the exponents of all reliable sequences whose rate is 
 \(\rate\) or higher. 
\begin{align*}
\TEX(\rate)&\DEF\sup_{\SC:\rate_{\SC}\geq \rate} \ex_{\SC}.
\end{align*}
\end{definition}

Burnashev  \cite{bur}  analyzed the performance of variable length block codes with feedback and 
established inner and outer bounds to their performance. Results of \cite{bur}  determine the 
reliability function of  variable length block codes on DMCs for all rates. According to \cite{bur}:
\begin{itemize}
\item If all entries of $\CTM$ are positive then \footnote{Problem is formulated somewhat differently in 
\cite{bur}, as a result \cite{bur} did not deal with the case \(\EX{\dt}=\infty\). The bounds in \cite{bur} does 
not guarantee that the error probability of a variable length code with infinite expected decoding time is  
greater than zero, however this is the case if all the transition probabilities are positive.  To see that consider  
a channel  with positive  minimum transition probability  \(\mtp\), i.e.,  
\(\mtp=\min_{\dinp,\dout}\CT{\dinp}{\dout}>0\). 
In such a channel  any variable length code  satisfies 
\(\Pe\geq \tfrac{|\mesS|-1}{|\mesS|}\EX{\left(\tfrac{\mtp}{1-\mtp}\right)^{\dt}}\), then \(\Pe>0\) as  \(\mtp>0\) and 
\(\PX{\dt<\infty}=1\). Consequently  both the rate and the error exponent  are zero for variable length block 
codes with  infinite expected decoding time. A more detailed discussion of this fact can be found in 
Appendix \ref{app:infdect-a}.} 
\begin{equation*}
  \ex(\rate)=\left(1-\frac{\rate}{\CX} \right)\DX \qquad \forall \rate \in [0,\CX]
\end{equation*}
where $ \DX$ is maximum Kullback Leibler divergence between  the output distributions of any two input letters:
\begin{equation}
\label{eq:dxdef}
\DX \DEF \max_{\dinp,\tilde{\dinp} \in {\inpS}} \KLD{\CTM_{\dinp}}{\CTM_{\tilde{\dinp}}}.
\end{equation}
\item If there are one or more zero entries\footnote{Note that in this situation  $\DX=\infty$.} in $\CTM$, 
i.e., if there are two input letters $\dinp$, $\tilde{\dinp}$ and an output letter $\dout$ such that, 
$\CT{\dinp}{\dout}=0$ and $\CT{\tilde{\dinp}}{\dout}>0$, then for all $\rate<\CX$, for
 large enough $\EX{\dt}$ there are rate $\rate$ variable length block codes which are error free, i.e.,
 $\Pe=0$.
\end{itemize}
When  $\Pe=0$ all error events can have zero probability at the same time. Consequently all  the 
\uep~problems are answered trivially when there is a zero probability transition.  This is why  we have assumed  
that  $\CT{\dinp}{\dout}>0$ for all $\dinp\in {\inpS}$ and $\dout \in {\outS}$.

We denote the input letters that get this maximum value of Kullback Leibler divergence by\footnote{This
 particular naming of letters is reminiscent of the use of these letters in Yamamoto Itoh scheme \cite{itoh}. 
Although  they are named differently in \cite{itoh},  \(\xa\) is used for  accepting and \(\xr\) is used for 
 rejecting the tentative decision in  Yamamoto Itoh scheme.}  \(\xa \) and \(\xr\):
\begin{equation}
  \label{eq:dm}
  \DX= \KLD{\CTM_{\xa}}{\CTM_{\xr}}.
\end{equation}

\section{Problem Statement and Main Results}\label{sec:psres}
\subsection{Problem Statement}\label{sec:ps}
For each $\dmes\in \mesS$, the conditional error probability is defined as,\footnote{Later in the paper we  
consider block codes with erasures. The conditional error  probabilities, \(\Pem{\dmes}\) for
 \(\dmes \in \mesS\),  are defined slightly differently for them, see equation (\ref{eq:defpe-feec}).}
\begin{equation}
\label{eq:n:defpem}
\Pem{\dmes}\DEF \PCX{\est \neq \mes}{\mes=\dmes}.
\end{equation}
In the conventional setting we are interested in either the average  or the maximum of the conditional 
error probability  of the messages.  The behavior of  the minimum conditional error probability is 
scarcely investigated.  Single message \emph{message-wise} \uep~problem attempts to answer  that
 question by  determining the trade-off between exponential decay rates of  \(\Pe\) and  
\(\min_{ \dmes \in \mesS} \Pem{\dmes}\).  The operational definition of the problem in terms of 
reliable sequences is as follows.
 \begin{definition}[Single Message \emph{Message-wise} \uep~Problem]
 \label{def:singlemes}
For any reliable sequence $\SC$ the missed detection exponent of the reliable sequence \(\SC\) is defined as
\begin{equation}
  {\Emd}_{,\SC}=\liminf_{\inx \rightarrow \infty} \frac{-\ln  
 \min\nolimits_{ \dmes \in \mesS^{(\inx)}} \Pem{\dmes}^{(\inx)}
 }{\EXS{\inx}{\dt^{(\inx)}}}
\end{equation}
where \(\Pem{\dmes}^{(\inx)}\) is  the conditional error probability of the message \(\dmes\)  for
 the  \(\inx^{th}\) code  of the reliable sequence \(\SC\).
 
For any rate \(\rate\in [0,\CX]\) and  error exponent\footnote{Burnashev's expression for error exponent  of 
variable length block codes is used explicitly  in the definition because we know, as a result of \cite{bur}, that 
the error exponents of all reliable sequences are upper  bounded by Burnashev's exponent. An alternative
 definition oblivious to Burnashev's result can simply define  \(\Emd(\rate, \ex)\) for all \revs~ that are 
achievable. That definition is equivalent to Definition \ref{def:singlemes}, because 
of  \cite{bur}.}
\(\ex\in [0,(1-\tfrac{\rate}{\CX})\DX]\), the missed detection exponent  $\Emd(\rate, \ex)$ is defined as,
\begin{equation}
\label{eq:def:singlemes}
    \Emd(\rate, \ex)\DEF\sup_{\SC: \substack{\rate_{\SC} \geq \rate \\ \ex_{\SC}  \geq \ex }} \Emd_{,\SC}.
\end{equation}
  \end{definition}

In  variable length block  codes with feedback, the single message \emph{message wise} \uep~problem 
not only answers a curious question about the  decay rate of the minimum conditional error probability
 of a code, but also  plays a key role  in  the \emph{bit-wise} \uep~problem, which is our main 
 focus in this manuscript.

Though they are central in the \emph{message-wise} \uep~problems, the conditional error 
probabilities of the messages are  not  relevant  in  the \emph{bit-wise} \uep~problems. In 
the \emph{bit-wise} \uep~problems  we analyze the error probabilities of groups of  sub-messages.  
In order to do that,   consider   a code with a message set \(\mesS\) of the form 
\vspace{-0.3cm}
\begin{align*}
\mesS&=\mesS_{1} \times \mesS_{2} \times \ldots \times \mesS_{\nlay} 
\end{align*}
Then the transmitted  message  \(\mes\) and decoded message \(\est\) of the code are of the form
\begin{align*}
\mes&=(\mes_1,\mes_2,\ldots,\mes_{\nlay})\\
\est&=(\est_1,\est_2,\ldots,\est_{\nlay})
\end{align*}
 where  \(\mes_j,\est_j\in\mesS_j\) for all \(j=1,2,\ldots, \nlay\).   Furthermore  \(\mes_j\) and \(\est_j \) 
 are called \(j^{th}\) transmitted sub-message  and \(j^{th}\) decoded sub-message,  respectively.
 
 The error probabilities we are interested in correspond to erroneous transmission of certain parts of
 the message. In order to define them succinctly let us define \( \mesS^{j}\), \(\mes^{j}\) and
\(\est^{j}\) for all \(j\) between one and \(\nlay\) as follows:
\vspace{-0.3cm}
\begin{align*}
\mesS^{j}
&\DEF\mesS_{1} \times \mesS_{2} \times \ldots \times \mesS_{j}  \\
\mes^{j}
&\DEF(\mes_1,\mes_2,\ldots,\mes_{j})\\
\est^{j}
&\DEF(\est_1,\est_2,\ldots,\est_{j}).
\end{align*}
Then \(\Peb{j}\) is defined\footnote{Similar to  the conditional error  probabilities, \(\Pem{\dmes}\)'s  
  for  \(\dmes \in \mesS\), error probabilities of sub-messages, \(\Peb{j}\)'s for \(j=1,2,\ldots, \nlay\), are 
  defined  slightly differently for  codes with erasures, see equation (\ref{eq:peb-def-we}).}   
  as the probability of the    event that \(\est^j \neq \mes^j\)
  \begin{equation}
\label{eq:peb-def}
\Peb{j}\DEF \PX{\est^j\neq \mes^j} \qquad \mbox{for~}j=1,2,\ldots, \nlay.
\end{equation}
Note that if \(\est^{j}\neq\mes^{j}\) then \(\est^{i}\neq\mes^{i}\) for all \(i\) greater than \(j\). Thus 
\begin{equation}
\label{eq:assume}
  \Peb{1}\leq   \Peb{2}\leq   \Peb{3}\leq \ldots \leq   \Peb{ \nlay}.
\end{equation}

\begin{definition}[\emph{Bit-wise} \uep~Problem For  Fixed \( \nlay\)]\label{def:bits}
 For any positive integer \(\nlay \) let \( \SC \) be a reliable sequence  whose message sets 
\(\mesS^{(\inx)} \)  are of the form
\(\mesS^{(\inx)}=\mesS_{1}^{(\inx)} \times \mesS_{2}^{(\inx)} \times \ldots \times \mesS_{ \nlay}^{(\inx)}\).
Then  the entries of the rate vector \(\vec{\rate}_{\SC}\) and the error exponent vector \(\vec{\ex}_{\SC}\) are defined as
\begin{align*}
 \rate_{\SC,j}&\DEF\liminf_{\inx \rightarrow \infty} \frac{\ln |\mesS_{j}^{(\inx)}|}{\EXS{\inx}{\dt^{(\inx)}}} &&\forall  j \in\{1,2,\ldots, \nlay\}\\
 \ex_{\SC,j}&\DEF\liminf_{\inx \rightarrow \infty} \frac{-\ln  \Peb{j}^{(\inx)} }{\EXS{\inx}{\dt^{(\inx)}}} &&\forall j \in\{1,2,\ldots, \nlay\}.
\end{align*}
A \rev~ \((\vec{\rate}, \vec{\ex})\) is achievable if and only if
there exists  a reliable sequence \(\SC\) such that  
\((\vec{\rate}, \vec{\ex})=(\vec{\rate}_{\SC}, \vec{\ex}_{\SC})\).
\end{definition}
This definition of the  \emph{bit-wise} \uep~ problem is slightly different than the one described in
 the introduction, because \(\Peb{j}\) is  defined as  \(\PX{\est^j\neq \mes^j} \) rather than 
 \(\PX{\est_j\neq \mes_j}\). 
Note that if \(\est_j\neq \mes_j\) then \(\est^j\neq \mes^j\); consequently    
\( \PX{\est^j\neq \mes^j}  \geq \PX{\est_j\neq \mes_j}\)
for all \(j\)'s. In addition, if we  assume without  loss of generality that  
\(\PX{\est_j\neq\!\mes_j}\!\geq\!\PX{\est_i\neq \mes_i}\) for all \(j\!\geq\!i\),  the  union bound 
implies that  \(\PX{\est^j\!\neq\!\mes^j}\!\leq\!j\PX{\est_j\!\neq\!\mes_j}\).
Thus for the case when \( \nlay\) is fixed, both formulations of the problem result in exactly the same 
achievable region of \revs.

The achievable region of  \revs~ could have been defined as the 
closure of the points of the form \((\vec{\rate}_{\SC}, \vec{\ex}_{\SC})\) for some reliable 
sequence \(\SC\). Using the definition of \((\vec{\rate}_{\SC}, \vec{\ex}_{\SC})\)'s  one can easily show
that, in this case too both definitions  result in exactly the same achievable region of \revs.

\subsection{Main Results}\label{sec:res}

For variable length block codes with feedback, the  results of both the single message  
\emph{message-wise} \uep~problem  and the \emph{bit-wise} \uep~problem  are given in  terms 
of the \(\JX{\rate}\)  function  defined below.  The  \(\JX{\rate}\) function is first introduced 
by\footnote{In \cite[equation (2.6)]{kud}  there is no optimization over the  parameter \(\tsc\). 
Thus strictly speaking,  what is introduced in  \cite[equation (2.6)]{kud}  is \(\jx{\rate}\) given in 
equation (\ref{eq:fx}) rather than \(\JX{\rate}\) given in (\ref{eq:FX}).}  
 Kudryashov \cite[equation (2.6)]{kud} while describing the performance of  non-block variable length 
 codes with feedback and  delay constraints.  Later the \(\JX{\rate}\) function is used 
 in \cite{bnz} for describing the performance of block codes in  single message 
 \emph{message-wise} \uep~problem. It is shown in \cite[Appendix D]{bnz} that for both fixed length 
 block codes without feedback  and variable length block codes with feedback on DMCs satisfy,
\begin{equation}
\label{eq:equida}
\Emd(\rate,0)=\JX{\rate}.
\end{equation}
Recently  Nazer, Shkel and Draper  obtained the closed form expression for 
\(\Emd(\rate,0)\) for fixed length block codes on  the Additive  White  Gaussian Noise channel, 
under certain average and peak power  constraints \cite[Theorem 1]{NSD}. 
Curiously equality given in (\ref{eq:equida}) 
holds in that case too.\footnote{Unlike DMC for  these channels it is possible to obtain a closed form 
expression in terms of the rate and the power  constraints.}

\begin{figure}[t]
\hspace{-.5cm}
\includegraphics[scale=.42]{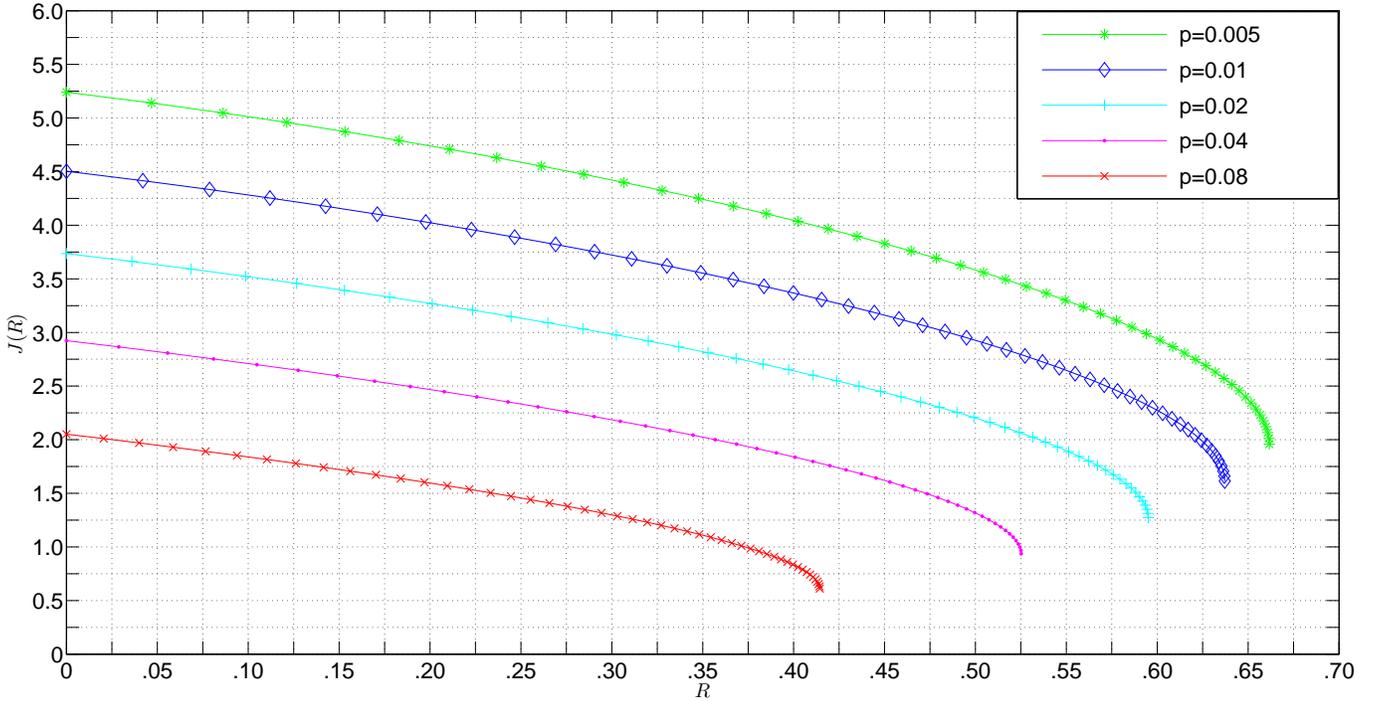}
~\vspace{-.4cm}
\caption{\label{fig:fig_J}
The \(\JX{\rate}\) function is drawn for Binary Symmetric  Channels  (BSCs) 
with cross over probabilities \(p \in \left\{0.005, 0.01, 0.02,0.04,0.08\right\}\).}
~\vspace{-.3cm}
\end{figure}
\begin{definition}
For any \(\rate \in [-\infty,\CX]\), \(\JX{\rate}\) is defined as
\begin{align}
\label{eq:FX}
 \JX{\rate}&   \DEF
 \max_{\tsc,\dinp_1,\dinp_2,\idis{1},\idis{2}:\substack{0\leq\tsc \leq 1\\
\dinp_1,\dinp_2 \in \inpS\\ \idis{1}, \idis{2} \in \PDS{\inpS}\\     \tsc \MI{\idis{1}}{\CTM}+(1-\tsc) \MI{\idis{2}}{\CTM}\geq \rate }} \tsc  \KLD{\odis{1}}{\CTM_{\dinp_1}}+(1-\tsc)  \KLD{\odis{2}}{\CTM_{\dinp_2}}
\end{align}
where \(\odis{i}(\dout)=\sum_{\dinp} \CT{\dinp}{\dout}  \idis{i}(\dinp)\) for \(i=1,2\).
\end{definition}

We have plotted the \(\JX{\rate}\) function for Binary Symmetric  Channels\footnote{Recall that in a 
binary symmetric channel with crossover probability probability \(p\),  \(\inpS=\left\{0,1\right\}\),   
\(\outS=\left\{0,1\right\}\) and  \(\CT{\dinp}{\dout}=(1-p)\IND{\dinp=\dout}+p\IND{\dinp\neq\dout}\).}  
(BSCs)  with various cross over probabilities  in  Figure \ref{fig:fig_J}.  Note  that as the channel becomes 
noisier, i.e., as the crossover probability becomes  closer to \(1/2\), the value of \(\JX{\rate}\) function 
decreases  at all values of rate where it is positive.   Furthermore  the highest value of rate where it is positive, i.e.,  the 
channel capacity, decreases.

\begin{lemma}
 \label{lem:fx}
The  function  \(\JX{\rate}\) defined in equation (\ref{eq:FX}) is a concave, decreasing function 
such that \(\JX{\rate}=\DX\) for \(\rate\leq 0\).   
\end{lemma}

Proof of Lemma \ref{lem:fx} is given in Appendix \ref{sec:appfx}. 

 Now let us  consider the singe message \emph{message-wise} \uep~problem given in 
Definition \ref{def:singlemes}.

\begin{figure}[t]
\hspace{-.5cm}
\includegraphics[scale=.42]{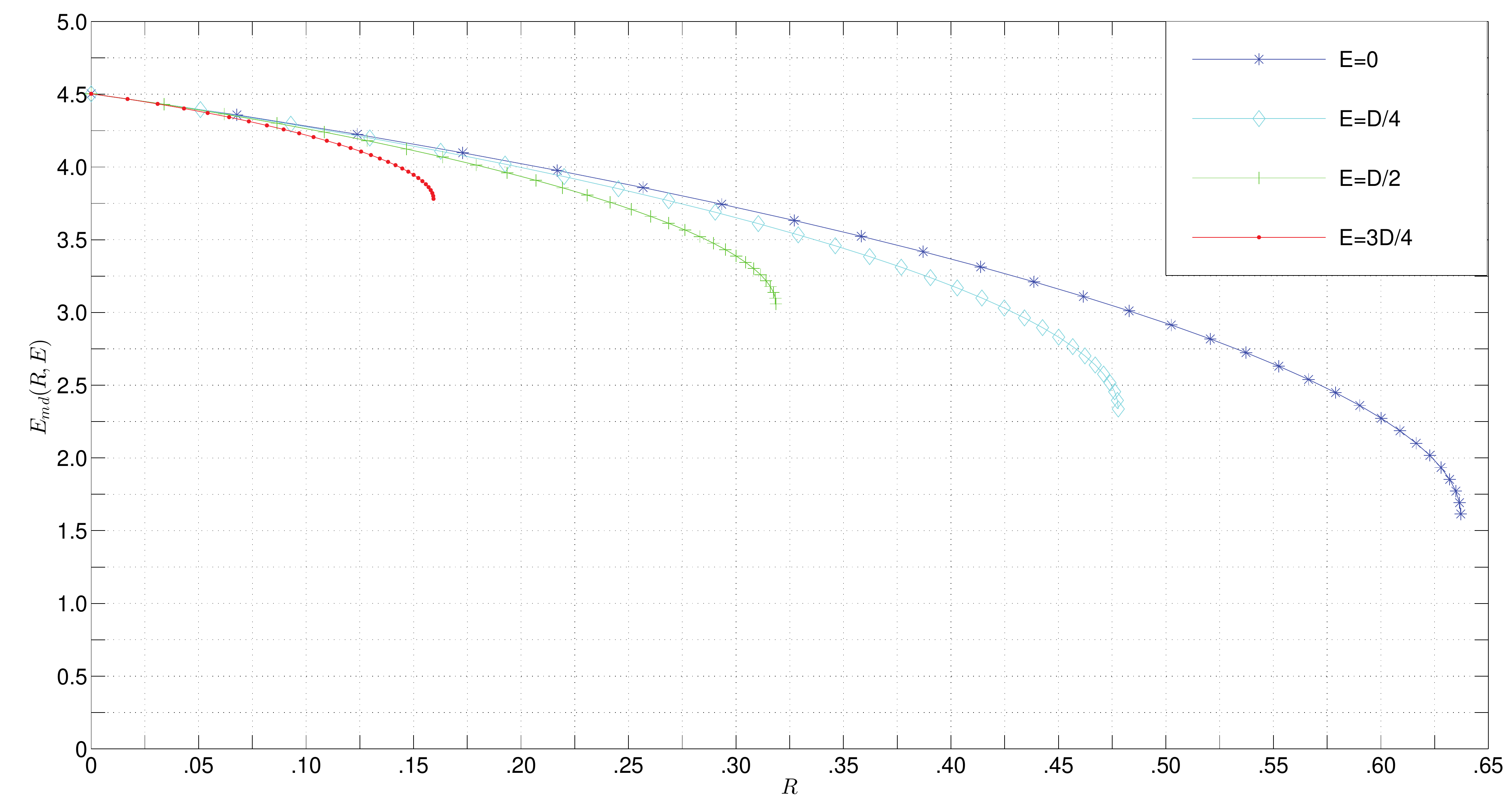}
\vspace{-.4cm}
\caption{\label{fig:fig_EMD} \(\Emd(\rate, \ex)\) is drawn at various values of the error exponent 
\(\ex\) as a function of rate \(\rate\) for 	 a BSC with crossover probability 
\(p=0.01\). Note that when \(p=0.01\), \(\CX=0.6371\) Nats per channel use and 
\(\DX=4.503\). As we increase the exponent of the average error probability, i.e., \(\ex\), the 
value of \(\Emd(\rate, \ex)\) decreases, as one would expect.}
\vspace{-.3cm}
\end{figure}
\begin{theorem}
\label{thm:md}
 For any rate $0\leq \rate \leq \CX$ and  error exponent $\ex\leq (1-\tfrac{\rate}{\CX})\DX$ the missed 
 detection exponent $\Emd(\rate, \ex)$ defined in equation (\ref{eq:def:singlemes})
 is equal to\footnote{For the case when \(\rate=0\) and \(\ex=\DX\) the 
\(\left(1-\tfrac{\ex}{\DX} \right) \JX{\tfrac{\rate}{1-\frac{\ex}{\DX}}} \)
   term should be interpreted as \(0\), i.e., 
 \(\left.\left(1-\tfrac{\ex}{\DX} \right) \JX{\tfrac{\rate}{1-\frac{\ex}{\DX}}} \right\vert_{\substack{\rate=0\\\ex=\DX}}=0\).}
\begin{equation}
  \Emd(\rate, \ex) = \ex+\left(1-\tfrac{\ex}{\DX} \right) \JX{\tfrac{\rate}{1-\frac{\ex}{\DX}}}
\end{equation}
 where \(\CX\), \(\DX\) and \(\JX{\cdot}\)  are given in equations  (\ref{eq:CX}),  (\ref{eq:dxdef}) 
 and (\ref{eq:FX}), respectively.
Furthermore \(\Emd(\rate, \ex)\) is jointly concave in \((\rate,\ex)\) pairs.
\end{theorem}
We have plotted \(\Emd(\rate, \ex)\) as a function of rate, for various values of \(\ex\) in Figure 
\ref{fig:fig_EMD}. When rate is  zero, the exponent of the average error probability can be made 
as high as \(\DX\). Thus all the curves meet at \((0,\DX)\) point. But for all positive rates the
 exponent of the average error probability makes a difference; as  \(\ex\) increases 
\(\Emd(\rate,\ex)\) decreases. Furthermore for any given rate \(\rate\) the exponent of the 
average error probability can only be as high as \((1-\tfrac{\rate}{\CX})\DX\). This is why
the curves corresponding  to higher values of 	\(\ex\) have smaller support on rate axis.

Proof of Theorem \ref{thm:md} is presented in Appendix \ref{sec:appmd}.

Similar to the single message \emph{message-wise} \uep~problem, the solution of 
the \emph{bit-wise} \uep~problem is given in terms of the \(\JX{\rate}\) function. 
\begin{theorem}
\label{thm:bits}
A \rev~  \((\vec{\rate},\vec{\ex})\)  is achievable if and only 
if there exists a \(\vec{\fr}\)  such that,\footnote{For the case when \(\rate_j=0\) and \(\fr_j=0\) 
the \(\fr_j\JX{\tfrac{\rate_j}{\fr_j}}\) term should be interpreted as \(0\), 
i.e., \(\left. \fr_j \JX{\tfrac{\rate_j}{\fr_j}}\right\vert_{\substack{\rate_j=0  \\ \fr_j=0}}=0\).}
 \begin{subequations}
\label{eq:bits}
   \begin{align}
\label{eq:bitsa}
\ex_{i} &\leq (1-\sum\nolimits_{j=1}^{ \nlay}\fr_j) \DX + \sum\nolimits_{j=i+1}^{ \nlay} \fr_j \JX{\tfrac{\rate_j}
{\fr_j}} && \forall i\in \{1,2,\ldots, \nlay\}\\
\label{eq:bitsb}
\rate_i &\leq \CX \fr_i && \forall i\in \{1,2,\ldots, \nlay\}\\
\label{eq:bitsbc}
\fr_i &\geq 0 && \forall i\in \{1,2,\ldots, \nlay\}\\
\label{eq:bitsc}
\sum\nolimits_{j=1}^{ \nlay}\fr_j&\leq 1     &&
   \end{align}
 \end{subequations}
 where \(\CX\), \(\DX\) and \(\JX{\cdot}\)  are given in equations  (\ref{eq:CX}),  (\ref{eq:dxdef}) and (\ref{eq:FX}), respectively.
 Furthermore the set of all achievable \revs~ is  convex. 
 \end{theorem}
Proof of Theorem \ref{thm:bits} is presented in Appendix \ref{sec:appbits}.

For the special case when there are only two sub-messages the condition  given in 
Theorem \ref{thm:bits} for the achievablity  of  a rate vector  error exponent vector pair 
can be turned  into an analytical expression for the optimal \(\ex_1\) in terms of 
\(\rate_1\), \(\rate_2\) and \(\ex_2\). In order to see why, note that revealing the region of achievable  
\((\rate_1,\rate_2,\ex_1,\ex_2)\)  vectors is equivalent to revealing the region of achievable 
\((\rate_1,\rate_2,\ex_2)\)'s and the value of the maximum achievable \(\ex_1\)  for all
 the \((\rate_1,\rate_2,\ex_2)\)'s in the achievable region.

\begin{figure}[t]
\hspace{-.4cm}
\includegraphics[scale=.42]{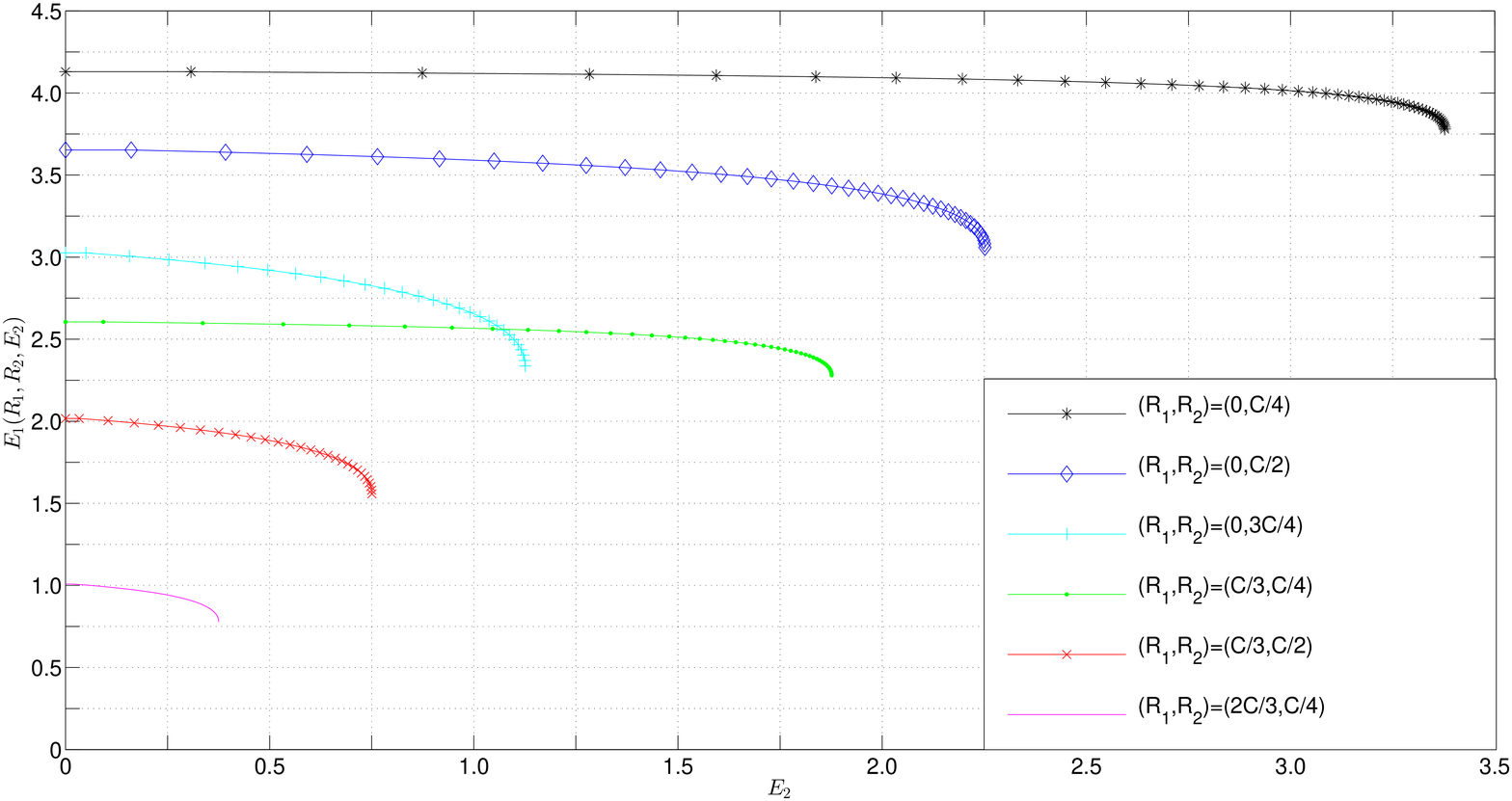}
\vspace{-.4cm}
\caption{\label{fig:fig_EB}\(\TEX_1 (\rate_1,\rate_2,\ex_2)\) is drawn for various  values rate pairs 
\((\rate_1,\rate_2)\) as a function error exponent \(\ex_2\) for  a BSC with 
crossover probability \(p=0.01\). Recall that when \(p=0.01\), \(\CX=0.6371\) Nats 
per channel use and \(\DX=4.503\).}
\vspace{-.6cm}
\end{figure}

\begin{corollary}
\label{cor:two}
For any rate pair $(\rate_1,\rate_2)$ such that $\rate_1+\rate_2 \leq \CX$ and  error exponent 
$\ex_2$ such that $\ex_2\leq (1-\tfrac{\rate_1+\rate_2}{\CX})\DX$, the  optimal value of $\ex_1$ 
is given  by\footnote{For the case when \(\rate_2=0\) and \(\ex_2=(1-\tfrac{\rate_1}{\CX})\DX\),
 the second term on the right hand side of equation (\ref{eq:alt-thm-x}) should be interpreted as zero, 
i.e., 
\(\left. \left(1-\tfrac{\rate_1}{\CX}-\tfrac{\ex_2}{\DX} \right) \JX{\tfrac{\rate_2}{1-\frac{\rate_1}{\CX} -\frac{\ex_2}{\DX}}} \right\vert_{\substack{\rate_2=0,~  \ex_2=(1-\tfrac{\rate_1}{\CX})\DX }}=0\) }
\begin{equation}
\label{eq:alt-thm-x}
\TEX_1 (\rate_1,\rate_2,\ex_2) = \ex_2+
\left(1-\tfrac{\rate_1}{\CX}-\tfrac{\ex_2}{\DX} \right) \JX{\tfrac{\rate_2}{1-\frac{\rate_1}{\CX} -\frac{\ex_2}{\DX}}}
\end{equation}
 where \(\CX\), \(\DX\) and \(\JX{\cdot}\)  are given in equations  (\ref{eq:CX}),  (\ref{eq:dxdef}) and (\ref{eq:FX}), respectively.
 Furthermore \(\TEX_1 (\rate_1,\rate_2,\ex_2)\) is concave in \((\rate_1,\rate_2,\ex_2)\).
\end{corollary}
Note that for the \(\TEX_1(\rate_1,\rate_2,\ex_2)\) given in equation (\ref{eq:alt-thm-x}),    \(\TEX_1(\rate_1,\rate_2,\ex_2)\geq \ex_2\) for all \((\rate_1,\rate_2,\ex_2)\) triples such that  \(\rate_1+\rate_2 \leq \CX\)   and  \(\ex_2\leq (1-\tfrac{\rate_1+\rate_2}{\CX})\DX\). Furthermore inequality is strict as long as \(\rate_2>0\). We have drawn \(\TEX_1(\rate_1,\rate_2,\ex_2)\) for various \((\rate_1,\rate_2)\) pairs as a function of \(\ex_2\) in Figure \ref{fig:fig_EB}.

\section{Achievablity}\label{sec:ach}
In both the single message \emph{message-wise} \uep~problem and the \emph{bit-wise}
 \uep~problem, the  codes that achieve the optimal performance employ a number of
  different ideas at the same time.  In order to avoid introducing all of those ideas at once, we
   first describe two families of codes and analyze the probabilities of various error events in
    those two families of codes. Later we use those two families of codes as the building blocks for the 
    codes that achieve the optimal performance in the \uep~problems we are interested in.  
    Before going into a more detailed description and analysis of those codes let us first give 
    a birds eye plan for this section.  
  \begin{enumerate}[(a)]
\item {\it A Single Message \emph{Message-wise}  \uep~Scheme without Feedback:} First
in Section \ref{sec:ach-smwof}, we  consider a family of fixed length codes without
 feedback. We prove that these codes can achieve any rate \(\rate\) less than channel
  capacity, with vanishing\footnote{Vanishing with increasing block length.} error probability 
  \(\Pe\) while having a minimum conditional error probability, 
  \(\min_{\dmes} \Pem{\dmes}\),   as low as \( e^{-\blx\JX{\rate}}\).  
The main drawback of this family of codes is that  the decay rate of the average error 
probability \(\Pe\) has to be subexponential in this  family of codes.

\item {\it Control Phase and Error-Erasure Decoding:} In   Section \ref{sec:ach-cont}  in order 
to obtain non-zero
 exponential decay for the average error probability, we use a method introduced by 
Yamamoto and Itoh in \cite{itoh}. We append the fixed length codes described in  
Section \ref{sec:ach-smwof}  with a control phase and use an error-erasure 
decoder.
%\footnote{\bf First a  code from the family of codes described in  Section \ref{sec:ach-smwof}   is used to transmit \(\mes\) and the receiver makes a tentative  decision \(\test\)  using its decoder. Transmitter knows  \(\test\)  because of the feedback link and  in the remaining time   instances, i.e., in the control   phase,  it transmits   the input letter \(\xa\), described in  equation     (\ref{eq:dm}),  if \(\test=\mes\).   If \(\test\neq\mes\) the input letter \(\xr\) is transmitted  instead. At the end of   the control phase, receiver checks whether the output sequence in the control phase is   typical with   \(\CTM_{\xa}\), if it is then \(\est=\test\) otherwise an erasure is   declared. }  
 This new family of codes with  control phase and error-erasure decoding  are     shown, in   Section \ref{sec:ach-cont}, to achieve any rate \(\rate\) less than the  channel capacity \(\CX\) with  exponentially decaying average error probability \(\Pe\),     exponentially decaying  minimum conditional error probability \(\min_{\dmes} \Pem{\dmes}\)        and        vanishing  erasure probability, \(\Per\). 
 
\item {\it Single Message \emph{Message-wise} \uep~for Variable Length Codes:} In Section \ref{sec:achsa} 
we obtain variable length codes for single message \emph{message-wise} \uep~problem using the codes 
described in   Section \ref{sec:ach-cont}. In order to do that we use the fixed length codes with feedback and 
erasures  described in   Section \ref{sec:ach-cont}, repetitively until a non-erasure decoding happens. This idea
 too,  was employed by Yamamoto and Itoh in \cite{itoh}. 

 \item {\it Bit-wise \uep~for Variable Length Codes:} In Section \ref{sec:achsb} we first use  the codes described in 
 Section \ref{sec:ach-smwof}  and the control phase discussed in Section \ref{sec:ach-cont}
  to obtain a family of  fixed length codes with feedback and erasures which has   
  \emph{bit-wise} \uep,  i.e., which has different bounds on error probabilities for 
  different sub-messages.  While using  the codes described in  Section \ref{sec:ach-smwof} 
  we employ an implicit acceptance explicit rejection  scheme first introduced  in
  \cite{kud}  by Kudrayshov.
%  \footnote{\bf The \emph{bit-wise} \uep~scheme with erasures  described in   Section \ref{sec:achsb} has three phases when there are two sub-messages,    \(\mes_1\)  and \(\mes_2\).  In the first phase \(\mes_1\) is transmitted using  the     \emph{ordinary  messages} of  a code like the one described in Section      \ref{sec:ach-smwof}.     In the second phase  \(\mes_2\) is transmitted  using the \emph{ordinary messages} of  a     code like the one described in Section \ref{sec:ach-smwof}     if the tentative decision   \(\test_1\) is correct,  the \emph{special message} is transmitted otherwise.   Throughout the the third phase  the input letter \(\xa\), described in  equation    (\ref{eq:dm}), is transmitted  if both of the tentative decisions are correct, the input letter     \(\xr\) is transmitted otherwise.      If the output distribution in the control phase is typical with \(\CTM_{\xa}\) and if neither of     the tentative decisions is equal to the \emph{special message} of the corresponding codes     then the tentative decisions become final i.e., \(\est_1=\test_1\) and     \(\est_2=\test_2\) , otherwise an erasure is declared. Note that during the second phase erroneous transmission of \(\mes_1\) is conveyed using  the \emph{special message}, hence the non-existence of the \emph{special message} is a  tacit approval of the tentative decision. This is why this scheme is called a implicit  acceptance explicit rejection scheme.} 
  Once we obtain a  fixed length code with erasures and \emph{bit-wise} \uep, we use a repeat at erasures scheme like the one described in Section \ref{sec:achsa}  to obtain a variable length code with \emph{bit-wise} \uep.
  \end{enumerate}
The achievablity results we derive in this section are revealed to be the optimal ones,
 in terms of the decay rates of error probabilities with expected decoding time $\EX{\dt}$,
 as a result of the  outer bounds we derive in Section \ref{sec:con}.

\subsection{A Single Message \emph{Message-wise}  \uep~Scheme without Feedback}\label{sec:ach-smwof}
In this subsection we describe a family of fixed length block codes without feedback  that 
achieves any rate \(\rate\) less then capacity with small error probability while having an 
exponentially small \(\min_{\dmes} \Pem{\dmes}\), for sufficiently large  block length \(\blx\). We
 describe these codes in terms of a time sharing constant \(\tsc\in [0,1]\), two
input letters \(\dinp_1,\dinp_2\in \inpS\) and two probability distributions on the input 
alphabet, \(\idis{1},\idis{2} \in \PDS{\inpS}\). 

In order to point out that certain sequence of input letters is a codeword or part of a 
codeword for message \(\dmes\), we put \((\dmes)\) after it. Hence we denote the codeword 
for \(\dmes\) by \(\dinp^{\blx}(\dmes)\) in a given  code and  by \(\inp^{\blx}(\dmes)\)   in a 
code ensemble, as a random quantity. 

Let us start with describing the encoding scheme. The codeword of the first message, i.e., 
\(\dinp^{\blx}(1)\), is  \(\dinp_1\) in first \(\blx_{\tsc}=\lfloor \tsc \blx \rfloor\) time instances  
and  \(\dinp_2\) in the rest,  i.e.,  \(\dinp_{\tin}(1)=\dinp_1\) for \(\tin=1,\ldots,\blx_{\tsc}\) and 
\(\dinp_{\tin}(1)=\dinp_2\) for \(\tin=(\blx_{\tsc}+1),\ldots,\blx\).  The codewords of the other messages
 are described via a random coding argument. In the ensemble of codes we are considering  
 all  entries of all codewords other than the first codeword, i.e.,
  \(\inp_{\tin}(\dmes)~ \forall \tin \in [1,\blx], \forall \dmes\neq 1\),  are generated independently 
  of other codewords and other entries of the same codeword.  In the first \(\blx_{\tsc}\) time 
  instances \(\inp_{\tin}(\dmes)\) is generated using \(\idis{1}\), in the rest using \(\idis{2}\), i.e., 
  \(\PX{\inp_{\tin}(\dmes)=\dinp}=\idis{1}(\dinp)\) for \(\tin=1,\ldots,\blx_{\tsc}\) and \(\PX{\inp_{\tin}(\dmes)=\dinp}=\idis{2}(\dinp)\) for \(\tin=(\blx_{\tsc}+1),\ldots,\blx\).

 Let us begin the description of the decoding scheme, by specifying the decoding region of 
 the first message \(\decs{1}\): it   is the set of all output sequences \(\dout^{\blx}\)  whose 
 the empirical distribution is  not typical with \((\tsc,\odis{1},\odis{2})\). More precisely, the 
 decoding region of the first message, \(\decs{1}\), is given by,
\begin{equation}
\label{eq:dec-reg-oo}
\decs{1}=\left\{\dout^{\blx}:
\blx_{\tsc}  \TV{\emp{\dout_1^{\blx_{\tsc}}}}{\odis{1}}+ (\blx-\blx_{\tsc}) \TV{\emp{\dout_{\blx_{\tsc}+1}^{\blx}}}{\odis{2}} \geq
|\inpS| |\outS|\sqrt{\blx \ln(1+\blx)}
 \right\}
\end{equation}
where  \(\TVO\) is  the total variation distance defined in  equation (\ref{eq:deftv}),
\(\emp{\dout_1^{\blx_{\tsc}}}\) and \(\emp{\dout_{\blx_{\tsc}+1}^{\blx}}\) are the empirical 
distributions of $\dout_1^{\blx_{\tsc}}$ and  \(\dout_{\blx_{\tsc}+1}^{\blx}\) defined in 
 equation (\ref{eq:defemp}) and    \(\odis{1}\) and \(\odis{2}\)  are probability distributions   
on \(\outS\), i.e.,  \(\odis{1}, \odis{2} \in \PDS{\outS}\),  such that 
\(\odis{i}(\dout)=\sum_{\dinp} \idis{i}(\dinp) \CT{\dinp}{\dout} \). 

 For other messages, \(\dmes\neq 1\), decoding regions  \(\decs{\dmes}\) are the set of all output 
sequences for which 
\(\emp{\dinp^{\blx}(\dmes),\dout^{\blx}}\) is typical with  
\((\tsc,\idis{1}\CTM,\idis{2}\CTM)\) and 
\(\emp{\dinp^{\blx}(\widetilde{\dmes}),\dout^{\blx}}\)  is not  typical with 
\((\tsc,\idis{1}\CTM,\idis{2}\CTM)\) for any \(\widetilde{\dmes} \neq \dmes\).
To be precise the decoding region of the messages other than the first  message  are
\begin{equation}
\label{eq:DECREGO}
\decs{\dmes}=
\decp{\dinp^{\blx}(\dmes)} \bigcap \left( \cap_{\widetilde{\dmes}\neq \dmes} \overline{\decp{\dinp^{\blx}(\widetilde{\dmes})}}  \right)
\qquad \forall \dmes \in \{2,3,\ldots,|\mesS|\}
\end{equation}
where for all $\dinp^{\blx} \in \inpS^{\blx}$,  
$\decp{\dinp^{\blx}}$  is the set of all $\dout^{\blx}$'s for which  $(\dinp^{\blx},\dout^{\blx})$ is typical  with $(\tsc,\idis{1}\CTM,\idis{2}\CTM)$:
\begin{equation}
\label{eq:dec-preg-r}
    \decp{\dinp^{\blx}}
=\{\dout^{\blx}:
\blx_{\tsc} 
\TV{\emp{\dinp_{1}^{\blx_{\tsc}},\dout_{1}^{\blx_{\tsc}}}}{\idis{1}\CTM}
+(\blx-\blx_{\tsc})
\TV{\emp{\dinp_{\blx_{\tsc}+1}^{\blx},\dout_{\blx_{\tsc}+1}^{\blx}}}{\idis{2}\CTM}
\leq |\inpS| |\outS|\sqrt{\blx \ln(1+\blx)}\}
\end{equation}
where  \(\TVO\)   is  the total variation distance defined in  equation (\ref{eq:deftv}), 
\(\emp{\dinp_{1}^{\blx_{\tsc}},\dout_{1}^{\blx_{\tsc}}}\) and
\(\emp{\dinp_{\blx_{\tsc}+1}^{\blx},\dout_{\blx_{\tsc}+1}^{\blx}}\)
 are the empirical distributions of \((\dinp_{1}^{\blx_{\tsc}},\dout_{1}^{\blx_{\tsc}})\) and
\((\dinp_{\blx_{\tsc}+1}^{\blx},\dout_{\blx_{\tsc}+1}^{\blx})\) 
  defined in equation (\ref{eq:defemp})  and 
  \(\idis{1}\CTM\) and \(\idis{2}\CTM\)  are probability distributions 
 on \(\inpS \times \outS\), i.e.,  \(\idis{1}\CTM \in \PDS{\inpS\times\outS}\) and
 \(\idis{2}\CTM \in \PDS{\inpS\times\outS}\). 

In Appendix \ref{sec:appach1} we have analyzed the conditional error probabilities, \(\Pem{\dmes}\) for the above described code and proved  Lemma \ref{lem:ach1} given below. 

\begin{lemma}
\label{lem:ach1}
For any block length $\blx$,   time sharing constant $\tsc\in[0,1]$, input letters \(\dinp_1,\dinp_2\in \inpS\) and input  distributions  \(\idis{1}, \idis{2} \in \PDS{\inpS}\)  there exists a length  \(\blx\)  block code such that
\begin{align*}
|\mesS|&\geq e^{\blx (\tsc \MI{\idis{1}}{\CTM} +(1-\tsc) \MI{\idis{2}}{\CTM}-\epsq{\blx})}
&& \\
 \Pem{1}& \leq  e^{-\blx (\tsc  \KLD{\odis{1}}{\CTM_{\dinp_1}}+(1-\tsc)  \KLD{\odis{2}}{\CTM_{\dinp_2}}-\epsq{\blx})}
&&\\
  \Pem{\dmes} &\leq \epsq{\blx}
&&  \dmes=2,3,\ldots, |\mesS|
\end{align*}
where \(\odis{1}(\dout)=\sum_{\dinp} \CT{\dinp}{\dout}  \idis{1}(\dinp)\), \(\odis{2}(\dout)=\sum_{\dinp} \CT{\dinp}{\dout}  \idis{2}(\dinp)\)
and  \(\epsq{\blx}=\tfrac{10|{\inpS}| |{\outS}|(\ln\frac{e}{\mtp}) \sqrt{\ln (1+\blx)}}{ \sqrt{\blx}}\).
\end{lemma}

Given the channel  $\CTM$, if we discard the error terms \(\epsq{\blx}\), for a given value of 
rate, $0\leq \rate\leq \CX $, we can  can optimize exponent of \(\Pem{1}\) over the  time
sharing constant  \(\tsc\), the input letters \(\dinp_1,\dinp_2\) and input distributions 
\(\idis{1}, \idis{2}\). Evidently the optimization problem we get is the one given for the 
definition of \(\JX{\rate}\), in equation (\ref{eq:FX}).  Thus Lemma \ref{lem:ach1} implies that  
for any \(\rate \in [0,\CX]\) and block length  \(\blx\) there exists a  length \(\blx\) code such
 that \(|\mesS|\geq e^{\blx (\rate-\epsq{\blx})}\), \(\Pem{\dmes} \leq \epsq{\blx}\) for 
 \(\dmes=2,3,\ldots, |\mesS|\) and \( \Pem{1} \leq  e^{-\blx (\JX{\rate}-\epsq{\blx})}\).

 One curious question is whether or not the exponent of \(\Pem{1}\) can be increased by including more than two phases.  Carath\'{e}odory's Theorem answers that question negatively, i.e., to obtain the largest value of \(\JX{\rate}\)	 one doesn't need  to do time sharing between more than two input-letter-input-distribution pairs.

\subsection{Control Phase and Error-Erasure Decoding:}\label{sec:ach-cont}
The family of codes described in Lemma \ref{lem:ach1} has  a large exponent for the conditional error 
 probability of the first message, i.e.,  \(\Pem{1}\). But the conditional error probabilities  of
  other messages, \(\Pem{\dmes}\) for \(\dmes\neq1\), decay subexponentially. In 
   order  to facilitate an exponential decay of \(\Pem{\dmes}\)  for \(\dmes\neq 1\) with block length, we append the codes described in Lemma \ref{lem:ach1} with a control phase and allow erasures.   The idea of using a control phase and an error-erasure decoding, in establishing achievablity results for  variable length code, was first employed by  Yamamoto and Itoh  in \cite{itoh}.

 In order explain what we mean by the control phase, let us describe our encoding scheme and 
 decoding rule briefly.   First a  code from the family of codes described in 
Section \ref{sec:ach-smwof}   is used to transmit \(\mes\) and the receiver makes a tentative  
decision \(\test\)  using the  decoder of the very same code. The transmitter knows  \(\test\)  
because of the feedback link. In the  remaining time   instances, i.e.,  in the control   phase,  
the transmitter sends    the input letter \(\xa\)  if \(\test=\mes\), 
the input letter \(\xr\) if \(\test \neq\mes\). The input letters \(\xa\) and \(\xr\) are  described 
in  equation     (\ref{eq:dm}).  At the end of  the control phase,  the receiver checks whether 
or not the output sequence in the control phase is  typical with   \(\CTM_{\xa}\), if it is 
then \(\est=\test\) otherwise an erasure is   declared.

  Lemma \ref{lem:ach2} given below states the results of the performance analysis of the 
  above described code. In order to understand what is stated in  Lemma \ref{lem:ach2} 
  accurately, let us make a brief digression and elaborate on the  codes with 
  erasure. We have  assumed in our models until now that \(\est\in \mesS\). However, there 
  are many interesting problems in which this might not hold. In codes with erasures for 
  example,  we replace \(\est\in \mesS\) with \(\est\in   \estS\) where  
  \(\estS=\mesS \cup \{\Era\}\) and \(\Era\) is the erasure symbol. Furthermore  in  codes with erasures for each  \(\dmes \in \mesS\)  the conditional error  probability \(\Pem{\dmes}\) and   conditional erasure  probability, \(\Perm{\dmes}\)  are defined  as follows. 
\begin{subequations}
\label{eq:defpe-feec}
    \begin{align}
\Pem{\dmes}&=\PCX{\est\notin \{\dmes, \Era\}}{\mes=\dmes} &
&\dmes=1,2,\ldots,|\mesS| \\
\Perm{\dmes}&=\PCX{\est=\Era}{\mes=\dmes} &
&\dmes=1,2,\ldots,|\mesS| 
\end{align}
\end{subequations}
Note that definitions of \(\Pem{\dmes}\) and \(\Perm{\dmes}\) given above can be seen as the generalizations of the corresponding definitions in block codes without erasures. In erasure free codes above definitions are equivalent to corresponding definitions there.

\begin{lemma}   \label{lem:ach2}
For any block length $\blx$, rate $0 \leq \rate \leq \CX$ and  error exponent $0\leq\ex \leq (1-\tfrac{\rate}{\CX})\DX$,  there exists a length \(\blx\) block code with  erasures such that,
\begin{align*}
|\mesS|
&\geq  e^{\blx (\rate-\epsq{\blx})} &
&\\
\Pem{1}\!~
& \leq  e^{-\blx \left(\ex+(1-\frac{\ex}{\DX}) \JX{\frac{\rate}{1-\ex/\DX}}-\epsq{\blx}\right)}&
&\\
\Pem{\dmes}
&\leq \epsq{\blx} \min\{1,e^{-\blx (\ex-\epsq{\blx})}\} &
& \dmes=2,3,\ldots, |\mesS|\\
\Perm{\dmes}
&\leq \epsq{\blx} +e^{-\blx\left((1-\frac{\ex}{\DX}) \JX{\frac{\rate}{1-\ex/\DX}}-\epsq{\blx}\right)}
 &&\dmes=1,2,\ldots, |\mesS|
\end{align*}
where  \(\epsq{\blx}=\tfrac{10|{\inpS}| |{\outS}|(\ln \frac{e}{\mtp}) \sqrt{\ln (1+\blx)}}{ \sqrt{\blx}} \).
 \end{lemma}
Proof of Lemma \ref{lem:ach2} is given in Appendix \ref{sec:appach2}.

Note that in  Lemma \ref{lem:ach2},  unlike \(\Pem{\dmes}\)'s which decrease exponentially  with \(\blx\),
\(\Perm{\dmes}\)'s decays as \(\tfrac{\ln \blx}{\sqrt{\blx}}\).  It is possible to tweak the proof so as to have a 
non-zero exponent for \(\Perm{\dmes}\)'s, see \cite{NZ}. But this can only be done at the expanse of 
\(\Pem{\dmes}\)'s. Our aim,  however,  is  achieving the optimal performance in variable length block codes.
As we will see in the following subsection, for that what matters is exponents of error probabilities and 
having vanishing  erasure probabilities.  The rate at which erasure probability decays does not effect 
the performance of variable length block codes in terms of error exponents.

\subsection{Single Message \emph{Message-Wise} \uep~Achievablity:}\label{sec:achsa}

In this section we construct  variable length block codes for the single message 
\emph{message-wise} \uep~problem using Lemma \ref{lem:ach2}. In first \(\blx\) time units
the variable length   encoding scheme uses  a fixed length block code with erasures  which has 
the  performance described in Lemma \ref{lem:ach2}.  If the decoded message of the 
fixed length   code is in the message set, i.e., if \(\est\in \mesS\) then decoded message 
of the  fixed length  code becomes the decoded message of the variable length code. If the 
decoded message of the fixed length code is the erasure symbol, i.e., if \(\est=\Era\), then the 
encoder uses the fixed length code again in the second \(\blx\) time units. By repeating this 
scheme until the decoded message of the fixed length code is in \(\mesS\), i.e.,  \(\est \in \mesS\),  we obtain a variable length code. 

Let \(\rep\) be the number of times  the fixed length code is used until a  \(\est \in \mesS\) is observed.  Then given the message \(\mes\),   \(\rep\) is a geometrically distributed random variable with success probability \((1-\Perm{\mes})\) where \(\Perm{\mes}\) is the conditional  erasure probability of the fixed length code given the message \(\mes\). Then the conditional probability distribution and the  conditional expected value of \(\rep\) given \(\mes\) are
\begin{subequations}
\begin{align}
\label{eq:mdr-a}
\PCX{\rep=\drep}{\mes}&= (1-\Perm{\mes})(\Perm{\mes})^{\drep-1}  
&&\drep=1,2,\ldots\\
\label{eq:mdr-b} 
\ECX{\rep}{\mes}&=(1-\Perm{\mes})^{-1}.
&&
\end{align}
\end{subequations}
Furthermore the conditional  expected value of decoding time   and the conditional  error probability  
given the message \(\mes\) are
\begin{subequations}
\label{eq:mdr}
\begin{align}
\ECX{\dt}{\mes}&=\blx\ECX{\rep}{\mes}\\
\PCX{\est\neq \mes}{\mes}&=\Pem{\mes} \ECX{\rep}{\mes} 
\end{align}
\end{subequations}
where \(\blx\) is  the block length of the fixed length code and \(\Pem{\mes}\) is the  conditional error probability given the message \(\mes\) for the fixed length code. 

Thus as result of equations (\ref{eq:mdr-b}),  (\ref{eq:mdr}) and Lemma \ref{lem:ach2} we know that for any rate \(\rate\in [0,\CX]\), error exponent \(\ex\in [0,(1-\tfrac{\rate}{\CX})\DX]\) there exists a reliable sequence \(\SC\) such that  \(\rate_{\SC}  =\rate\), \(\ex_{\SC} =\ex\) and 
\begin{align}
\label{eq:mdr-z}
\Emd_{\SC}&=\ex+(1-\tfrac{\ex}{\DX}) \JX{\tfrac{\rate}{1-\ex/\DX}}.
\end{align}
We show in Section (\ref{sec:con-2}) that for any reliable sequence \(\SC\) with rate \(\rate_{\SC}=\rate\) and 
error exponent \(\ex_{\SC}=\ex\), \(\Emd_{,\SC}\) is upper bounded by the expression on the right hand side 
of equation (\ref{eq:mdr-z}). 

 \subsection{\emph{Bit-Wise} \uep~Achievablity:}\label{sec:achsb}
In this section we first  use  the  family of codes described in Section \ref{sec:ach-smwof}
and the control phase idea described in Section \ref{sec:ach-cont} to
 construct  fixed length block codes with erasures which have  \emph{bit-wise} 
\uep. Then we  use them with a  repeat until non-erasure decoding scheme, similar to the one 
described in Section \ref{sec:achsa}, to obtain variable length block codes with 
\emph{bit-wise} \uep.

Let us start with describing the encoding scheme for the fixed length block code with 
  \emph{bit-wise} \uep. If there are  \( \nlay\) sub-messages, i.e.,  if 
 \(\mesS=(\mesS_1 \times \mesS_2\times \cdots \times \mesS_ \nlay)\), then  the encoding 
 scheme has  \( \nlay+1\) phases  with lengths \(\blx_1\), \(\blx_2\), \(\ldots\), \(\blx_{ \nlay+1}\) 
   such that \(\blx_1+\blx_2+\ldots+\blx_{ \nlay+1}=\blx\)
  \begin{itemize}
\item   In the first phase a  length \(\blx_1\)	code from the  family of codes described in 
Section \ref{sec:ach-smwof}  is used. The  message  set of the code \(\tmesS_{1}\) is    
\(\mesS_1\cup \{|\mesS_1|+1 \}\) and the message \(\tmes_{1}\) of the code   is determined 
by the first sub-message:  \(\tmes_1=\mes_1+1\).  At the end of first phase receiver uses the 
decoder of the length  \(\blx_1\)  code   to get a tentative decision \(\test_1\) which is known 
by the transmitter at  the beginning of the second phase because of the feedback link.
\item  In the second phase  a length \(\blx_2\) code   from the  family of codes described in  
Section \ref{sec:ach-smwof},   with the  message set  
\(\tmesS_{2}=\mesS_2\cup \{|\mesS_2|+1 \}\), is used.  If \(\test\) is decoded correctly at the 
end of the first phase then   the message \(\tmes_{2}\) of the    code  used in the second phase 
is  determined by the second sub-message as    \(\tmes_2=\mes_2+1\), else \(\tmes_2=1\). 
At the end of the second phase  the  receiver uses the decoder of the second phase code to 
get the tentative decision \(\test_2\)  which is known by the transmitter at  the beginning of
 the third phase because of the feedback link..
\item  In  phases \(3\) to \( \nlay\) above described scheme is used.  In phase \(i\),  a length 
\(\blx_i\) code,    with the message  set    \(\tmesS_{i}=\mesS_i\cup \{|\mesS_i|+1 \}\),  from 
the  family of codes described in  Section \ref{sec:ach-smwof}  is used. The  message of the  
 length \(\blx_i\) code   \(\tmes_i\) is \(\mes_i+1\)  if \(\test_{i-1}=\tmes_{i-1}\),   \(1\) 
 otherwise  for \(i=3,4,\ldots, \nlay\). 
\item The last phase is a  \(\blx_{ \nlay+1}\) long control phase, i.e., a \(\blx_{ \nlay+1}\) long 
 code with 
the message set  \(\tmesS_{ \nlay+1}=\{1,2\}\)   is used in the last phase.   The codewords for the 
first and second  messages are \(\blx_{ \nlay+1}\) long sequences of input letters \(\xr\) and 
\(\xa\) respectively,  where  \(\xr\) and \(\xa\) are  described in equation (\ref{eq:dm}).  The  
tentative decision in  the last phase  \(\test_{ \nlay+1}\)  is  equal to  the  first message if the   
output sequence in the last phase is not  typical with  \(\CTM_{\xa}\), the second message 
otherwise. The message of the \(\blx_{ \nlay+1}\) long code \(\tmes_{ \nlay+1}\) is  equal
 to \(2\) if  \(\test_{ \nlay}=\tmes_{ \nlay}\),  \(1\) otherwise. 
 \end{itemize}
Note that if we define \(\test_{0}\), \(\tmes_{0}\) and \(\mes_{ \nlay+1}\) all to be \(1\), i.e.,
\(\test_{0} =\tmes_{0} =\mes_{ \nlay+1} =1\) we  can  write the following rule for 
determining the \(\tmes_{i}\)'s for \(i=1\) to \( \nlay+1\).
\begin{align}
\label{eq:r-bituep-enc}
\tmes_i
&=1+\IND{\test_{i-1}=\tmes_{i-1}}\mes_i   &
i
&=1,2,\ldots, ( \nlay+1) 
 \end{align}
It is important however to keep in mind that the last phase is a control phase and the codes in the first \(\nlay\) 
phases are from the  family of codes described in  Section \ref{sec:ach-smwof}. 

Note that during the phases \(i=2\) to \( \nlay\) erroneous transmission of \(\tmes_{i-1}\) is conveyed using  
\(\tmes_{i}=1\), hence the transmission of  \(\mes_i\) through  \(\tmes_{i}\), i.e., \(\tmes_{i}=1+\mes_i\), is a  
tacit approval of the tentative decision  \(\test_{i-1}\).  Because of this, the above encoding  scheme is said 
to have  an implicit  acceptance explicit rejection property. The idea of  implicit  acceptance explicit 
rejection was first  introduced by Kudryashov in \cite{kud} in the context of non-block  variable length 
codes with feedback and  delay constraints.

After finishing the  description of  the encoding scheme, we are ready to describe the decoding 
scheme. The receiver determines the decoded message using the tentative decisions,  
\(\test_i\) for \(i=1\) to \( \nlay+1\). If  one or more of the   tentative decisions are equal to \(1\), 
then  an erasure is declared. If all \( \nlay+1\) tentative decision are different from \(1\) then 
\(\est_{i}=\test_{i}-1\) for all \(i=1,2,\ldots, \nlay\). Hence the decoding rule is
\begin{align}
\label{eq:r-bituep-dec}
(\est_1,\est_2,\ldots,\est_ \nlay)
=
\begin{cases}
(\test_1-1,\test_2-1,\ldots,\test_{\nlay}-1)    
& \mbox{if~} \prod_{i=1}^{ \nlay+1} (\test_{i}-1) >0\\
\Era
& \mbox{if~} \prod_{i=1}^{ \nlay+1} (\test_{i}-1) =0\\
\end{cases}.
\end{align}

For \emph{bit-wise} \uep~codes with erasure, the definition of \(\Peb{i}\) is slightly different 
from   the original one given in equation (\ref{eq:peb-def})
\begin{equation}
\label{eq:peb-def-we}
\Peb{i}=\PX{ \{\est^{i}\neq  \dmes^{i},\est\neq \Era\}}.
\end{equation}

With this alternative definition in mind  let us define \(\Pemb{\dmes}{i}\) as the conditional probability
of the erroneous  transmission any one of the first \(i\) sub-message when \(\mes=\dmes\):
\begin{equation}
\label{eq:peb-def-wef}
\Pemb{\dmes}{i}=\PCX{ \{\est^{i}\neq  \dmes^{i},\est\neq \Era\}}{\mes=\dmes}
\end{equation}

 The error analysis of the above described fixed length codes, presented in  Appendix \ref{sec:appach3},
 leads to Lemma \ref{lem:ach3} given below.

%{\bf The \emph{bit-wise} \uep~scheme with erasures   described in   Section \ref{sec:achsb} has three phases when there are two sub-messages,    \(\mes_1\)  and \(\mes_2\).  In the first phase \(\mes_1\) is transmitted using  the  \emph{ordinary  messages} of  a code like the one described in Section      \ref{sec:ach-smwof}.     In the second phase  \(\mes_2\) is transmitted  using the \emph{ordinary messages} of  a     code like the one described in Section \ref{sec:ach-smwof}     if the tentative decision     \(\test_1\) is correct,  the \emph{special message} is transmitted otherwise.    Throughout the the third phase  the input letter \(\xa\), described in  equation     (\ref{eq:dm}), is transmitted  if both of the tentative decisions are correct, the input letter     \(\xr\) is transmitted otherwise.      If the output distribution in the control phase is typical with \(\CTM_{\xa}\) and if neither of     the tentative decisions is equal to the \emph{special message} of the corresponding codes     then the tentative decisions become final i.e., \(\est_1=\test_1\) and     \(\est_2=\test_2\) , otherwise an erasure is declared. Note that during the second phase erroneous transmission of \(\mes_1\) is conveyed using  the \emph{special message}, hence the non-existence of the \emph{special message} is a  tacit approval of the tentative decision. This is why this scheme is called a implicit  acceptance explicit rejection scheme.}

\begin{lemma}\label{lem:ach3}
 For  block length \( \blx \), any integer \( \nlay \leq \tfrac{\blx}{\ln (1+\blx)} \), rate vector $\vec{\rate}$, and time sharing vector $\vec{\fr}$ such that
 \begin{subequations}
   \label{eq:as1}
\begin{align}
\rate_i 
&\leq \CX \fr_i  
& \forall i\in\{1,2,\ldots, \nlay\}\\
\fr_{i}
&\geq 0 
& \forall i\in\{1,2,\ldots, \nlay\}\\
\sum\nolimits_{i=1}^{ \nlay}\fr_i
&\leq 1
&
\end{align}
\end{subequations}
there exists a  length \(\blx\) block  code  such that:
\begin{align*}
|\mesS_i|&\geq  e^{\blx (\rate_i-\epsq{\blx,\nlay})} &&\forall i \in \{1,2,\ldots  \nlay\}\\
|\mesS^i|&\geq  e^{\blx (-\epsq{\blx,\nlay}+\sum_{j=1}^{i}\rate_j )} &&\forall i \in \{1,2,\ldots  \nlay\}\\
\Pemb{\dmes}{i}
& \leq \epsq{\blx,\nlay}  e^{-\blx \left(-\epsq{\blx,\nlay}+
\sum_{j=i+1}^{ \nlay+1}\fr_{j}\JX{\frac{\rate_{j}}{\fr_{j}}} \right)} 
 &&\forall \dmes\in \mesS,~ i \in \{1,2,\ldots  \nlay\}\\
\Perm{\dmes}
&\leq \epsq{\blx,\nlay}  && \forall \dmes \in \mesS
\end{align*}
where  \(\fr_{ \nlay+1}=1-\sum_{i=1}^{ \nlay}\fr_i\), \(\rate_{ \nlay+1}=0\), 
\(\epsq{\blx,\nlay}=\tfrac{10|{\inpS}| |{\outS}| \ln (\frac{e}{\mtp}) \sqrt{\ln (1+\blx)}}{\sqrt{\blx}} \sqrt{1+\nlay}\).
\end{lemma}

Recall the repeat at erasures scheme described in Section \ref{sec:achsa}. If we use that scheme 
to obtain a variable length code from the fixed length  \emph{bit-wise} \uep~code described in
Lemma \ref{lem:ach3}, we obtain a variable length code  with \uep~such that
\begin{subequations}
\label{eq:bitsr}
\begin{align}
  \ECX{\dt}{\mes}
&=\tfrac{\blx}{1-\Perm{\mes}}&
& \\
\PCX{\est^{i}\neq \mes^{i}}{\mes}
&\leq \tfrac{\Pemb{\mes}{i}}{1-\Perm{\mes}}&
& ~i=1,2,\ldots, \nlay.		
\end{align}
\end{subequations}
As result of equation (\ref{eq:bitsr}) and Lemma \ref{lem:ach3}  we know that for any rate vector \(\vec{\rate}\),  error exponent vector \(\vec{\ex}\) and time sharing vector $\vec{\fr}$ such that
 \begin{subequations}
\label{eq:bitsr-z}
   \begin{align}
\label{eq:bitsa-z}
\ex_{i} &\leq (1-\sum\nolimits_{j=1}^{ \nlay}\fr_j) \DX + \sum\nolimits_{j=i+1}^{\nlay} \fr_j \JX{\tfrac{\rate_j}
{\fr_j}} && \forall i\in \{1,2,\ldots, \nlay\}\\
\label{eq:bitsb-z}
\rate_i &\leq \CX \fr_i&& \forall i\in \{1,2,\ldots, \nlay\}\\
\label{eq:bitsbc-z}
\fr_i &\geq 0 && \forall i\in \{1,2,\ldots, \nlay\}\\
\label{eq:bitsc-z}
\sum\nolimits_{j=1}^{ \nlay}\fr_j&\leq 1     &&
   \end{align}
\end{subequations}
there exists a reliable sequence \(\SC\) such that 
\((\vec{\rate}_{\SC},\vec{\ex}_{\SC})=(\vec{\rate},\vec{\ex})\).
Thus the existence of the time sharing vector \(\vec{\fr}\) satisfying  the constraints given 
in (\ref{eq:bitsr-z}) is a sufficient condition for the  achievablity of a \rev~  \((\vec{\rate},\vec{\ex})\).  We show in Section (\ref{sec:con-3}) that the existence of  a 
time sharing vector \(\vec{\fr}\)  satisfying  the constraints given in (\ref{eq:bitsr-z})   is also a 
necessary condition for the achievablity of a \rev~  \((\vec{\rate},\vec{\ex})\).

\section{Converse}\label{sec:con}
Berlin et. al. \cite{bnrt} used the error probability  of a random binary query posed  at a  
stopping time  for bounding the error probability of a  
variable length block code.  Later similar techniques have been applied in  \cite{bnz} for 
establishing  outer bounds in \uep~problems. Our approach is  similar to  that of \cite{bnrt} 
and \cite{bnz}; we, too,  use error probabilities of  random  queries posed  at stopping times
 for establishing outer bounds. Our approach,  nevertheless,   is novel because of the 
 error events  we choose to  analyze and the bounding techniques we use. Furthermore,  the relation 
 we establish in Lemma \ref{lem:con} between the  error probabilities and the decay rate of the 
 conditional entropy of  the messages with time is a  brand new tool for \uep~problems.
 
 For rigorously and unambiguously generalizing the technique used in \cite{bnrt} 
and \cite{bnz} we  introduce the concept of  anticipative list decoders in Section \ref{sec:con-0}. 
Then in   Section \ref{sec:con-1} we  bound the  probabilities of certain  error events  associated with  
anticipative list decoders  from below. This bound, i.e., Lemma \ref{lem:con}, is used in  Sections
 \ref{sec:con-2}   and \ref{sec:con-3}  to derive  tight outer bounds  for the performance of variable 
 length block codes in  the  single message {\it message-wise}  \uep~problem and  in 
 the {\it bit-wise} \uep~problem, respectively. 

\subsection{Anticipative  List Decoders}\label{sec:con-0}
In this section we first introduce the concepts of anticipative list decoders and non-trivial anticipative list 
decoders. After  that  we  show that for a given variable length code, any  non-trivial anticipative list 
decoder \((\wdt,\ldsf)\)   can be used to define   a probability distribution, \(\PXAD{\ldsf}\), on
 \(\mesS\times \outS^{\dt*}\). Finally we use \(\PXAD{\ldsf}\) to define the probability measure 
 \(\PXA{\ldsf}{\cdot}\)  for the events in \(\PWS{\mesS \times \outS^{\dt}}\).  Both the 
 non-trivial anticipative list  decoders  \((\wdt,\ldsf)\) and the  probability measures     \(\PXA{\ldsf}{\cdot}\) 
 associated with  them  play key roles in Lemma \ref{lem:con} of Section \ref{sec:con-1}.

An anticipative list decoder for a variable length code is a list decoder \(\ldsf\) that decodes at a stopping
 time  \(\wdt\)	that is  always less than or equal to the decoding time of the code \(\dt\).  The
 anticipative list decoders  are used  to formulate questions about the transmitted message or the decoded 
 message, in terms of a subset  of the message set \(\mesS\) that  is chosen at a stopping time  
 \(\wdt\). For example let \(\ldsf\) be the set of all \(\dmes \in \mesS\) whose posterior probability at time 
 one is larger than \(1/{|\mesS|}\). Evidently for all values of \(\out_1\),  \(\ldsf\) is a subset of  \(\mesS\), but 
 it is not necessarily the same subset for all values of \(\out_1\).  Indeed \(\ldsf\) is a function from 
 \(\outS_1\) to the power set of \(\mesS\) and \((\wdt,\ldsf)\) is  an anticipative list decoder, for   which 
 \(\wdt=1\). Formal definition, for  anticipative list decoders, is given below. In order to avoid separate 
 treatment in  certain special cases we include the case when \(\wdt=0\) and \(\ldsf\) is fixed subset of 
 \(\mesS\), in the definition.

\begin{definition}[Anticipative List Decoder]\label{def:ald}
For a variable  length code with decoding time \(\dt\),   a pair \((\wdt,\ldsf)\) is called  an
anticipative  list decoder (\ald)  if 
\begin{itemize}
\item either  \(\wdt\) is the constant random variable  \(0\) and \(\ldsf\) is a fixed subset of \(\mesS\), i.e.,
\begin{align*}
\wdt&=0\\
\ldsf&\in \PWS{\mesS}
\end{align*}
\item  or    \(\wdt\) is a  stopping time, which is smaller than \(\dt\) with probability one, and \(\ldsf\) is
a  \(\PWS{\mesS}\) valued function defined on \(\outS^{\wdt}\), i.e.,
\begin{align*}
&\PX{\wdt \leq \dt }=1\\
 &\ldsf: \outS^{\wdt} \to \PWS{\mesS}.
 \end{align*}
\end{itemize}
 \end{definition}	
Definition of  \ald~ does not require \(\ldsf\) to be of some fixed size, nor it requires \(\ldsf\) to include more likely or less likely messages. Thus for certain values of  \(\out^{\wdt}\),  \(\ldsf\) might not include any \(\dmes\in \mesS\) with positive posterior  probability. In other words for some values of \(\out^{\wdt}\) we might have
\begin{equation*}
\PCX{ \mes \in  \lds{\out^{\wdt}} }{\out^{\wdt}=\dout^{\dwdt}}=0.
\end{equation*} 
The \ald's in which such \(\dout^{\dwdt}\)'s have zero probability are called nontrivial \ald's.

\begin{definition}[Nontrivial \ald]\label{def:nald}
An anticipative  list decoder  \((\wdt,\ldsf)\) is called a  nontrivial anticipative  list decoder (\nald) if 
\(\PCX{\mes \in  \lds{\out^{\wdt}}}{\out^{\wdt}}>0\) with probability one, i.e.,
\begin{equation}
\PX{\PCX{ \mes \in  \lds{\out^{\wdt}} }{\out^{\wdt}}>0 }=1.
\end{equation}
\end{definition}	

  Below, for any variable length code and an associated nontrivial anticipative  list decoder  	\((\wdt,\ldsf)\)  we define a probability distribution \(\PXAD{\ldsf}\)  on \(\mesS \times \outS^{\dt*}\)
and a probability measure \(\PXA{\ldsf}{\cdot}\) for the events in \(\PWS{\mesS \times \outS^{\dt}}\).
For doing that first note that the probability measure generated by the code, i.e., \(\PX{\cdot}\), can be used to  define  a probability distribution \(\PXD\) on \(\mesS \times \outS^{\dt*}\)
 as follows:
\begin{align}
\label{eq:defpxd-a}
\PXD(\dmes, \dout^{\ddt})
&\DEF 
\PX{\mes=\dmes,\out^{\dt}=\dout^{\ddt}}
&
& \forall \dmes\in \mesS,\dout^{\ddt}\in \outS^{\dt*} 
\end{align}
where \(\outS^{\dt*}\) is a countable set  for any  stopping time, given in equation (\ref{eq:def-stop-dom-b}).

As \(\dt\) is a stopping time,   the probability of any event \(\event\) in \( \PWS{\mesS \times \outS^{\dt}}\)
under  \(\PX{\cdot}\), i.e., \(\PX{\event}\), is equal to  
\begin{align}
\label{eq:defpxd-b}
\PX{\event}
&=\sum_{(\dmes,\dout^{\ddt}) \in  \event \cap (\mesS \times \outS^{\dt*}) }\PXD(\dmes, \dout^{\ddt}).
 \end{align}
 Evidently we can  extend the definition of \(\PXD\)	and  assume that \(\PXD\) is zero whenever 
 \(\dout^{\ddt}\) is in \(\outS^{\infty}_{ \left\{ \dt=\infty \right\}}\), i.e.,
 \begin{align}
\label{eq:defpxd-c}
\PXD(\dmes, \dout^{\ddt})
&\DEF 0
&
& \forall \dmes\in \mesS,\dout^{\ddt}\in \outS^{\infty}_{\left\{ \dt=\infty \right\}}.
\end{align}
This   extension is neither necessary nor relevant for  calculating    the probabilities of the events  in
   \(\PWS{\mesS \times \outS^{\dt}}\), because \(\dt\) is a stopping time, i.e., \(\PX{\dt<\infty}=1\).

\begin{definition}\label{def:eq:defpxa-a}
Given a variable length code with decoding time \(\dt\),  for any  \nald~  \((\wdt,\ldsf)\) let \(\PXAD{\ldsf}\) be\footnote{There is a slight abuse of notation in equation (\ref{eq:defpxa-a}); if \(\wdt\) is not a stopping time but rather a constant random variable \(\dt=0\),  \(\PXAD{\ldsf}(\dmes, \dout^{\ddt})\) should be interpreted as 
\begin{align*}
\PXAD{\ldsf}(\dmes, \dout^{\ddt})
&\DEF \frac{\IND{\dmes \in \ldsf}}{|\ldsf|}
\PXD(\dout^{\ddt}|\dmes)
& \forall \dmes\in \mesS,\dout^{\ddt}\in \outS^{\dt*}. 
\end{align*}}
\begin{align}
\label{eq:defpxa-a}
\PXAD{\ldsf}(\dmes, \dout^{\ddt})
&\DEF 
\PXD(\dout^{\dwdt})
\frac{\PXD(\dmes|\dout^{\dwdt}) \IND{\dmes \in \lds{\dout^{\dwdt}}}}{\sum_{\widetilde{\dmes} \in \mesS}\PXD(\widetilde{\dmes}|\dout^{\dwdt}) \IND{\widetilde{\dmes} \in \lds{\dout^{\dwdt}}}}
\PXD(\dout_{\dwdt+1}^{\ddt}|\dout^{\dwdt},\dmes)
& \forall \dmes\in \mesS,\dout^{\ddt}\in \outS^{\dt*} 
\end{align}
\end{definition}
Note that  Definition \ref{def:eq:defpxa-a} is a parametric definition in the sense that it  assigns 
a \(\PXAD{\ldsf}\) for all nontrivial  anticipative list decoders \((\wdt,\ldsf)\). While proving outer bounds
we will employ not one but multiple \nald's and use them  in conjunction with our new result, i.e., 
Lemma \ref{lem:con}. But before introducing  Lemma \ref{lem:con},  let us elaborate on the relations 
between marginal and conditional distributions 
of  \(\PXAD{\ldsf}\) and \(\PXD\).

For \(\PXAD{\ldsf}\)  defined in equation (\ref{eq:defpxa-a}) we have
\begin{equation*}
\sum_{\dmes\in \mesS,\dout^{\ddt}\in \outS^{\dt*}}\PXAD{\ldsf}(\dmes, \dout^{\ddt})=1.
\end{equation*}
Hence   \(\PXAD{\ldsf}\) is a probability distribution on \(\mesS\times \outS^{\dt*}\), i.e.,
 \(\PXAD{\ldsf} \in \PDS{\mesS\times \outS^{\dt*}}\).

Note that the marginal  distributions of \(\PXAD{\ldsf}\)  and \(\PXD\) are the same on \(\outS^{\wdt*}\). 
Furthermore for all 	\(\dout^{\dwdt} \in \outS^{\wdt*}\) and \(\dmes\in \mesS\) the conditional 
distributions of  \(\PXAD{\ldsf}\)  and \(\PXD\) are the same on \(\outS_{\wdt*}^{\dt*}\).  
The probability distributions  \(\PXAD{\ldsf}\)  and \(\PXD\) differ only in their conditional distributions on \(\mesS\) given \(\dout^{\dwdt} \). More specifically,
\begin{subequations}
\label{eq:defpxa-c}
\begin{align}
\PXAD{\ldsf}(\dout^{\dwdt})
&=\PXD(\dout^{\dwdt})
&&\forall \dout^{\dwdt} \in \outS^{\wdt*}\\
\PXAD{\ldsf}(\dmes|\dout^{\dwdt})
&=
\frac{\PXD(\dmes|\dout^{\dwdt}) \IND{\dmes \in \lds{\dout^{\dwdt}}}}{\sum_{\widetilde{\dmes} \in \mesS}\PXD(\widetilde{\dmes}|\dout^{\dwdt}) \IND{\widetilde{\dmes} \in \lds{\dout^{\dwdt}}}}
&&\forall \dout^{\dwdt} \in \outS^{\wdt*}, \forall \dmes\in \mesS\\
\PXAD{\ldsf}(\dout_{\dwdt+1}^{\ddt}|\dout^{\dwdt},\dmes)
&=\PXD(\dout_{\dwdt+1}^{\ddt}|\dout^{\dwdt},\dmes)
&&\forall \dout^{\ddt} \in \outS^{\dt*}, \forall \dmes\in \mesS.
\end{align}
\end{subequations}

Using the  parametric definition of probability distribution \(\PXAD{\ldsf}\)  on  
\(\mesS\times \outS^{\dt*}\)  we define a probability measure \(\PXA{\ldsf}{\cdot}\) for the events 
in \(\PWS{\mesS \times \outS^{\dt}}\) as follows:
\begin{align}
\label{eq:defpxa-b}
\PXA{\ldsf}{\event}
&\DEF
\sum_{(\dmes,\dout^{\ddt}) \in \event \cap (\mesS \times \outS^{\dt*}) }
\PXAD{\ldsf}(\dmes, \dout^{\ddt})
&&
\forall \event\in  \PWS{\mesS \times \outS^{\dt}}.
 \end{align}
Evidently we can extend the definition of \(\PXAD{\ldsf}\) to \(\mesS \times \outS^{\dt}\) by defining it to be zero   on  \(\mesS \times \outS^{\infty}_{\left\{\dt=\infty\right\}}\), i.e.,
 \begin{align}
\label{eq:defpxa-c}
\PXAD{\ldsf}(\dmes, \dout^{\ddt})
&\DEF 0
&
& \forall \dmes\in \mesS,\dout^{\ddt}\in \outS^{\infty}_{\left\{\dt=\infty\right\}}.
\end{align}
As in the case of \(\PXD\),  this extension is neither necessary nor relevant for calculating the probabilities 
\(\PXA{\ldsf}{\event}\)  given in  equation (\ref{eq:defpxa-b}).

\subsection{Error Probability and Decay Rate of Entropy:}\label{sec:con-1}
In this section we lower bound the probability of the event that the decoded message \(\est\) is in \(\ldsf\)  under the probability measure  \(\PXA{\ldsf}{\cdot}\), i.e.,  \(\PXA{\ldsf}{\est \notin \lds{\out^{\wdt}} }\). The bounds we derive depend on the
    decay rate   of the  conditional entropy of the messages in the interval between \(\wdt\) and  \(\dt\).

Before even stating our bound,  we  need to specify what we mean by the conditional entropy of the 
messages.  While defining the conditional entropy, many authors do take an average over the 
sample values of the conditioned random variable and obtain a constant. We, however, do not take an 
average over the conditioned random variable and define conditional entropy as a random variable  itself, 
which is a function of the random variable that is conditioned on:\footnote{Recall the standard notation in 
probability theory about the conditional  expectations and conditional probabilities: Let \(\hv\) be a real 
valued random variable and \(\gv\) be a random quantity that takes values from a finite set \(\gvS\), such 
that  \(\PX{\gv=\dgv}>0\) for all \(\dgv\in \gvS\). Then unlike \(\EX{\hv}\), which is constant, 
\( \ECX{\hv}{\gv}\) is a  random variable.  Thus  an equation  of the form \(\zv=\ECX{\hv}{\gv}\), 
implies not the equality of two  constants but the equality of two random variables, i.e. it means that    
\(\dzv=\ECX{\hv}{\gv=\dgv}\) for all  \(\dgv \in \gvS\).  Similarly let \(\hvS_1\) be a set of sample values  of 
the random variable \(\hv\) then,  unlike  \( \PX{\hv\in \hvS_1} \), which is a constant,
 \(\PCX{\hv \in \hvS_1}{\gv}\) is a random variable.   Equations (\ref{eq:n:entdef}) and 
(\ref{eq:n:entdef-alt})  are such equations. Explaining conditional
expectations and conditional probabilities are beyond the scope of this paper, readers who are not
sufficiently fluent  with these concepts are encouraged to read \cite[Chapter I, Section 8]{shiryaev}
which deal the case where random variables can take finitely many values. Appropriately generalized 
formal treatment of the subject in terms of sigma fields is presented
 in  \cite[Chapter II, Section 7]{shiryaev}.}
\begin{equation}
\label{eq:n:entdef}
\HX(\mes|\out^{\tin})
\DEF \sum_{\dmes \in \mesS} \PCX{\mes=\dmes}{\out^{\tin}} \ln \tfrac{1}{\PCX{\mes=\dmes}{\out^{\tin}}}. 
\end{equation}
Using the  probability distribution \(\PXD\) defined in equation (\ref{eq:defpxd-a})  we see that the conditional 
entropy defined in (\ref{eq:n:entdef}) is equal to, 
\begin{equation}
\label{eq:n:entdef-alt}
\HX(\mes|\out^{\tin})= \ECX{\ln \tfrac{1}{\PXD(\mes|\out^{\tin})}}{\out^{\tin}}.
\end{equation}

\begin{lemma}\label{lem:con}
For any  variable length block code with finite expected decoding time, \(\EX{\dt}<\infty\), 
let \((\dt_1,\ldsf_1)\), \((\dt_2,\ldsf_2)\), \(\ldots\),\((\dt_k,\ldsf_k)\) be \(k\) 
\nald's\footnote{Recall \ald's and \nald's are defined in Definitions \ref{def:ald} 
and \ref{def:nald}, respectively.}   such that
\begin{equation}
\label{eq:lemcon-c1}
\PX{\left\{0\leq\dt_1\leq \dt_2\leq \cdots \leq \dt_k\leq \dt\right\}}=1.
\end{equation}
Then for all \(i\) in  \(\{1,2,\ldots, k\}\) such that    \(\left(\PX{ \mes\in\ldsn{i}{\out^{\dt_i}} }+\Pe \right) \leq 1/2\)
we have
\begin{align}
\label{eq:lemcon}
\PXA{\ldsf_i}{\est \notin \ldsn{i}{\out^{\dt_i}}}
&\geq 
\exp
\left(
 \frac{-\bent{\Pe+\PX{ \mes\in\ldsn{i}{\out^{\dt_i}} }}- 
 \sum\nolimits_{j=i+1}^{k+1}\!\EX{\dt_{j}\!-\!\dt_{j-1}}\JX{\arate{j} }}{1-\Pe-\PX{  \mes\in\ldsn{i}{\out^{\dt_i}} }}
 \right)
 \end{align}
% {\bf \(\PXA{\ldsf_i}{\cdot}\) is defined in equation (\ref{eq:defpxa-b})}
where  \(\dt_{0}=0\),  \(\dt_{k+1}=\dt\) and
for  all \(j\) in \(\left\{1,2,\ldots,(k+1)\right\}\), \(\arate{j}\)'s are given by
\begin{align}
\label{eq:defarate}
 \arate{j}
 &=
\begin{Bmatrix}
0 
&\mbox{if~} \PX{\dt_{j} =\dt_{j-1}}=1\\
\frac{\EX{\HX(\mes|\out^{\dt_{j-1}})-\HX(\mes|\out^{\dt_{j}})}}{\EX{\dt_{j}-\dt_{j-1}}}
&\mbox{if~} \PX{\dt_{j} =\dt_{j-1}}<1
\end{Bmatrix}.
\end{align}
\end{lemma}
Proof of Lemma \ref{lem:con} is presented in Appendix \ref{sec:appcon}.

Before presenting the  application of Lemma \ref{lem:con} in  \uep~problems, let us elaborate on its
hypothesis and ramifications. We assumed that  \((\dt_i,\ldsf_i)\) are all \nald. Thus  for each   \((\dt_i,\ldsf_{i})\)    the set of all \(\dout^{\ddt_i} \in \outS^{\dt_i}\)  such that  the transmitted message is  guaranteed to be outside  
\(\ldsn{i}{\dout^{\ddt_i}}\), has zero probability and   there is an associated probability measure 
\(\PXA{\ldsf_i}{\cdot}\) given in equation (\ref{eq:defpxa-b}).   Furthermore 
 \(\PXA{\ldsf_i}{\est \notin \ldsn{i}{\out^{\dt_i}} }\) is  the probability of the event that  
 decoded message \(\est\)	is not in  \(\ldsf_{i}\) under the probability measure  \(\PXA{\ldsf_i}{\cdot}\).

Condition  given in equation (\ref{eq:lemcon-c1}) ensures that the decoding times  of the \(k\) \nald's we are 
considering, \(\dt_1\), \(\dt_2\), \(\ldots\),\(\dt_k\), are reached in their indexing order  and  before the
 decoding time  of  the variable length code \(\dt\). Any   \(\dt_1\), \(\dt_2\), \(\ldots\), \(\dt_k\) satisfying 
equation (\ref{eq:lemcon-c1})  divides the time interval between \(0\) and \(\dt\) into \(k+1\) disjoint intervals.  
The  duration of these  intervals as well as the decrease of the conditional entropy during them   are random.
For the \(j^{th}\) interval the expected values of the duration  and   the  decrease in  the conditional entropy   
are given by  \(\EX{\dt_{j}-\dt_{j-1}}\) and \(\EX{\HX(\mes|\out^{\dt_{j-1}})-\HX(\mes|\out^{\dt_{j}})}\), 
respectively. Hence \(\arate{j}\)'s defined in equation (\ref{eq:defarate}) are rate of decrease of the 
conditional  entropy of the messages per unit time in different intervals. 

  Lemma \ref{lem:con}  bounds the  probability of \(\est\) being outside \(\ldsf_i\) under 
 \(\PXA{\ldsf_i}{\cdot}\) from below in terms of \(\arate{j}\)'s and \(\EX{\dt_{j}-\dt_{j-1}}\)'s for \(j>i\).
  The bound on  \(\PXA{\ldsf_i}{ \est \notin \ldsn{i}{\out^{\dt_i}}  }\) also  depends on 
\(\PX{   \mes \in \ldsn{i}{\out^{\dt_i}} }\) and \(\Pe\).  But the  particular choice of $\ldsf_j$'s  for $j\neq i$ 
has   no effect on the bound.  This feature  of the bound is its main merit over bounds resulting from the 
  previously suggested techniques.

\subsection{Single Message \emph{Message-Wise} \uep~Converse:}\label{sec:con-2}
In this section we  bound the conditional error probabilities of the messages,
i.e., \(\Pem{\dmes}\)'s,  from below uniformly over the message set \(\mesS\)  in a variable length 
block code with  average error probability \(\Pe\), using Lemma \ref{lem:con}.
Resulting outer bound reveals that  the inner bound we obtained in Section \ref{sec:achsa} for the single 
message \emph{message-wise} \uep~problem is tight.

Consider a variable length block code with finite expected decoding time, i.e., \(\EX{\dt}<\infty\).
In order to bound \(\Pem{\dmes}\),  defined in equation (\ref{eq:n:defpem}), from below we apply Lemma \ref{lem:con}  for \(k=2\) with \((\dt_1,\ldsf_1)\), \((\dt_2,\ldsf_2)\) given below.
\begin{itemize}
\item Let  \(\dt_1\) be zero and \(\ldsf_{1}\)   be  \(\{\dmes\}\), i.e.,
\begin{align}
\label{eq:smmuep-c0}
\dt_1
&=0\\
\label{eq:smmuep-c1}
\ldsf_{1}
&=\{\dmes\}.
\end{align}
\item Let \(\dt_2\)   be  the first time instance before \(\dt\) such that one message, not necessarily the one  
chosen  for \(\ldsf_{1}\),  i.e., \(\dmes\), has a posteriori probability \(1-\delta\) or higher,
\begin{equation}
\label{eq:smmuep-c2}
        \dt_2 \DEF \min\{\tin :\max_{\widetilde{\dmes}} \PCX{\mes=\widetilde{\dmes}}{\out^{\tin}} \geq (1-\delta) \mbox{~or~} \tin=\dt \}.
\end{equation}
Let \(\ldsf_2\)  be the set of  all messages whose posterior probability at time \(\dt_2\) is less then 
\((1-\delta)\),
\begin{equation}
\label{eq:smmuep-c3}
\ldsn{2}{\out^{\dt_2}} \DEF \{\widetilde{\dmes} \in \mesS: \PCX{\mes=\widetilde{\dmes}}{\out^{\dt_2}}< (1-\delta)\}.
\end{equation}
\end{itemize}
We apply Lemma \ref{lem:con} for \((\dt_1,\ldsf_1)\) and \((\dt_2,\ldsf_2)\) given in equations
(\ref{eq:smmuep-c0}), (\ref{eq:smmuep-c1}), (\ref{eq:smmuep-c2}) and  (\ref{eq:smmuep-c3}).
Then using the fact that \(\JX{\cdot} \leq \DX\) we get,
\begin{subequations}
\label{eq:n:a2b0}
  \begin{align}
\label{eq:n:a2b1}
\ln \Pem{\dmes} 
&\geq 
 \tfrac{-\bent{\Pe+ |\mesS|^{-1} }- 
 \EX{\dt_2}
 \JX{\frac{\EX{\HX(\mes)-\HX(\mes|\out^{\dt_{2}})}}{\EX{\dt_{2}}}}
-\EX{\dt-\dt_2}
\DX
 }{1-\Pe- |\mesS|^{-1}}   \\
 \label{eq:n:a2b2}
\ln  \PXA{\ldsf_2}{\est \notin \ldsn{2}{\out^{\dt_2}}} 
&\geq 
 \tfrac{-\bent{\Pe+ \PX{\mes \in \ldsn{2}{\out^{\dt_2}}}}- 
\EX{\dt-\dt_2}\DX }{1-\Pe- \PX{\mes \in \ldsn{2}{\out^{\dt_2}}}}.
\end{align}
\end{subequations}
If \(\delta<1/2\) one can show \(\PXA{\ldsf_2}{\est \notin \ldsn{2}{\out^{\dt_2}}}\) is roughly equal 
to \(\Pe/\delta \).  Thus inequality in 
(\ref{eq:n:a2b2}) becomes a lower bound on \(\EX{\dt-\dt_2}\) in terms of \(\Pe\).  It can be shown that the 
lower bound (\ref{eq:n:a2b1}) takes its  smallest  value for the smallest value of  \(\EX{\dt-\dt_2}\).  Then 
using Fano's inequality for \(\EX{\HX(\mes|\out^{\dt_2})}\) we obtain Lemma \ref{lem:consm} given below. 

A complete   proof  of  Lemma \ref{lem:consm} for variable length block codes with finite expected decoding 
time is presented  in  Appendix  \ref{sec:appconsm}. For variable length block codes with infinite expected 
decoding time, Lemma  \ref{lem:consm} follows from the lower bounds  on \(\Pe\) and \(\Pem{\dmes}\) 
derived in  Appendix \ref{app:infdect-a}  and Appendix \ref{app:infdect-b}.
\begin{lemma}
  \label{lem:consm}
For any  variable length block code
and  positive  \(\delta\)  such that \(\Pe+\delta+\tfrac{\Pe}{\delta}+|\mesS|^{-1}\leq 1/2\)
\begin{equation}
  \label{eq:consm}
-\tfrac{\ln \Pem{\dmes }}{\EX{\dt}}
\leq \ex+
(1-\tfrac{\ex-\epst{}}{\DX})
\JX{\tfrac{\rate-\epst{}}{1-\frac{\ex-\epst{}}{\DX}}}
\qquad \forall \dmes  \in \mesS
\end{equation}
where \(\rate=\frac{|\mesS|}{\EX{\dt}}\),  
\(\ex=\tfrac{-\ln \Pe}{\EX{\dt}}\),
\(\epst{}=\tfrac{\epst{1}\DX+\epst{2}}{1-\epst{1}}\),
\(\epst{1}=\Pe+\delta+\tfrac{\Pe}{\delta}+|\mesS|^{-1}\)
and
\(\epst{2}=\tfrac{\bent{\epst{1}} -\ln \mtp \delta}{\EX{\dt}}\).
\end{lemma}

  Lemma \ref{lem:consm} is a  generalization of \cite[Theorem 8]{bnz}  and  \cite[Lemma 1]{bnz}.
While  deriving bounds given in  \cite[Theorem 8]{bnz}  and  \cite[Lemma 1]{bnz}, no attention is payed 
to  the fact that the rate of decrease of the conditional entropy of the messages 
can be different in different time intervals. As result both   \cite[Theorem 8]{bnz}  and  \cite[Lemma 1]{bnz}
are tight only when the error   exponent  is very close to zero.
While deriving the bound  given in  Lemma \ref{lem:consm},  on the other hand, 
the variation in the rate the conditional entropy decreases in different intervals 
 is taken into account. Hence   the outer bound given in   Lemma \ref{lem:consm}  matches 
 the inner  bound given in Section \ref{sec:achsa} for all achievable values of 
 error exponent, \(0\leq\ex \leq (1-\frac{\rate}{\CX})\DX\). 

Consider a reliable sequence of codes \(\SC\) with rate \(\rate_\SC\) and error exponent \(\ex_\SC\). Then if we apply Lemma \ref{lem:consm} with \(\delta=\tfrac{1}{\ln (1/\Pe)}\) we get 
\begin{align}
\label{eq:mdr-z-con}
\Emd_{,\SC}&\leq\ex_{\SC}+(1-\tfrac{\ex_{\SC}}{\DX}) \JX{\tfrac{\rate_{\SC}}{1-\ex_{\SC}/\DX}}.
\end{align}
Note that the upper bound  on \(\Emd_{,\SC}\)'s given in equation (\ref{eq:mdr-z-con}) is achievable by at least one \(\SC\) described  in Section \ref{sec:achsa}.

\subsection{\emph{Bit-Wise} \uep~Converse:}\label{sec:con-3} 
In this section we apply  Lemma \ref{lem:con} to  a variable length block code 
with a  message set \(\mesS\) of the form \(\mesS=\mesS_1\times\mesS_2\times\ldots \times\mesS_{ \nlay}\), 
in order to  obtain lower bounds on \(\Peb{i}\)'s for \(i=1,2,\ldots, \nlay\)  in terms of
the  sizes of the sub-message sets   \(|\mesS_1|\),  \(|\mesS_2|\), \(\ldots\), \(|\mesS_ \nlay|\) 
and  the  expected decoding time \(\EX{\dt}\).
 When applied to reliable code sequences these bounds on  \(\Peb{i}\)'s in terms of  \(|\mesS_i|\)'s and   
\(\EX{\dt}\)  gives  a necessary condition for  the  achievablity of  a rate vector  and error exponent vector  pair \((\vec{\rate},\vec{\ex})\) that matches the  sufficient condition for  the  achievablity  derived in  Section \ref{sec:achsb}.

In order to bound \(\Peb{i}\)'s we use Lemma \ref{lem:con} with \( \nlay\) \nald's, 
 \((\dt_1,\ldsf_1)\),\(\ldots\),\((\dt_{\nlay},\ldsf_{ \nlay})\). 
 Let us start with defining  \(\dt_i\)'s and \(\ldsn{i}{\out^{\dt_{i}}}\)'s.
\begin{itemize}
\item For any \(i\) in \(\left\{1,2,\ldots, \nlay\right\}\), let \(\dt_i\) be  the first time instance 
that  a member of \(\mesS^{i}\) gains a posterior probability  larger than or equal to 
\((1-\delta)\) if it happens before  \(\dt\), \(\dt\) otherwise:
\begin{align}
  \label{eq:dti}
\dt_i
&\DEF\min \{\tin:\max_{\dmes^{i}} \PCX{\mes^{i}=\dmes^{i}}{\out^{\tin}}\geq 1-\delta \mbox{~or~} \tin=\dt\}.
\end{align}
\item  For any \(i\) in \(\left\{1,2,\ldots, \nlay \right\}\), let  \(\ldsn{i}{\out^{\dt_{i}}}\) be the 
set of all messages of the form  \(\dmes=(\dmes^{i},\dmes_{i+1},\ldots,\dmes_{\nlay})\) 
for which posterior probability of \(\dmes^{i}\) is less than \((1-\delta)\) at \(\dt_i\):
\begin{align}
\label{eq:ati}
\ldsn{i}{\out^{\dt_i}}
&\DEF\{(\dmes^{i},\dmes_{i+1},\ldots,\dmes_{ \nlay}) \in \mesS: 
\PCX{\mes^{i}=\dmes^{i}}{\out^{\dt_i}}< 1-\delta \}.
\end{align}
\end{itemize}
If  we apply Lemma \ref{lem:con} for  \((\dt_1,\ldsf_1)\),\(\ldots\),\((\dt_{\nlay},\ldsf_{\nlay})\)
defined in equations (\ref{eq:dti}) and (\ref{eq:ati}),
we obtain lower bounds on  \(\PXA{\ldsf_i}{\est \notin \ldsn{i}{\out^{\dt_i}}}\)'s in terms of 
 \(\PX{\mes \in \ldsn{i}{\out^{\dt_i}}}\)'s and  
\(\arate{j}\)'s and  \(\EX{\dt_{j}-\dt_{j+1}}\)'s for \(j>i\).
   In order to  turn these bounds into  bounds on \(\Peb{i}\)'s we bound
    \(\PXA{\ldsf_i}{\est \notin \ldsn{i}{\out^{\dt_i}}}\)'s and \(\PX{\mes \in \ldsn{i}{\out^{\dt_i}}}\)'s
     from above.
\begin{itemize}
\item The posterior probability of a message at time \(\tin+1\) can not be smaller than \(\mtp\) times 
its value at time \(\tin\)  
because   \(\min\nolimits_{\dinp \in \inpS, \dout\in \outS} \CT{\dinp}{\dout}=\mtp\). 
Thus if \(\delta<1/2\) one can   bound  \(\PXA{\ldsf_i}{\est \notin \ldsn{i}{\out^{\dt_i}}}\)'s from above:
 \begin{align}
\label{eq:n:bits-a}
\PXA{\ldsf_i}{\est \notin \ldsn{i}{\out^{\dt_i}}}
&< \tfrac{1}{\mtp \delta} \Peb{i}
& \forall i\in \left\{1,2,\ldots, \nlay\right\}.
\end{align}
\item Note that if at \(\dt_i\) there is a \(\dmes^{i}\) with posterior probability \((1-\delta)\) then \(\PCX{\mes\in\ldsn{i}{\out^{\dt_i}}}{\out^{\dt_i}} \leq \delta\). If at \(\dt_i\) there is no  \(\dmes^{i}\) with posterior probability \((1-\delta)\) then \(\PCX{\est^{i} \neq \mes^{i}}{\out^{\dt_i}} \geq \delta\). Using these facts one can bound \(\PX{\mes \in \ldsn{i}{\out^{\dt_i}}}\) from above:
\begin{align}
\label{eq:n:bits-b}
\PX{\mes \in \ldsn{i}{\out^{\dt_i}}}
&\leq \tfrac{\Pe}{\delta}+\delta
& \forall i\in \left\{1,2,\ldots, \nlay\right\}.
\end{align}
\end{itemize}
More detailed derivations of the inequalities given in  (\ref{eq:n:bits-a}) and (\ref{eq:n:bits-b}) can be found in  Appendix \ref{sec:appconbits}. 

Using equations  (\ref{eq:n:bits-a}) and (\ref{eq:n:bits-b}) together with Lemma \ref{lem:con} we can conclude that,
\begin{align}
\label{eq:n:bits}
\ln \Peb{i} 
&\geq  \ln (\mtp \delta) +
 \tfrac{-\bent{\Pe+ \delta+\Pe/\delta}- \sum_{j=i+1}^{ \nlay+1}
 \EX{\dt_{j}-\dt_{j-1}}
 \JX{\arate{j}}}{  1-\Pe- \delta-\Pe/\delta}  
& \forall i\in \left\{1,2,\ldots, \nlay\right\}.
  \end{align}
provided that \(\Pe+ \delta+ \Pe/\delta\leq 1/2\), where \(\arate{j}\)'s are defined in (\ref{eq:defarate}).

Note that the lower bound on \(\Peb{i}\)'s given in equation (\ref{eq:n:bits}) takes different values depending 
on the rate of decrease of the conditional entropy of the messages in different  intervals, i.e.,
  \(\arate{j}\)'s,   and   the expected duration of different  intervals, i.e., \(\EX{\dt_{j}-\dt_{j-1}}\)'s. 
Making a worst case 
assumption on the rate of decrease of entropy and the durations of the intervals one can obtain Lemma 
\ref{lem:conbits} given below.

A complete   proof  of  Lemma \ref{lem:conbits} for variable length block codes with finite expected decoding 
time is presented  in  Appendix  \ref{sec:appconbits}. For variable length block codes with infinite expected 
decoding time, Lemma \ref{lem:conbits} follows from the lower bounds  on \(\Pe\) and \(\Peb{i}\)'s 
derived in  Appendix \ref{app:infdect-a}  and Appendix \ref{app:infdect-c}.

\begin{lemma}
  \label{lem:conbits}
For any variable length block code with feedback with a message set \(\mesS\)  of the form\footnote{We
tacitly assume, without loss of generality,  that \(|\mesS_1|\geq 2\).}
  \(\mesS=\mesS_1\times\mesS_2\times\ldots,\mesS_{\nlay}\)
and  for any positive  \(\delta \) such that \(\Pe+\delta+\tfrac{\Pe}{\delta}\leq\tfrac{1}{5}\), 
we have 
\begin{subequations}
\label{eq:conbits-a}
  \begin{align}
\label{eq:conbits1}
(1-\epst{3}) \ex_i - \epst{5}
&\leq  (1- \sum\nolimits_{j=1}^{\nlay}\fr_j) \DX+ \sum\nolimits_{j=i+1}^{\nlay}\fr_j \JX{\tfrac{(1-\epst{3})\rate_{j}}{\fr_j}}&&i=1,2,\ldots, \nlay\\
\label{eq:conbits2}
(1-\epst{3})\rate_{i}-\epst{4}\IND{i=1}
&\leq \CX \fr_i &&i=1,2,\ldots, \nlay
\end{align}
\end{subequations}
for some time sharing vector \(\vec{\fr}\) such that
\begin{subequations}
\label{eq:conbits-b}
  \begin{align}
\label{eq:conbits3a}
\fr_i
&\geq 0 &&i=1,2,\ldots, \nlay\\
\label{eq:conbits3}
\sum\nolimits_{j=1}^{ \nlay} \fr_j
&\leq1 &&
\end{align}
\end{subequations}
where \(\rate_i=\tfrac{|\ln \mesS_{i}|}{\EX{\dt}}\), 
\(\ex_i=\tfrac{-\ln \Peb{i}}{\EX{\dt}}\),  \(\epst{3}=\Pe+\delta+\tfrac{\Pe}{\delta}\), \(\epst{4}=\tfrac{\bent{\epst{3}}}{\EX{\dt}}\)
\(\epst{5}=\tfrac{\bent{\epst{3}}-\ln \mtp \delta}{\EX{\dt}}\).
\end{lemma}

For any reliable sequence \(\SC\) whose message sets \(\mesS^{(\inx)}\) are of the form 
\(  \mesS^{(\inx)}=\mesS_{1}^{(\inx)} \times \mesS_{2}^{(\inx)} \times \ldots \times \mesS_{ \nlay}^{(\inx)}\)
if we set \(\delta \) to \(\delta=\tfrac{1}{-\ln \Pe }\) Lemma \ref{lem:conbits} implies that there exists a \(\vec{\fr}\) 
such that\footnote{This fact is far from trivial, yet it is intuitive to all who has worked with sequences of vectors 
in a bounded subset of \(\real^{\nlay}\) where \(\real^{\nlay}\) is the \(\nlay \) dimensional real vector space  with 
the norm \(\lVert \vec{X} \rVert =\sup_{j} |x_j|\) For details see Appendix  \ref{sec:appbits}}
  \begin{subequations}
\label{eq:bitsr-z-con}
   \begin{align}
\label{eq:bitsa-z-con}
\ex_{\SC,i} &\leq (1-\sum\nolimits_{j=1}^{k}\fr_j) \DX + \sum\nolimits_{j=i+1}^{ \nlay} \fr_j \JX{\tfrac{\rate_{\SC,j}}
{\fr_j}} && \forall i\in \{1,2,\ldots, \nlay\}\\
\label{eq:bitsb-z-con}
\rate_{\SC,i}&\leq \CX \fr_{i} && \forall i\in \{1,2,\ldots, \nlay\}\\
\label{eq:bitsbc-z-con}
\fr_i &\geq 0 && \forall i\in \{1,2,\ldots, \nlay\}\\
\label{eq:bitsc-z-con}
\sum\nolimits_{j=1}^{ \nlay}\fr_j&\leq 1.     &&
   \end{align}
\end{subequations}
Recall  that a \rev~  \((\vec{\rate},\vec{\ex})\) is achievable 
only if there exists a reliable code sequence \(\SC\) such that
 \((\vec{\rate}_{\SC},\vec{\ex}_{\SC})=(\vec{\rate},\vec{\ex})\).
 Thus a \rev~  \((\vec{\rate},\vec{\ex})\) is achievable  only 
 if there exists a  time sharing vector \(\vec{\fr}\) satisfying equation  (\ref{eq:bitsr-z}). In other words 
 the  sufficient condition for the achievablity of \((\vec{\rate},\vec{\ex})\) we have  derived in Section \ref{sec:achsb} is also a necessary condition.

 \section{Conclusions}\label{sec:conc}

We have considered the single message \emph{message-wise} and the fixed \( \nlay\)  \emph{bit-wise} 
\uep~problems and characterized the achievable rate  error exponent regions completely for both of
 the problems.

In  \emph{bit-wise}  \uep~ problem
we have observed that encoding schemes decoupling the   communication and bulk of the 
error correction both at the transmitter and  at the receiver  can achieve optimal performance.
This result is extending the similar observations made for conventional variable length block 
coding schemes without \uep.    However, for doing that one needs to go beyond the idea of 
communication phase and  control phase introduced in \cite{itoh}, and 
harness the implicit confirmation explicit rejection schemes,  introduced by 
 Kudryashov in \cite{kud}.

For the converses results, we have introduced  a new technique for establishing outer bounds to 
the performance of the variable length block codes, that can be use in 
both \emph{message-wise} and \emph{bit-wise} \uep~problems.\footnote{We have not employed 
the bound in any hybrid problem but it seems result is abstract enough to be employed even in those
problems with judicious choice of \nald's.}

We were only interested in \emph{bit-wise} \uep~problem in this paper. We have analyzed 
single-message \emph{message-wise} \uep~problem, because it is closely related to \emph{bit-wise} 
\uep~problem and its analysis allowed us to introduce the ideas we use  for  \emph{bit-wise} \uep, gradually.  
However it seems using the technique employed in \cite[Theorem 9]{bnz} on the achievablity side
and Lemma \ref{lem:con} on the converse side, one might be able to determine the achievable region
of rates-exponents vectors for variable length block codes in \emph{message-wise} \uep~problem. 
Such a work would allow us to determine the gains of feedback and variable length decoding, because 
Csisz\'ar \cite{csiszar1} had already solved the problem for fixed length block codes.

Arguably, the most important  shortcoming of our \emph{bit-wise} \uep~ result is that it only addresses the 
case when the number of groups of bits \(\nlay\) is a fixed integer. However this has more to do with the 
formal definition of the problem we have chosen  in Section \ref{sec:psres} than our analysis and 
non-asymptotic  results given in Sections  \ref{sec:ach} and  \ref{sec:con}, i.e.,  Lemma  \ref{lem:ach3} 
and  Lemma  \ref{lem:conbits}.

Using the \revs~ for representing the performance of a reliable sequence 
with \emph{bit-wise} \uep,  is apt only when the number of groups of bits are fixed or bounded. When  the
 number of groups of bits \( \nlay\) in a reliable sequence diverge  with increasing \(\inx\),  i.e., when
\(\lim_{\inx \to \infty} \nlay_{\inx}=\infty\),  the \rev~ formulation becomes 
fundamentally inapt.    Consider, for example, a reliable sequence in which  
\( |\mesS_{i}^{(\inx)}|= \lceil e^{\EX{\dt^{(\inx)}}  \frac{\rate}{\nlay_{\inx}}} \rceil\). The rate of this reliable 
sequence is \( \rate\), yet  the rate of all of the sub-messages are zero.  Thus  when \( \nlay_{\inx}\) diverges 
the rate vector  does not have the same operational relevance or meaning it has when \( \nlay_{\inx}\) is
fixed or bounded. In order to characterize  the change of error performance among sub-messages in the case
when \( \nlay_{\inx}\) diverges, one  needs to come up with an alternative formulation of the  problem, in terms 
of  cumulative rate of sub-messages. 

Our non-asymptotic results are useful to some extend even when \(\nlay_{\inx}\) diverges. Although
infinite dimensional \revs~ falls short of representing all achievable 
performances one can still use Lemma \ref{lem:ach3} of Section  \ref{sec:ach} and 
Lemma \ref{lem:conbits} of Section \ref{sec:con} to characterize the set of achievable rate vector 
error exponent vector pairs.     
\begin{itemize}
\item As a result of Lemma \ref{lem:conbits} the necessary condition given in equation (\ref{eq:bits}) is 
still a  necessary condition for the achievablity of \rev. 
\item Using   Lemma \ref{lem:ach3}  we see that  the  sufficient condition given in equation 
(\ref{eq:bits}) is still a  sufficient  condition as long as the number of sub-messages in the reliable sequence 
satisfy  \(\limsup\nolimits_{\blx \to \infty} \frac{\nlay_{\blx} }{\blx/\ln \blx}=0.\)
\end{itemize}
Thus for the case when   \(\nlay \sim \ord{ \tfrac{\EX{\dt}}{\ln \EX{\dt}}}  \), i.e.,
 \(\limsup\limits_{\inx \to \infty} \frac{\nlay_{\inx}}{\EXS{\inx}{\dt^{(\inx)}}/\ln \EXS{\inx}{\dt^{(\inx)}}}=0\),   
 the condition given in equation (\ref{eq:bits}) is still a necessary and sufficient condition for the achievablity of a \rev.

\section*{Acknowledgment}
Authors would like to thank Gerhard Kramer and Reviewer II for pointing out the importance 
of the case  of varying the number of groups of bits with expected block length. Reviewer II 
has also coined the term \rev~ which replaced authors' previous attempt: rate vector error 
exponent vector pair. Authors would also like to 
thank all of the reviewers for their  meticulous reviews and penetrating questions.
\appendix

\subsection{Proof of Lemma \ref{lem:fx}}
\label{sec:appfx}
\begin{proof}
Note that \(\JX{\rate}\) defined in equation (\ref{eq:FX}) is also equal to 
\begin{align}
\notag
 \JX{\rate}
% &=\max_{\tsc,\dinp_1,\dinp_2,\idis{1},\idis{2}:\substack{
 %0\leq\tsc \leq 1\\
%\dinp_1,\dinp_2 \in \inpS\\ 
%\idis{1}, \idis{2} \in \PDS{\inpS}\\
 %\tsc \MI{\idis{1}}{\CTM}+(1-\tsc) \MI{\idis{2}}{\CTM}\geq \rate }} 
 %\tsc  \KLD{\odis{1}}{\CTM_{\dinp_1}}+(1-\tsc)  \KLD{\odis{2}}{\CTM_{\dinp_2}}\\
 %\notag
&=\max_{\tsc,\dinp_1,\dinp_2,\idis{1},\idis{2},\rate_1,\rate_2:\substack{
0\leq\tsc \leq 1\\
\dinp_1,\dinp_2 \in \inpS\\ 
\idis{1}, \idis{2} \in \PDS{\inpS}\\  
\rate_1,\rate_2\in [0,\CX]\\
\MI{\idis{1}}{\CTM}\geq \rate_1\\
\MI{\idis{2}}{\CTM}\geq \rate_2\\
\tsc \rate_1+(1-\tsc) \rate_2 = \rate }} 
\tsc  \KLD{\odis{1}}{\CTM_{\dinp_1}}+(1-\tsc)  \KLD{\odis{2}}{\CTM_{\dinp_2}}\\
%\notag
%&=\max_{\tsc,\rate_1,\rate_2:\substack{
%0\leq\tsc \leq 1\\
%\tsc \rate_1+(1-\tsc) \rate_2= \rate }} 
%   \left(
 %     \tsc   \max_{\dinp_1,\idis{1}:\substack{
%\MI{\idis{1}}{\CTM}\geq \rate_1\\
%\dinp_1\in \inpS\\ 
%\idis{1} \in \PDS{\inpS}}}
 % \KLD{\odis{1}}{\CTM_{\dinp_1}}   
%+   
 %  (1-\tsc) \max_{\dinp_2,\idis{2}:\substack{
%\MI{\idis{2}}{\CTM}\geq \rate_2\\
%\dinp_2 \in \inpS\\
%\idis{2} \in \PDS{\inpS}}}  
%   \KLD{\odis{2}}{\CTM_{\dinp_2}}
% \right)\\
\label{eq:FXfx}
&=\max_{\tsc,\rate_1,\rate_2:\substack{
0\leq\tsc \leq 1\\
\rate_1,\rate_2\in [0,\CX]\\
\tsc \rate_1+(1-\tsc) \rate_2= \rate }} 
   \tsc   \jx{\rate_1} +    (1-\tsc) \jx{\rate_2} 
        \end{align}
where \(\jx{\rate}\) is given by
\begin{equation}
\label{eq:fx}
  \jx{\rate}\DEF
  \max_{\tsc,\dinp,\idis{}:\substack{
  \dinp \in \inpS\\ 
  \idis{} \in \PDS{\inpS}\\
   \MI{\idis{}}{\CTM} \geq \rate }}
     \KLD{\odis{}}{\CTM_{\dinp}} 
     \qquad 
     \forall \rate \in \CX.
  \end{equation}
 Note that \(\jx{\rate}\) is a bounded  real valued  function of a real variable. Therefore,  Carath\'{e}odory's Theorem implies that considering   two point convex combinations suffices in order make \(\jx{\rate}\) a concave function. In other words for any \(k\) we have,
 \begin{equation}
\label{eq:fxconcave}
\max_{\tsc,\rate_1,\rate_2:\substack{
0\leq\tsc \leq 1\\
\rate_1,\rate_2 \in [0,\CX]\\
\tsc \rate_1+(1-\tsc) \rate_2= \rate }} 
   \tsc   \jx{\rate_1} +    (1-\tsc) \jx{\rate_2} 
=\max_{
\substack{\tsc_1,\ldots,\tsc_k,\\\rate_1,\ldots,\rate_k}:
\substack{
0\leq \tsc_i  \leq1~\forall i\\
0 \leq \rate_i\leq \CX ~\forall i\\
\sum_{i}\tsc_i       = 1\\
\sum_{i} \tsc_i \rate_i = \rate }} 
\sum\nolimits_{i=1}^{k} \tsc_i    \jx{\rate_{i}}.  
\end{equation}
Then the concavity of \(\JX{\rate}\) follows from the  equations (\ref{eq:FXfx}), (\ref{eq:fx}) and (\ref{eq:fxconcave}).

Evidently if the constraint set in a maximization is curtailed than resulting maximum value 
can not increase. Hence \(\JX{\rate}\) function defined in equation (\ref{eq:FX}) is a  decreasing function of \(\rate\).

As a result of the definition of \(\DX\) given in equation (\ref{eq:dxdef}) and the convexity of Kullback-Leibler  divergence, we have \(\DX \geq \JX{0}\). On the other hand  
\( \KLD{\odis{}}{\CTM_{\dinp}} =\DX\) and \( \MI{\idis{}}{\CTM} \geq 0 \) for \(\dinp=\xr\) and \(\idis{}(\cdot)=\IND{\cdot=\xa}\)
where \(\xa\) and \(\xr\) described in equation (\ref{eq:dm}).
 Therefore we have \(\jx{0} \geq \DX \). Using the fact that 	\(\JX{\rate}\geq \jx{\rate}\) we conclude that 
\(\JX{0}=\jx{0}=\DX\).
\end{proof}

\subsection{Proof of Lemma \ref{lem:ach1}}
\label{sec:appach1}
\begin{proof}
We prove the lemma  for a slightly more general  setting and establish a result that will be easier to 
make use of in the proofs of other achievablity results.  Let \(\decsg{1}\),  \(\decsg{\dmes}\)
and \(\decpg{\dinp^{\blx}}\)  be
\begin{subequations}
\begin{align}
\label{eq:dec-reg-oo-a}
\decsg{1}
&=\left\{\dout^{\blx}:
\blx_{\tsc}  \TV{\emp{\dout_1^{\blx_{\tsc}}}}{\odis{1}}+ (\blx-\blx_{\tsc}) \TV{\emp{\dout_{\blx_{\tsc}+1}^{\blx}}}{\odis{2}} \geq  \dtr \right\}
\\
\label{eq:DECREGO-a}
\decsg{\dmes}
&=
\decpg{\dinp^{\blx}(\dmes)} \bigcap \left( \cap_{\widetilde{\dmes}\neq \dmes} \overline{\decpg{\dinp^{\blx}(\widetilde{\dmes})}}  \right)
 \qquad \forall \dmes \in \{2,3,\ldots,|\mesS|\}\\
 \label{eq:dec-preg-r-a}
    \decpg{\dinp^{\blx}}
&=\{\dout^{\blx}:
\blx_{\tsc} 
\TV{\emp{\dinp_{1}^{\blx_{\tsc}},\dout_{1}^{\blx_{\tsc}}}}{\idis{1}\CTM}
+(\blx-\blx_{\tsc})
\TV{\emp{\dinp_{\blx_{\tsc}+1}^{\blx},\dout_{\blx_{\tsc}+1}^{\blx}}}{\idis{2}\CTM}
< \dtr
\}.
\end{align}
\end{subequations}	
Note that  \(\decs{1}\),  \(\decs{\dmes}\) and  \(\decp{\dinp^{\blx}} \) given equations 
(\ref{eq:dec-reg-oo}),  (\ref{eq:DECREGO}) and  (\ref{eq:dec-preg-r}) are simply the 
\(\decsg{1}\), \(\decsg{\dmes}\) and \(\decpg{\dinp^{\blx}} \)  for \(\dtr=|\inpS| |\outS|\sqrt{\blx \ln(1+\blx)} \).

For all \(\dout^{\blx}\notin \decsg{1}\) we have,
\begin{align*}
\blx_{\tsc}  \KLD{\emp{\dout_1^{\blx_{\tsc}}}}{\CTM_{\dinp_1}}+ (\blx-\blx_{\tsc})
& \KLD{\emp{\dout_{\blx_{\tsc}+1}^{\blx}}}{\CTM_{\dinp_2}}\\
&=
\blx_{\tsc}  
\KLD{\emp{\dout_1^{\blx_{\tsc}}}}{\odis{1}}+ 
(\blx-\blx_{\tsc}) \KLD{\emp{\dout_{\blx_{\tsc}+1}^{\blx}}}{\odis{2}}\\
&\qquad +\blx_{\tsc} \sum_{\dout} \emp{\dout_{1}^{\blx_{\tsc}}}(\dout)       
  \ln \tfrac{\odis{1}(\dout)}{\CT{\dinp_1}{\dout}}
+ (\blx-\blx_{\tsc}) \sum_{\dout} \emp{\dout_{\blx_{\tsc}+1}^{\blx}}(\dout) \ln \tfrac{\odis{2}(\dout)}{\CT{\dinp_2}{\dout}}\\
&\mathop{\geq}^{(a)}
\blx_{\tsc} \sum_{\dout} \emp{\dout_1^{\blx_{\tsc}}}(\dout) \ln \tfrac{\odis{1}(\dout)}{\CT{\dinp_1}{\dout}}
+ (\blx-\blx_{\tsc}) \sum_{\dout} \emp{\dout_{\blx_{\tsc}+1}^{\blx}}(\dout) \ln \tfrac{\odis{2}(\dout)}{\CT{\dinp_2}{\dout}}\\
&\mathop{\geq}^{(b)}
 \blx_{\tsc} \KLD{\odis{1}}{\CTM_{\dinp_1}}+(\blx-\blx_{\tsc}) \KLD{\odis{2}}{\CTM_{\dinp_2}} +2\dtr \ln \mtp.
\end{align*}
Inequality  \((a)\) follows from the non-negativity of the Kullback Leibler divergence. In order to
 see why \((b)\) holds, first recall that \(\min_{\dinp,\dout} \CT{\dinp}{\dout} =\mtp\). Hence
\(\lvert \ln \tfrac{\odis{1}(\dout)}{\CT{\dinp_1}{\dout}} \rvert \leq \ln \tfrac{1}{\mtp} \)
and
\(\lvert \ln \tfrac{\odis{2}(\dout)}{\CT{\dinp_2}{\dout}} \rvert \leq \ln \tfrac{1}{\mtp} \).
Then the inequality 
\((b)\) follows from the definitions of total variation \(\TVO\) and \(\decsg{1}\), given in 
equations (\ref{eq:deftv}) and (\ref{eq:dec-reg-oo-a}) and the fact
 that  \(\dout^{\blx}\notin \decsg{1}\).

Note that the conditional error probability of the first message is given by
\begin{align*}
\Pem{1}
&=\PCX{\hat{\mes}\neq 1}{\mes=1}\\
&=\sum\limits_{\dout^{\blx}\notin \decsg{1}} \PCX{\out^{\blx}=\dout^{\blx}}{\mes=1}.
\end{align*}

Recall that,  the codeword of the message $\mes=1$ is the  concatenation of $\blx_{\tsc}$ 
$\dinp_1$'s and $(\blx-\blx_{\tsc})$ $\dinp_2$'s where $\blx_{\tsc}=\lfloor \blx \tsc \rfloor$.
 Hence  the probability of all \(\dout^{\blx}\)'s whose empirical  distribution in first \(\blx_{\tsc}\)
 times  instances is \(\emp{\dout_1^{\blx_{\tsc}}}\) and whose empirical distribution in 
\([(\blx_{\tsc}+1),\blx]\) is \(\emp{\dout_{\blx_{\tsc}+1}^{\blx}}\) is upper bounded by 
\(e^{-\blx_{\tsc}  \KLD{\emp{\dout_1^{\blx_{\tsc}}}}{\CTM_{\dinp_1}}-(\blx-\blx_{\tsc}) 
 \KLD{\emp{\dout_{\blx_{\tsc}+1}^{\blx}}}{\CTM_{\dinp_2}}}\).
Furthermore, there are less than $(\blx_{\tsc}+1)^{|{\outS}|}$ distinct empirical distributions 
in the first phase and  there are less than $(\blx- \blx_{\tsc}+1)^{|{\outS}|}$ distinct empirical 
distributions in the second phase. Thus
\begin{align*}
{\Pem{1}}
&\mathop{\leq}  (\blx_{\tsc}+1)^{|{\outS}|} (\blx-\blx_{\tsc}+1)^{|{\outS}|}  
e^{-\blx_{\tsc} \KLD{\odis{1}}{\CTM_{\dinp_1}}+(\blx-\blx_{\tsc}) \KLD{\odis{2}}{\CTM_{\dinp_2}} -2\dtr \ln \mtp}\\
&\mathop{\leq} e^{-\blx(\tsc \KLD{\odis{1}}{\CTM_{\dinp_1}}+(1-\tsc)\KLD{\odis{2}}{\CTM_{\dinp_2}} -\eps{2}{\dtr,\blx})}
\end{align*}
where $\eps{2}{\dtr,\blx}=\tfrac{-2\dtr \ln \mtp+\DX+2|{\outS}| \ln (\blx+1)}{\blx}$.

The codewords and the decoding  regions of the remaining messages are specified using a random  
coding argument together with an empirical typicality decoder.  Consider an ensemble of codes in
 which first  $\blx_{\tsc}$ entries of all the codewords are independent and identically distributed 
(i.i.d.) with  input distribution $\idis{1}$ and the rest of the entries are i.i.d. with the input 
distribution $\idis{2}$.

For any message \(\dmes\) other than the first one,  i.e., \(\dmes\neq1\), the decoding region 
is  \(\decsg{\dmes}\) given in   (\ref{eq:DECREGO-a}). In other words for any message \(\dmes\),
 other than the  first one,  the decoding region  is the set of output sequences for which   
\((\dinp^{\blx}(\dmes),\dout^{\blx})\)  is typical   with \((\tsc,\idis{1}\CTM,\idis{2}\CTM)\), i.e.,  
\(\dout^{\blx} \in \decpg{\dinp^{\blx}(\dmes)}\),  and \((\dinp^{\blx}(\widetilde{\dmes}),\dout^{\blx})\) is not  typical  with \((\tsc,\idis{1}\CTM,\idis{2}\CTM)\),
  i.e.,  \(\dout^{\blx} \in \decpg{\dinp^{\blx}(\widetilde{\dmes})}\),   for any \(\widetilde{\dmes}\neq \dmes\).

Since the decoding regions of different messages are disjoint, above described code does not decode to 
more than one message. Disjointness of decoding regions of messages \(2\),\(3\),\(\ldots\),\(|\mesS|\) follows 
from the  definitions of \(\decsg{2}\),\(\decsg{3}\),\(\ldots\),\(\decsg{|\mesS|}\), given in equation 
 (\ref{eq:DECREGO-a}). In order  to see why 
  \(\decsg{1} \cap (\cup_{\dmes \neq 1}\decsg{\dmes}) =\emptyset\)  holds, note that  for any pair 
probability of distributions, the total variation between them is lower bounded by the total variation  
between their marginals. In particular, 
\begin{align*}
\TV{\emp{\dinp_{1}^{\blx_{\tsc}}(\dmes),\dout_{1}^{\blx_{\tsc}}}}{\idis{1}\CTM}
&\geq \TV{\emp{\dout_{1}^{\blx_{\tsc}}}}{\odis{1}}\\
\TV{\emp{\dinp_{\blx_{\tsc}+1}^{\blx}(\dmes),\dout_{\blx_{\tsc}+1}^{\blx}}}{\idis{2}\CTM}
&\geq \TV{\emp{\dout_{\blx_{\tsc}+1}^{\blx}}}{\odis{2}}.
\end{align*}
 Then as results of definitions of \(\decsg{1}\),  \(\decpg{\dinp^{\blx}}\) and \(\decsg{\dmes}\) 
 for \(\dmes\neq 1\)     given in equations  (\ref{eq:dec-reg-oo-a}), (\ref{eq:dec-preg-r-a}) 
 and (\ref{eq:DECREGO-a}) we have
\begin{equation*}
 \decsg{1}
 \cap
 \decsg{\dmes} 
 =\emptyset
 \qquad 
 \dmes=2,3,\ldots |\mesS|.
\end{equation*}

Then for \(\dmes \in \{2,3,\ldots,|\mesS|\}\) the average of the conditional error probability of $\dmes^{{th}}$ message over the ensemble is upper bounded as
\begin{equation}
\label{eq:q-sch0}
\EX{\Pem{\dmes}}\leq \PCX{\out^{\blx} \notin \decpg{\inp^{\blx}(\dmes)}}{\mes=\dmes}+\sum_{\widetilde{\dmes}\neq \dmes}
\PCX{\out^{\blx} \in \decpg{\inp^{\blx}(\widetilde{\dmes})}}{\mes=\dmes}.
\end{equation}
Let us start with bounding $\PCX{\out^{\blx} \notin \decpg{\inp^{\blx}(\dmes)}}{\mes=\dmes}$.
Let \(\tvs_1(\dinp,\dout)\) and \( \tvs_2(\dinp,\dout) \) be
\begin{subequations}
\label{eq:sch1}
\begin{align}
\label{eq:sch1a}
\tvs_1(\dinp,\dout)
&\DEF \blx_{\tsc}|\emp{\inp_{1}^{\blx_{\tsc}}(\dmes),\out_{1}^{\blx_{\tsc}}}(\dinp,\dout)-\idis{1}(\dinp)\CT{\dinp}{\dout}|,\\
\label{eq:sch1b}
\tvs_2(\dinp,\dout)
&\DEF (\blx-\blx_{\tsc}) |\emp{\inp_{\blx_{\tsc}+1}^{\blx}(\dmes),\out_{\blx_{\tsc}+1}^{\blx}}(\dinp,\dout) -\idis{2}(\dinp)\CT{\dinp}{\dout}|.
\end{align}
\end{subequations}
As a result of the definition of total variation distance given in equation (\ref{eq:deftv})  
and above definitions we have
\begin{align*}
\blx_{\tsc} 
\TV{\emp{\inp_{1}^{\blx_{\tsc}}(\dmes),\out_{1}^{\blx_{\tsc}}}}{\idis{1}\CTM}
+(\blx-\blx_{\tsc})
\TV{\emp{\inp_{\blx_1+1}^{\blx}(\dmes),\out_{\blx_1+1}^{\blx}}}{\idis{2}\CTM}
=\tfrac{1}{2} \sum\nolimits_{\dinp,\dout} 
[\tvs_1(\dinp,\dout)+\tvs_2(\dinp,\dout)]
\end{align*}
 Thus the definition of \(\decpg{\dinp^{\blx}(\dmes)}\) given in equation  (\ref{eq:dec-preg-r-a}) implies that
\begin{align}
\label{eq:sch2-new0} 
\PCX{\out \notin \decpg{\inp^{\blx}(\dmes)}}{\mes=\dmes}
&= \PCX{ \sum_{\dinp,\dout} [\tvs_1(\dinp,\dout)+\tvs_2(\dinp,\dout)] \geq   2\dtr}{\mes=\dmes}.
\end{align}
If for all \(\dinp \in \inpS\), \(\dout \in \outS\) and \(j\in\{1,2\}  \),
 \( \tvs_j(\dinp,\dout)\leq    \dtr |\inpS|^{-1} |\outS|^{-1} \) 
then   \( \sum_{\dinp,\dout} [\tvs_1(\dinp,\dout)+\tvs_2(\dinp,\dout)] \leq   2\dtr \). 
Thus if \( \out \notin \decpg{\inp^{\blx}(\dmes)} \) then for at least one \((\dinp,\dout,j)\) triple 
 \( \tvs_j(\dinp,\dout)\geq   \dtr |\inpS|^{-1} |\outS|^{-1}\).  Using the union bound we get 
\begin{align}
\label{eq:sch2-new1} 
 \PCX{ \sum\nolimits_{\dinp,\dout} [\tvs_1(\dinp,\dout)+\tvs_2(\dinp,\dout)] \geq   2\dtr}{\mes=\dmes}
&\leq \sum\nolimits_{\dinp,\dout,j}  \PCX{  \tvs_j(\dinp,\dout) \geq   \tfrac{\dtr}{|\inpS| |\outS|}}{\mes=\dmes}.
\end{align}
For bounding \(  \PCX{  \tvs_j(\dinp,\dout) \geq  \tfrac{\dtr}{|\inpS| |\outS|}}{\mes=\dmes} \), we can simply use Chebyshev's inequality,  however in order to get better error terms we use a standard concentration result about the sums of bounded random variables, \cite[Theorem 5.3]{chunglu}.
\begin{lemma}
 Let  \(\dum_1,\dum_2, \ldots, \dum_{k}\) be independent random variables satisfying \( |\dum_{i} -\EX{\dum_i}| \leq c_i \) for all \( 1\leq i \leq k \). Then 
 \begin{equation*}
 \PX{ \left\lvert \sum\nolimits_{i=1}^{k} (\dum_i- \EX{\dum_i}) \right\rvert > \dtr} 
\leq  2 e^{-\frac{\dtr^2}{2\sum_{i=1}^{k}c_i}}.
\end{equation*}
\end{lemma}
For all \(\idis{1} \in  \PDS{\inpS} \), \(\dinp\in \inpS\), \(\dout \in \outS\)  we have  \(c_i=1\) for all \( i=1,2,\ldots, \blx_{\tsc}\);  thus
 \begin{align}
\notag
\PCX{ \tvs_1(\dinp,\dout) \geq   \tfrac{\dtr}{|\inpS| |\outS|}}{\mes=\dmes}
&\leq 2 e^{-\tfrac{\dtr^2}{2|\inpS|^2 |\outS|^2 \blx_{\tsc}}}\\  
\label{eq:sch2-new2} 
&\leq  2 e^{-\tfrac{\dtr^2}{2|\inpS|^2 |\outS|^2 \blx}}.  
\end{align}
Similarly,
 \begin{align}
\PCX{  \tvs_2(\dinp,\dout) \geq   \tfrac{\dtr}{|\inpS| |\outS|}}{\mes=\dmes}
\label{eq:sch2-new3} 
&\leq  2 e^{-\tfrac{\dtr^2}{2|\inpS|^2 |\outS|^2 \blx}}  
\end{align}
Using equations  (\ref{eq:sch2-new0}),  (\ref{eq:sch2-new1}), (\ref{eq:sch2-new2}) and (\ref{eq:sch2-new3}) we get
\begin{align}
\label{eq:sch2-new4} 
\PCX{\out \notin \decpg{\inp^{\blx}(\dmes)}}{\mes=\dmes}
&\leq  4 |\inpS| \outS| e^{-\tfrac{\dtr^2}{2|\inpS|^2 |\outS|^2 \blx}}
%{eq:q-sch1}
\end{align}

Now we focus on $\PCX{\out^{\blx} \in \decpg{\inp^{\blx}(\widetilde{\dmes})}}{\mes=\dmes}$ terms. Note that all  \(\dout^{\blx}\) in \(\decpg{\dinp^{\blx}(\widetilde{\dmes})}\) satisfy 
 \begin{equation}
\label{eq:ach-ty1}
\blx_{\tsc} 
\TV{\emp{\dinp_{1}^{\blx_{\tsc}}(\widetilde{\dmes}),\dout_{1}^{\blx_{\tsc}}}}{\idis{1}\CTM}
+(\blx-\blx_1)
\TV{\emp{\dinp_{\blx_{\tsc}+1}^{\blx}(\widetilde{\dmes}),\dout_{\blx_{\tsc}+1}^{\blx}}}{\idis{2}\CTM}
\leq \dtr.
\end{equation}
 On the other hand, when \( \mes=\dmes \),    \(\inp^{\blx}(\widetilde{\dmes})\)
and \(\out^{\blx}\) are  independent and their distribution is given by,
\begin{align}
\notag
\hspace{-0.3cm}\PCX{(\inp^{\blx}(\widetilde{\dmes}),\out^{\blx})\!=\!(\dinp^{\blx}(\widetilde{\dmes}),\dout^{\blx})}{\mes\!=\!\dmes}  
&=
\prod\nolimits_{i=1}^{\blx_{\tsc}} \idis{1}(\dinp_i(\widetilde{\dmes})) \odis{1}(\dout_i) 
\prod\nolimits_{j=\blx_{\tsc}+1}^{\blx}  \idis{2}(\dinp_j(\widetilde{\dmes})) \odis{2}(\dout_j)\\   
\notag
&=e^{-\blx_{\tsc}\KLD{\emp{\dinp_{1}^{\blx_{\tsc}}(\widetilde{\dmes}),\dout_{1}^{\blx_{\tsc}}}}{\idis{1}\odis{1}}}
e^{-\blx_{\tsc}\ENT{\emp{\dinp_{1}^{\blx_{\tsc}}(\widetilde{\dmes}),\dout_{1}^{\blx_{\tsc}}}}}\\
&\qquad 
e^{-(\blx-\blx_{\tsc})\KLD{\emp{\dinp_{\blx_{\tsc}+1}^{\blx}(\widetilde{\dmes}),\dout_{\blx_{\tsc}+1}^{\blx}}}{\idis{2}\odis{2}}}
e^{ -(\blx-\blx_{\tsc}) \ENT{\emp{\dinp_{\blx_{\tsc}+1}^{\blx}(\widetilde{\dmes}),\dout_{\blx_{\tsc}+1}^{\blx}}}}.
\label{eq:ach-ty2}
\end{align}
Furthermore the number of \((\dinp_{1}^{\blx_{\tsc}}(\widetilde{\dmes}),\dout_{1}^{\blx_{\tsc}})\) sequences
 with an empirical  distribution  \(\emp{\dinp_{1}^{\blx_{\tsc}}(\widetilde{\dmes}),\dout_{1}^{\blx_{\tsc}}}\) is  
upper bounded as 
\(e^{\blx_{\tsc} \ENT{\emp{\dinp_{1}^{\blx_{\tsc}}(\widetilde{\dmes}),\dout_{1}^{\blx_{\tsc}}}}}\). 
In addition there are at most \((\blx_{\tsc}+1)^{|\inpS| |\outS|}\) different empirical  distributions. 
Using these two bounds and their counter parts for  
\((\dinp_{\blx_{\tsc}+1}^{\blx}(\widetilde{\dmes}),\dout_{\blx_{\tsc}+1}^{\blx})\) together with 
equations (\ref{eq:ach-ty1}) and (\ref{eq:ach-ty2}) we get 
\begin{align}
\notag
\PCX{\out^{\blx} \in  \decpg{\inp^{\blx}(\widetilde{\dmes})}}{\mes=\dmes}
&\leq (\blx_{\tsc}+1)^{|{\inpS}| |{\outS}|} (\blx-\blx_{\tsc}+1)^{|{\inpS}| |{\outS}|} 
e^{-\blx_{\tsc} 
\KLD{\idis{1}\CTM}{\idis{1}\odis{1}}
 - (\blx-\blx_{\tsc}) 
\KLD{\idis{2}\CTM}{\idis{2}\odis{2}} 
- 2\dtr \ln \mtp} \\
\notag
&= (\blx_{\tsc}+1)^{|{\inpS}| |{\outS}|} (\blx-\blx_{\tsc}+1)^{|{\inpS}| |{\outS}|} 
e^{-\blx_{\tsc} \MI{\idis{1}}{\CTM} - (\blx-\blx_{\tsc}) \MI{\idis{2}}{\CTM} - 2\dtr  \ln \mtp} \\
\label{eq:ach-ty2-ff}
&\leq e^{-\blx (\tsc \MI{\idis{1}}{\CTM} + (1-\tsc) \MI{\idis{2}}{\CTM})} e^{\CX+2|{\inpS}||{\outS}| \ln (\blx+1) -2\dtr \ln \mtp}.
\end{align}
Hence if \(|\mesS \setminus \{1\}|=4 |\inpS|\!~|\outS| e^{-\tfrac{\dtr^2}{2|\inpS|^2 |\outS|^2 \blx}}   e^{\blx \left(\tsc \MI{\idis{1}}{\CTM} + (1-\tsc) \MI{\idis{2}}{\CTM}\right)} e^{-\CX-2|{\inpS}||{\outS}| \ln (\blx+1) + 2\dtr \ln \mtp}\)  then
\begin{equation}
\label{eq:q-sch2}
\sum\nolimits_{\widetilde{\dmes}\neq \dmes} 
\PCX{\out^{\blx} \in  \decpg{\inp^{\blx}(\widetilde{\dmes})}}{\mes=\dmes}
\leq 4 |\inpS| \outS| e^{-\tfrac{\dtr^2}{2|\inpS|^2 |\outS|^2 \blx}}.
\end{equation}
Thus the  average $\Pe$ over the ensemble  can be bounded using  (\ref{eq:q-sch0}), (\ref{eq:sch2-new1}) and(\ref{eq:q-sch2}) as
\begin{equation*}
  \EX{\Pe}\leq  8 |\inpS|\!~|\outS| e^{-\tfrac{\dtr^2}{2|\inpS|^2 |\outS|^2 \blx}}.
\end{equation*}
But if the ensemble average of the error probability is upper  bounded like this, there is at least one code that has this low error probability. Furthermore half of its messages have conditional  error probabilities less then twice this average.  
Thus  for any block length $\blx$, time sharing constant $\tsc\in[0,1]$,  input letters \(\dinp_1,\dinp_2 \in \inpS\),   input distributions \(\idis{1},\idis{2} \in \PDS{\inpS} \)   there exists a length \(\blx\)  code such that
\begin{subequations}
\label{eq:lem-ach1-g}
\begin{align}
|\mesS \setminus \{1\}|&\geq e^{\blx (\tsc \MI{\idis{1}}{\CTM} +(1-\tsc) \MI{\idis{2}}{\CTM}-\eps{1}{\dtr,\blx})}
&& \\
 \Pem{1}& \leq  e^{-\blx (\tsc  \KLD{\odis{1}}{\CTM_{\dinp_1}}+(1-\tsc)  \KLD{\odis{2}}{\CTM_{\dinp_2}}-\eps{2}{\dtr,\blx})}
&&\\
 \Pem{\dmes} &\leq \eps{3}{\dtr,\blx}
&&  \dmes=2,3,\ldots, |\mesS|
\end{align}
\end{subequations}	
where 
\begin{subequations}
\label{eq:lem-ach1-gc}
\begin{align}
\eps{1}{\dtr,\blx}
&=\tfrac{\CX-\ln (2 |{\inpS}||{\outS}|)+2|{\inpS}||{\outS}|\ln (\blx+1) - 2\dtr \ln \mtp) }{\blx}+\tfrac{\dtr^2}{2|\inpS|^2 |\outS|^2 \blx^2}\\
\eps{2}{\dtr,\blx}
&=\tfrac{\DX+2|{\outS}| \ln (\blx+1) -2\dtr \ln \mtp}{\blx}\\
\eps{3}{\dtr,\blx}
&=16 |\inpS|\!~|\outS| e^{-\tfrac{\dtr^2}{2|\inpS|^2 |\outS|^2 \blx}}.
\end{align}
\end{subequations}
Lemma \ref{lem:ach1} follows from equation  (\ref{eq:lem-ach1-g}) and the fact that 
\begin{align}
\label{eq:lem-ach1-gc-new}
\left.\eps{i}{\dtr,\blx}\right\vert_{\dtr=|\inpS| |\outS|\sqrt{\blx \ln(1+\blx)}}
&\leq  \tfrac{9|{\inpS}| |{\outS}|(1- \ln \mtp) \sqrt{\ln (1+\blx)}}{ \sqrt{\blx}} 
&& i=1,2,3
\end{align}
for  \(\eps{1}{\dtr,\blx}\), \(\eps{2}{\dtr,\blx}\) and \(\eps{3}{\dtr,\blx}\) given in equation (\ref{eq:lem-ach1-gc}).
\end{proof}

\subsection{Proof of Lemma \ref{lem:ach2}}
\label{sec:appach2}
\begin{proof}
Let \(\blx_1\) be   \(\blx_1=\lceil (1-\tfrac{\ex}{\DX}) \blx \rceil\).  Recall that  we have assumed \(\ex\leq (1-\tfrac{\rate}{\CX})\DX\), then we have  $\tfrac{\rate}{\CX} \leq 1- \frac{\ex}{\DX}$. Consequently $\tfrac{\rate}{\CX}\leq \tfrac{\blx_1}{\blx}$ and \(\tfrac{\blx}{\blx_1} \rate \leq \CX\). On the other hand as a result of equation (\ref{eq:lem-ach1-g})  and the definition of $\JX{\cdot}$ given in equation (\ref{eq:FX}), for any  positive integer \(\blx_1\), positive  real number \(\dtr_1\), rate \(\widetilde{\rate} \leq \CX\) there exists a length $\blx_1$ code such that,
\begin{subequations} 
\label{eqn:ach1-a}
\begin{align}
|\mesS|-1&\geq e^{\blx_1 [ \widetilde{\rate}-\eps{1}{\dtr_1,\blx_1}]}  && \\
\Pem{1}
& \leq  e^{-\blx_1 [ \JX{\widetilde{\rate}}-\eps{2}{\dtr_1,\blx_1}]}&
&\\
\Pem{\dmes}
 &\leq \eps{3}{\dtr_1,\blx_1} &  
 \dmes&=2,3,\ldots, |\mesS|
 \end{align}
\end{subequations}
where \(\eps{1}{\dtr_1,\blx_1}\), \(\eps{2}{\dtr_1,\blx_1}\), \(\eps{3}{\dtr_1,\blx_1}\)  are given 
in equation (\ref{eq:lem-ach1-gc}). 

We use such a code in the first phase with \(\widetilde{\rate}=\tfrac{\blx}{\blx_1} \rate\)
and call its  decoded message \(\test\), the tentative decision. Then as a result of equation (\ref{eqn:ach1-a}) and the fact that\footnote{Recall that \(\blx_1\geq (1-\ex/\DX)\blx\) and \(\JX{\cdot}\) is a non-increasing and positive function. } \(\blx_1 \JX{\frac{\blx}{\blx_1}  \rate} \geq \blx  (1-\tfrac{\ex}{\DX}) \JX{\frac{\rate}{1-\ex/\DX}}\) we get
\begin{subequations} \label{eqn:ach1}
\begin{align}
|\mesS|-1&\geq e^{\blx \rate- \blx_1\eps{1}{\dtr_1,\blx_1}}  && \\
\PCX{\test \neq \dmes}{\mes=1}
& \leq  e^{-\blx(1-\frac{\ex}{\DX}) \JX{\frac{\rate}{1-\ex/\DX}}+\blx_1\eps{2}{\dtr_1,\blx_1}}&
&\\
\PCX{\test \neq \dmes}{\mes=\dmes}
 &\leq \eps{3}{\dtr_1,\blx_1} &  
 \dmes&=2,3,\ldots, |\mesS|.
\end{align}
\end{subequations}

The transmitter knows what the tentative decision is and determines the channel inputs in the 
last \((\blx-\blx_1)\) time instances depending on   its correctness.  If  \(\test=\mes\)  the channel inputs in 
the last \((\blx-\blx_1)\) time instances are all \(\xa\), if 	\(\test \neq \mes\)  the channel inputs in the 
last \((\blx-\blx_1)\) time instances are all \(\xr\).

 After observing \(\out^{\blx}\), receiver checks whether the  empirical distribution of the channel 
 output in the last \((\blx-\blx_1)\) time units is  typical with  $\CTM_{\xa}$, if it is then \(\est=\test\)
  otherwise \(\est=\Era\).  Hence the decoding region for erasures is given by
\begin{equation*}
\decsg{\Era}=\{\dout^{\blx}:  (\blx-\blx_1) 
\TV{\emp{\dout_{\blx_1+1}^{\blx}}}{\CTM_{\xa}}\geq \dtr_2 \}.
\end{equation*}
Let us start  with bounding \(\PCX{\est=\Era}{\test=\dmes,\mes=\dmes}\), i.e., the probability of 
erasure for correct tentative decision. First note that
\begin{align*}
(\blx-\blx_1)\TV{\emp{\out_{\blx_1+1}^{\blx}}}{\CTM_{\xa}}
=\tfrac{1}{2} \sum\nolimits_{\dout}  \tvs(\dout)
\end{align*}
where \(\tvs(\dout)=(\blx-\blx_1) |\emp{\out_{\blx_1+1}^{\blx}}(\dout) -\CT{\xa}{\dout}|\).
Then following an analysis similar to that one presented between equations (\ref{eq:sch2-new0})  and (\ref{eq:sch2-new4}) we get
\begin{align}
\notag
\PCX{\est=\Era}{\test=\dmes,\mes=\dmes}
&\leq 2|\outS| e^{-\tfrac{\dtr_2^2}{2|\outS|^2 (\blx-\blx_1)}}
&&\\
&= \eps{3}{\dtr_2,\blx-\blx_1}
\label{eqn:ames:eqset1a}
&&\forall \dmes \in \mesS.
\end{align}
In order to bound  the probability of non-erasure decoding when tentative decision is incorrect, note that
\begin{align}
\notag
\PCX{\out_{\blx_1+1}^{\blx}=\dout_{\blx_1+1}^{\blx}}{\test\neq \dmes, \mes=\dmes}  
&=\prod\nolimits_{j=\blx_1+1}^{\blx}  \CT{\xr}{\dout_j}\\   
\notag
&=
e^{-(\blx-\blx_1)\KLD{\emp{\dout_{\blx_1+1}^{\blx}}}{\odis{2}}}
e^{ -(\blx-\blx_1) \ENT{\emp{\dout_{\blx_1+1}^{\blx}}}}.
\end{align}
Then following an analysis similar to the one between (\ref{eq:ach-ty2}) and (\ref{eq:ach-ty2-ff}) we get
\begin{align}
\notag
\PCX{\est\neq\Era}{\test\neq\dmes,\mes=\dmes}
&\leq \min\{(\blx-\blx_1+1)^{|{\outS}|} e^{-(\blx-\blx_1)\DX- 2\dtr_2 \ln \mtp},1\}\\
\notag
&\leq  \min\{ e^{-\blx\ex+|{\outS}| \ln(\blx-\blx_1)+\DX -2\dtr_2 \ln \mtp},1\}\\
\label{eqn:ames:eqset1b}
&\leq  \min\{ e^{-\blx\ex+(\blx-\blx_1) \eps{2}{\dtr_2,\blx-\blx_1}},1\}
&&\forall \dmes \in \mesS.
\end{align}
Furthermore  the conditional error and erasure probabilities can be bounded in terms of 
$\PCX{\test \neq \dmes}{\mes=\dmes}$,  
$ \PCX{\est\neq\Era}{\test\neq\dmes,\mes=\dmes}$  and 
$\PCX{\est=\Era}{\test=\dmes,\mes=\dmes}$  as follows.
\begin{subequations}
\label{eq:st-pec-q}
\begin{align}
\label{eq:st-pec-qa}
\Pem{\dmes} & =  \PCX{\test \neq \dmes}{\mes=\dmes}
 \PCX{\est\neq\Era}{\test\neq\dmes,\mes=\dmes} 
&&\forall \dmes \in \mesS\\
\label{eq:st-pec-qb}
\Perm{\dmes}&\leq
\PCX{\test \neq \dmes}{\mes=\dmes}
+\PCX{\est=\Era}{\test=\dmes,\mes=\dmes}
&&\forall \dmes \in \mesS.
\end{align}
\end{subequations}
Using the equations   (\ref{eqn:ach1}), (\ref{eqn:ames:eqset1a}), (\ref{eqn:ames:eqset1b}) and     (\ref{eq:st-pec-q})     we get
\begin{subequations}
\label{eq:st-pec-qq}
\begin{align}
|\mesS|-1&\geq e^{\blx \rate- \blx_1\eps{1}{\dtr_1,\blx_1}}  && \\
\Pem{1} &\leq
 e^{-\blx(1-\frac{\ex}{\DX}) \JX{\frac{\rate}{1-\ex/\DX}}+\blx_1\eps{2}{\dtr_1,\blx_1}}
\min\{ e^{-\blx\ex+\blx_2\eps{2}{\dtr_2,\blx_2}},1\}
&& \\
\Perm{1}&\leq
 e^{-\blx(1-\frac{\ex}{\DX}) \JX{\frac{\rate}{1-\ex/\DX}}+\blx_1\eps{2}{\dtr_1,\blx_1}}
+ \eps{3}{\dtr_2,\blx_2}
&&\\
\Pem{\dmes} &\leq
\eps{3}{\dtr,\blx_1}
\min\{ e^{-\blx\ex+|{\outS}| \ln(\blx+1)+\blx_2\eps{2}{\dtr_2,\blx_2}},1\}
&& \dmes \neq 1\\
\Perm{\dmes}&\leq
\eps{3}{\dtr_1,\blx_1}
+ \eps{3}{\dtr_2,\blx_2}
&&\dmes \neq 1.
\end{align}
\end{subequations}
where \( \blx_2=\blx-\blx_1 \).  

We set  \(\dtr_j=|\inpS| |\outS| \sqrt{ 5 \blx_j \ln  (1+\blx)} \)  for \(j=1,2\) and obtain
\begin{subequations}
\label{eq:st-pec-qqq}
\begin{align}
\blx_j \eps{1}{\dtr_j,\blx_j} 
 &\leq  2 |\inpS| |\outS| ( \ln (\blx+1) -\sqrt{5\blx \ln(1+\blx)}   \ln \mtp) +(5/2)\ln(1+\blx)\\
\blx_j \eps{2}{\dtr_j,\blx_j}
&\leq  2 |\inpS| |\outS| ( \ln (\blx+1) -\sqrt{5\blx \ln(1+\blx)}   \ln \mtp) +\DX \\
\eps{3}{\dtr_j,\blx_j}
&\leq  16 |\inpS| |\outS|  /(1+ \blx)^{5/2}  
\end{align}
\end{subequations}	
Lemma \ref{lem:ach2}  follows from  the identities \(|\inpS|\!\geq\!2\), \(|\outS|\!\geq\!2\), 
 \(\DX\!\leq\!\ln(\frac{1}{\mtp})\), \(\blx\!\geq\!1\) and the equations  (\ref{eq:st-pec-qq}) and  (\ref{eq:st-pec-qqq}).
\end{proof}

\subsection{Proof of Lemma \ref{lem:ach3}} 
\label{sec:appach3}
\begin{proof}
Note that given  the encoding scheme summarized in equation (\ref{eq:r-bituep-enc}) and the decoding rule given 
in equation (\ref{eq:r-bituep-dec}), if  \(\est=\Era\) then there is a \(i\leq  \nlay+1\) such that \(\test_{j}=\tmes_{j}\) 
for all \(j<i\) and \(\test_{i}\neq \tmes_{i}\). Thus the conditional erasure probability \(\Perm{\dmes} \) is upper 
bounded as 
  \begin{align}
\Perm{\dmes}
\notag
&\leq \sum\nolimits_{i=1}^{ \nlay+1}  \PCX{\test_{i} \neq (1+\dmes_{i})}{\mes=\dmes, 
\test_{1}=\tmes_{1},\ldots,\test_{i-1}=\tmes_{i-1}}\\
\label{eqn:abits:eqset3a}
&=\sum\nolimits_{i=1}^{ \nlay+1}  
\PCX{\test_{i} \neq  \tmes_{i}}{\tmes_{i}=1+\dmes_i}
\end{align}
Similarly if  \(\est\neq \Era\) and \(\est^i\neq \mes^i \) then   for all \(j>i\), \(\tmes_j=1\) and \(\test_j\neq 1\); 
furthermore  there  is a  \(k\leq i\)  such that \(\test_{j}=\tmes_{j}\) for all \(j<k\) and 
\(\test_k \neq \tmes_k\). Hence one can bound \(\Pemb{\dmes}{i}\) as
\begin{align}
\label{eqn:abits:eqset3b}
\Pemb{\dmes}{i}
& \leq \left[\sum\nolimits_{j=1}^{i}  
\PCX{\test_{j} \neq  \tmes_{j}}{\tmes_{j}=1+\dmes_j} 
\right] 
\prod\nolimits_{j=i+1}^{ \nlay+1} \PCX{\test_{j} \neq 1}{\tmes_{j} =1}.
\end{align}

 In the first \( \nlay\) phases, we use   \( \blx_i=\lfloor  \fr_i \blx \rfloor \) long codes with rate  
 \(\tfrac{\rate_i}{\fr_i} \) with the performance given in equation (\ref{eqn:ach1-a}). Thus for 
 \(1\leq i \leq  \nlay\) we have,
 \begin{subequations} 
\label{eqn:ach1-a-bwu}
\begin{align}
|\tmesS_{i}|
&\geq 1+ e^{\blx \rate_i  -\CX- \blx_i\eps{1}{\dtr_i,\blx_i}}  && \\
\PCX{\test_{i} \neq 1}{\tmes_{i} =1}
& \leq  e^{-\blx \fr_i \JX{\tfrac{\rate_i}{\fr_i}}+\DX -\blx_i \eps{2}{\dtr_i,\blx_i}}&
&\\
\PCX{\test_{i} \neq  \tmes_{i}}{\tmes_{i}=1+\dmes_i}
 &\leq \eps{3}{\dtr_i,\blx_i} &  
 \dmes_i&=1,2,3,\ldots, (|\tmesS_i|-1)
 \end{align}
\end{subequations}
where \(\eps{1}{\dtr_i,\blx_i}\), \(\eps{2}{\dtr_i,\blx_i}\), \(\eps{3}{\dtr_i,\blx_i}\)  are given 
in equation (\ref{eq:lem-ach1-gc}).

In order derive bounds corresponding to the ones given in equation (\ref{eqn:ach1-a-bwu}) for the last phase let us 
give the decoding regions for \(1\) and \(2\) for the length \(\blx_{ \nlay+1}\)	code employed between 
\((\blx+1-\blx_{ \nlay+1})\) and \(\blx\).  
\begin{align*}
 \decsg{1}&=\{\dout_{\blx+1-\blx_{ \nlay+1}}^{\blx}:\blx_{ \nlay+1}
  \TV{\emp{\dout_{\blx+1-\blx_{ \nlay+1}}^{\blx}}}{\CTM_{\xa}}\geq \dtr_{\nlay+1} \}\\
 \decsg{2}&=\{\dout_{\blx+1-\blx_{ \nlay+1}}^{\blx}: \blx_{ \nlay+1}
  \TV{\emp{\dout_{\blx+1-\blx_{ \nlay+1}}^{\blx}}}{\CTM_{\xa}}< \dtr_{\nlay+1} \}.
\end{align*}
Following an analysis similar to the one leading to  equations   (\ref{eqn:ames:eqset1a}) 
and  (\ref{eqn:ames:eqset1b}) we get
 \begin{subequations} 
\label{eqn:ach1-a-bwv}
\begin{align}
\PCX{\test_{ \nlay+1} \neq 1}{\tmes_{ \nlay+1} =1}
& \leq e^{-\blx_{ \nlay+1} \DX+\blx_{ \nlay+1} \eps{2}{\dtr_{\nlay+1},\blx_{ \nlay+1}}}\\
\PCX{\test_{ \nlay+1} \neq  2}{\tmes_{ \nlay+1}=2}
 &\leq \eps{3}{\dtr_{\nlay+1},\blx_{ \nlay+1}}
 \end{align}
\end{subequations}

Using  equations  (\ref{eqn:abits:eqset3a}), (\ref{eqn:abits:eqset3b}), (\ref{eqn:ach1-a-bwu}) 
and  (\ref{eqn:ach1-a-bwv})  we get,
\begin{subequations}
\label{eqn:ach1-a-bwy}
 \begin{align}
|\mesS_{i}|
&\geq e^{\blx \rate_i  - \blx_i\eps{1}{\dtr_i,\blx} -\CX} 
&& \forall i=1,2,\ldots, \nlay\\
|\mesS^{i}|
&\geq e^{\blx \sum_{j=1}^{i}\rate_j}e^{- \sum_{j=1}^{i}(\blx_j\eps{1}{\dtr_j,\blx_j} +\CX)} 
&& \forall i=1,2,\ldots, \nlay\\
\Pemb{\dmes}{i}
 &\leq
 \sum_{j=1}^{i}\eps{3}{\dtr_j,\blx_j}
 \min\left\{1,
    e^{-\blx\sum\limits_{j=i+1}^{ \nlay+1} \fr_i \JX{\tfrac{\rate}{\fr_i}}}
    e^{  \blx\sum\limits_{j=i+1}^{ \nlay+1} \blx_j \eps{2}{\dtr_j, \blx_j}+\DX}
  \right\}
&& \forall i=1,2,\ldots, \nlay, \forall \dmes\in\mesS\\
\Perm{\dmes} 
 &\leq  \sum_{j=1}^{\nlay+1}\eps{3}{\dtr_j,\blx_j}
&& \forall \dmes\in\mesS
\end{align}
\end{subequations}
If we set  \(\dtr_i=|\inpS| |\outS| \sqrt{4\blx_i \ln (1+\blx)} \)  for \(i=1,2,\ldots,(\nlay+1)\)
for \(\eps{1}{\dtr_i,\blx_i}\), \(\eps{2}{\dtr_i,\blx_i}\) and \(\eps{3}{\dtr_i,\blx_i}\) given in equation (\ref{eq:lem-ach1-gc}) we have
\begin{align*}
\blx_i \eps{1}{\dtr_i,\blx_i}+\CX
 &\leq  2 |\inpS| |\outS| ( \ln (\blx_i+1)-\sqrt{4\blx_i \ln (1+\blx)} \ln  \mtp)+ 2 \ln (1+\blx)+\CX\\
\blx_i \eps{2}{\dtr_i,\blx_i}+\DX
&\leq  2 |\inpS| |\outS| ( \ln (\blx_i+1)-\sqrt{4\blx_i \ln (1+\blx)}  \ln \mtp) +2\DX\\
\eps{3}{\dtr_i,\blx_i}
&\leq  16 |\inpS| |\outS|/(1+\blx)^2
\end{align*}
Using the concavity of  \( \sqrt{\ddum} \) function we can conclude that, 
\begin{subequations}
\label{eqn:ach1-a-bwy-new}
\begin{align}
\sum_{i=1}^{\nlay+1}\tfrac{\blx_i \eps{1}{\dtr_i,\blx_i}+\CX}{\blx} 
&\leq 2 |\inpS| |\outS| 
\left( \tfrac{(\nlay+1)\ln (1+\blx)}{\blx} -\tfrac{(\nlay+1)}{\blx}2\sqrt{\tfrac{\blx}{\nlay+1} \ln (1+\blx)}\ln \mtp \right)
 +\tfrac{2(\nlay+1)\ln (1+\blx)}{\blx}+ \tfrac{(\nlay+1)}{\blx} \CX\\
\sum_{i=1}^{\nlay+1}\tfrac{\blx_i \eps{2}{\dtr_i,\blx_i}+\DX}{\blx} 
&\leq 2 |\inpS| |\outS| 
\left(\tfrac{(\nlay+1)\ln (1+\blx)}{\blx} -\tfrac{(\nlay+1)}{\blx}2\sqrt{\tfrac{\blx}{\nlay+1} \ln (1+\blx)}\ln \mtp \right)
+\tfrac{\nlay+1}{\blx} 2 \DX\\
\sum_{i=1}^{\nlay+1}  \eps{3}{\dtr_i,\blx_i}
&\leq 8 |\inpS| |\outS|  \tfrac{\nlay+1}{1+\blx}
\end{align}
\end{subequations}
Then  Lemma \ref{lem:ach3} follows from equations (\ref{eqn:ach1-a-bwy}) and  (\ref{eqn:ach1-a-bwy-new})
 for any \(\nlay \leq \tfrac{\blx}{\ln (\blx+1)}\).
\end{proof}

\subsection{Proof of Lemma \ref{lem:con}}
\label{sec:appcon}
\begin{proof}
For \(\PXD\) defined in equation (\ref{eq:defpxd-a}) as a result of equation (\ref{eq:defpxd-b}) we have
\begin{subequations}
\label{eqn:con:decReg-0a}
\begin{align}
\PX{\est \in \ldsn{i}{\out^{\dt_i}}}
&=\sum\nolimits_{\dout^{\ddt}\in\left\{\dout^{\ddt}: \est \in \ldsn{i}{\out^{\dt_i}} \right\} \cap  \outS^{\dt*}}
\PXD(\dout^{\ddt})
\\
\PX{\est \notin \ldsn{i}{\out^{\dt_i}}}
&=\sum\nolimits_{\dout^{\ddt}\in\left\{\dout^{\ddt}:\est\notin\ldsn{i}{\out^{\dt_i}} \right\}\cap \outS^{\dt*}}
\PXD(\dout^{\ddt})
\end{align}
\end{subequations}
For \(\PXAD{\ldsf}\) defined in equation  (\ref{eq:defpxa-a}) as  a result of equation  (\ref{eq:defpxa-b}) we  have
\begin{subequations}
\label{eqn:con:decReg-0b}
\begin{align}
\PXA{\ldsf_i}{\est \in \ldsn{i}{\out^{\dt_i}}}
&=\sum\nolimits_{\dout^{\ddt} \in \left\{\dout^{\ddt}: \est \in \ldsn{i}{\out^{\dt_i}} \right\} \cap  \outS^{\dt*}}
\PXAD{\ldsf}(\dout^{\ddt})\\
\PXA{\ldsf_i}{\est \notin \ldsn{i}{\out^{\dt_i}}}
&=\sum\nolimits_{\dout^{\ddt} \in\left\{\dout^{\ddt}:\est\notin\ldsn{i}{\out^{\dt_i}}\right\}\cap \outS^{\dt*}}
\PXAD{\ldsf}(\dout^{\ddt}).
\end{align}
\end{subequations}
Using equations (\ref{eqn:con:decReg-0a}) and (\ref{eqn:con:decReg-0b})  together with the   data processing inequality for Kullback-Leibler divergence, we get
\begin{equation*}
\sum_{\dout^{\ddt} \in \outS^{\dt*}} 
\PXD(\dout^{\ddt})  \ln \tfrac{\PXD(\dout^{\ddt})}{\PXAD{\ldsf_i}(\dout^{\ddt})}
\geq
 \PX{ \est \in \ldsn{i}{\out^{\dt_i}} } \ln 
 \tfrac{\PX{ \est \in \ldsn{i}{\out^{\dt_i}} }}
 { \PXA{\ldsf_i}{ \est \in \ldsn{i}{\out^{\dt_i}} }}
+ \PX{ \est \notin \ldsn{i}{\out^{\dt_i}} } \ln 
\tfrac{\PX{ \est \notin \ldsn{i}{\out^{\dt_i}} }}
{\PXA{\ldsf_i}{ \est \notin \ldsn{i}{\out^{\dt_i}} }}.
\end{equation*}
Since \(0\leq \PXA{\ldsf_i}{ \!\est\!\in\!\ldsn{i}{\out^{\dt_i}} } \leq 1\)  we have
\begin{equation}
\label{eqn:con:decReg-1}
\sum_{\dout^{\ddt} \in \outS^{\dt*}}
\PXD(\dout^{\ddt})  \ln \tfrac{\PXD(\dout^{\ddt})}{\PXAD{\ldsf_i}(\dout^{\ddt})}
\geq -\bent{\PX{\est \in \ldsn{i}{\out^{\dt_i}}  }} +\left(1-\PX{\est \in \ldsn{i}{\out^{\dt_i}}}\right) 
\ln \tfrac{1}{\PXA{\ldsf_i}{\est \notin \ldsn{i}{\out^{\dt_i}} }}.
\end{equation}

Note that if \(\est \in \ldsn{i}{\out^{\dt_i}}\) and \(\mes \notin \ldsn{i}{\out^{\dt_i}}\) then \(\est\neq \mes\). 
Consequently
\begin{align}
\PX{\est \in \ldsn{i}{\out^{\dt_i}}   }
\notag
&=
\PX{\left\{
\est \in \ldsn{i}{\out^{\dt_i}},
\mes \notin \ldsn{i}{\out^{\dt_i}}
\right\}}
+
\PX{\left\{
\est \in \ldsn{i}{\out^{\dt_i}},
\mes \in \ldsn{i}{\out^{\dt_i}}
\right\}}\\
&\leq  \Pe+\PX{ \mes \in \ldsn{i}{\out^{\dt_i}} }.   
\label{eqn:con:decReg-2}
\end{align}
Since the   binary entropy function \(\bent{\cdot}\)	is increasing on the interval \([0,1/2]\) if \(\Pe+\PX{ \mes \in \ldsn{i}{\out^{\dt_i}}} \leq 1/2\)   equations \eqref{eqn:con:decReg-1} and \eqref{eqn:con:decReg-2} imply 
\begin{align}
\label{eq:con:Exp1}
\sum_{\dout^{\ddt} \in \outS^{\dt*}}  
\PXD(\dout^{\ddt}) \ln \tfrac{\PXD(\dout^{\ddt})}{\PXAD{\ldsf_i}(\dout^{\ddt})}
\geq  
-\bent{\Pe\!+\!\PX{ \mes\! \in\! \ldsn{i}{\out^{\dt_i}}  }}
+ \left(1\!-\!\Pe\!-\!\PX{ \mes\! \in\! \ldsn{i}{\out^{\dt_i}}  } \right)
\ln \tfrac{1}{ \PXA{\ldsf_i}{ \est \notin \ldsn{i}{\out^{\dt_i}} }}.
\end{align}
Let \(\mar{B}\), \(\mar{B}^{*}\) and \(\mar{B}_{\tin}\) be
\begin{subequations}
\label{eq:n-domcon}
\begin{align}
\label{eq:n-domcon-a}
\mar{B}
&\DEF\ln \tfrac{\PXD(\out^{\dt})}{\PXAD{\ldsf_i}(\out^{\dt})}
&&\\
\label{eq:n-domcon-b}
\mar{B}^{*}
&\DEF\ln \tfrac{\PXD(\out^{\dt})}{\PXAD{\ldsf_i}(\out^{\dt})} \IND{\dt<\infty}
&&\\
\label{eq:n-domcon-c}
\mar{B}_{\tin}
&\DEF\ln \tfrac{\PXD(\out^{\dt \wedge \tin})}{\PXAD{\ldsf_i}(\out^{\dt \wedge \tin})}
&\forall \tin \in \{1,2,\ldots\}&
\end{align}
\end{subequations}
where \(\dt\wedge \tin\) is the minimum of \(\dt\) and \(\tin\). 

Note that as \(\tin\)  goes to infinity,  \(\mar{B}_{\tin} \to \mar{B}\) and \(\mar{B}_{\tin} \to \mar{B}^{*}\) 
with  probability one. Since   \(|\mar{B}_{\tin}| \leq  \dt \ln \tfrac{1}{\mtp}\) and \(\EX{\dt}<\infty\),   we can apply the dominated convergence  theorem \cite[Theorem 3 p 187]{shiryaev} to obtain
\begin{equation}
\label{eq:domcon}
\EX{\mar{B}}=\EX{\mar{B}^{*}}=\lim_{\tin \rightarrow \infty} \EX{\mar{B}_{\tin}}.
\end{equation}
Finally  for  \(\mar{B}\)  and  \(\mar{B}^{*}\) defined in equation (\ref{eq:n-domcon}) we have
\begin{subequations}
\label{eq:nn-domcon}
\begin{align}
\EX{\mar{B}}
&=\EX{\ln \tfrac{\PXD(\out^{\dt})}{\PXAD{\ldsf_i}(\out^{\dt})}}\\
\EX{\mar{B}^{*}}
&=\sum_{\dout^{\ddt} \in \outS^{\dt*}}  \PXD(\dout^{\ddt})
\ln \tfrac{\PXD(\dout^{\ddt})}{\PXAD{\ldsf_i}(\dout^{\ddt})}.
\end{align}
\end{subequations}
Thus as a result of equations  (\ref{eq:domcon}) and (\ref{eq:nn-domcon}) we have
\begin{equation}
\label{eq:domcon-p}
\EX{\ln \tfrac{\PXD(\out^{\dt})}{\PXAD{\ldsf_i}(\out^{\dt})}}
=\sum_{\dout^{\ddt} \in \outS^{\dt*}} \PXD(\dout^{\ddt})  \ln \tfrac{\PXD(\dout^{\ddt})}{\PXAD{\ldsf_i}(\dout^{\ddt})}.
\end{equation}
Furthermore using the definition of \(\PXAD{\ldsf_i}\)  given in equation  (\ref{eq:defpxa-a}) we get
\begin{align}
\notag
\EX{\ln \tfrac{\PXD(\out^{\dt})}{\PXAD{\ldsf_i}(\out^{\dt})}}
&=\EX{\ln \tfrac{\PXD(\out_{\dt_i+1}^{\dt}|\out^{\dt_i})}{\PXAD{\ldsf_i}(\out_{\dt_i+1}^{\dt}|\out^{\dt_i})}}\\
\label{eq:domcon-p-x1}
&=\sum\nolimits_{j=i}^{k} \pkld{i}{j}
\end{align}
where for all \(i\geq 1\) and \(j>i\)
\begin{equation}
\label{eq:domcon-p-x2}
\pkld{i}{j}
\DEF 
\begin{Bmatrix}
0 &\mbox{if~} \PX{\dt_{j+1}=\dt_{j}}=1\\
\EX{\ln
\frac{\PXD(\out_{\dt_{j}+1}^{\dt_{j+1}}|\out^{\dt_{j}})}{\PXAD{\ldsf_{i}}(\out_{\dt_{j}+1}^{\dt_{j+1}}|\out^{\dt_{j}})}}
 & \mbox{if~}  \PX{\dt_{j+1}=\dt_{j}}<1\\
\end{Bmatrix}.
\end{equation}

Assume for the moment that,
\begin{equation}
\label{3eq:sita}
\pkld{i}{j} \leq \EX{\dt_{j+1}-\dt_{j}} \JX{\arate{j+1}}
\end{equation}
where \(\dt_{k+1}=\dt\) and \(\arate{j}\) is defined in equation (\ref{eq:defarate}).

Then Lemma  (\ref{lem:con}) follows from  equations (\ref{eq:con:Exp1}), 
(\ref{eq:domcon-p}), (\ref{eq:domcon-p-x1}) and (\ref{3eq:sita}).

Above, we have proved  Lemma \ref{lem:con} by assuming that the inequality given in (\ref{3eq:sita}) 
holds for all  \(i\) in \(\{1,2,\ldots,k\}\) and \(j\) in \(\left\{(i+1),\ldots,(k+1)\right\}\); below we  prove 
that fact.

First note that if \(\PX{\dt_{j+1}=\dt_{j}}=1\) then as result of equations  (\ref{eq:defarate}) and  (\ref{eq:domcon-p-x1}) equation  (\ref{3eq:sita}) is equivalent to \(0\leq 0 \JX{0}\) which holds 
trivially. Thus we assume hence forth that  \(\PX{\dt_{j+1}=\dt_{j}}<1\), which  implies
 \(\EX{\dt_{j+1}-\dt_{j}}>0\).

Let us consider the stochastic sequence
\begin{equation}
  \label{eq:sdef}
   \mar{U}_{\tin} = \left[
-\ln \tfrac{\PXD(\out_{\dt_{j}+1}^{\tin}|\out^{\dt_{j}})}{\PXAD{\ldsf_i}(\out_{\dt_{j}+1}^{\tin}|\out^{\dt_{j}})}
+\sum\nolimits_{k=\dt_{j}+1}^{\tin} \JX{\CMIX{\mes}{\out_{k}}{\out^{k-1}}}
\right]\IND{\tin>\dt_{j}}
\end{equation}
where \(  \CMIX{\mes}{\out_{k}}{\out^{k-1}} \) is the  conditional mutual information between \(\mes\) and \(\out_k\) given \(\out^{k-1}\), defined as 
\begin{equation*}
  \label{eq:cmidef}
  \CMIX{\mes}{\out_k}{\out^{k-1}} \DEF 
  \ECX{\ln \tfrac{\PXD(\out_k|\mes,\out^{k-1})}{\PXD(\out_k|\out^{k-1})}}{\out^{k-1}}.
\end{equation*}
Note that  as it was the case for conditional entropy, while defining the conditional mutual information 
we do not take the average over the conditioned random variable. Thus 
\(  \CMIX{\mes}{\out_k}{\out^{k-1}} \) is itself a random variable. 

For \(\mar{U}_{\tin}\)  defined in equation (\ref{eq:sdef})  we have
\begin{align}
\label{3eq:newdifeqa}
\mar{U}_{\tin+1}-\mar{U}_{\tin}
&=\left(-\ln \tfrac{\PXD(\out_{\tin+1}|\out^{\tin})}{\PXAD{\ldsf_{i}}(\out_{\tin+1}|\out^{\tin})}+\JX{\CMIX{\mes}{\out_{\tin+1}}{\out^{\tin}}}\right)\IND{\tin\geq \dt_{j}}.
\end{align}
Conditioned on $\out^{\tin}$ random variables $\mes-\inp_{\tin+1}-\out_{\tin+1}$ form a Markov chain: 
thus as a result of the data processing inequality for the mutual information  we have  
$\CMIX{\inp_{\tin+1}}{\out_{\tin+1}}{\out^{\tin}}>\CMIX{\mes}{\out_{\tin+1}}{\out^{\tin}}$.  
Since $\JX{\cdot}$ is a decreasing function this  implies that
\begin{align}
\label{3eq:newdifeqb}
\JX{\CMIX{\mes}{\out_{\tin+1}}{\out^{\tin}}}\geq
\JX{\CMIX{\inp_{\tin+1}}{\out_{\tin+1}}{\out^{\tin}}}.
\end{align}
Furthermore, because of the definitions  of \(\JX{\cdot}\), \(\PXD\) and \(\PXAD{\ldsf}\) given in equations 
(\ref{eq:FX}), (\ref{eq:defpxd-a}) and (\ref{eq:defpxa-a}),   the convexity of Kullback Leibler divergence 
and Jensen's inequality we have
\begin{align}
\label{3eq:newdifeqc}
\JX{\CMIX{\inp_{\tin+1}}{\out_{\tin+1}}{\out^{\tin}}}
\geq \ECX{\ln \frac{\PXD(\out_{\tin+1}|\out^{\tin})}{\PXAD{\ldsf_i}(\out_{\tin+1}|\out^{\tin})}}{\out^{\tin}}.
\end{align}
Using equations (\ref{3eq:newdifeqa}), (\ref{3eq:newdifeqb}) and (\ref{3eq:newdifeqc}) we get
\begin{align}
\label{3eq:newdifeqd}
\ECX{\mar{U}_{\tin+1}}{\out^{\tin}}\geq \mar{U}_{\tin}.
\end{align}
Recall that $\min_{\dinp,\dout} \CT{\dinp}{\dout} = \mtp$ and $|\JX{\cdot}|\leq \DX$.
Thus as a result of equation (\ref{3eq:newdifeqa}) we have
\begin{align}
\label{3eq:newdifeqe}
\ECX{|\mar{U}_{\tin+1}-\mar{U}_{\tin}|}{\out^{\tin}} \leq \ln \tfrac{1}{\mtp}+\DX.
\end{align}
As a result of (\ref{3eq:newdifeqd}), (\ref{3eq:newdifeqe}) and the fact that $\mar{U}_0=0$,
 $\mar{U}_{\tin}$ is a submartingale.
 
Recall that we have assumed that \(\PX{\dt_{j+1}\leq \dt}=1\) and \(\EX{\dt}<\infty\); consequently
\begin{align}
\label{3eq:newdifeqexx}
\EX{\dt_{j+1}}<\infty.
\end{align}
Because of (\ref{3eq:newdifeqe}) and (\ref{3eq:newdifeqexx}) we can apply a version of  Doob's 
 optional stopping theorem \cite[Theorem 2, p 487]{shiryaev}  to the submartingale \(\mar{U}_{\tin}\) and 
 the  stopping time \(\dt_{j+1}\) to obtain $\EX{\mar{U}_{\dt_{j+1}}} \geq \EX{\mar{U}_{0}} = 0$. 
 Consequently,
\begin{equation}
  \label{eq:sdoob}
\EX{\ln\tfrac{\PXD(\out_{\dt_{j}+1}^{\dt_{j+1}}|\out^{\dt_{j}})}{\PXAD{\ldsf_{i}}(\out_{\dt_{j}+1}^{\dt_{j+1}}|\out^{\dt_{j}})}}
\leq\EX{\sum\nolimits_{\tin=\dt_{j}+1}^{\dt_{j+1}} \JX{\CMIX{\mes}{\out_{\tin}}{\out^{\tin-1}}}}.
\end{equation}
Note that as a result of the concavity of $\JX{\cdot}$ and Jensen's inequality we have
\begin{align}
\EX{\sum\nolimits_{\tin=\dt_{j}+1}^{\dt_{j+1}} \JX{\CMIX{\mes}{\out_{\tin}}{\out^{\tin-1}}}}
\notag
&=\EX{\dt_{j+1}-\dt_{j}}\EX{
\sum\nolimits_{\tin\geq 1}   \tfrac{\IND{\dt_{j+1}\geq \tin>\dt_{j}}\JX{\CMIX{\mes}{\out_{\tin}}{\out^{\tin-1}}}}{\EX{\dt_{j+1}-\dt_{j}}}} \\
&\stackrel{~}{\leq}
\EX{\dt_{j+1}-\dt_{j}}  \JX{\tfrac{\EX{  \sum_{\tin\geq 1} \IND{\dt_{j+1}\geq \tin> \dt_{j}} \CMIX{\mes}{\out_{\tin}}{\out^{\tin-1}}}}{\EX{\dt_{j+1}-\dt_{j}}}}.
\label{eq:con:JandD}
\end{align}
In order to calculate the argument of \(\JX{\cdot}\) in (\ref{eq:con:JandD})  consider the  stochastic sequence
\begin{equation}
  \label{eq:vdef}
\mar{V}_{\tin} = \HX(\mes|\out^{\tin}) + \sum\nolimits_{j=1}^{\tin} \CMIX{\mes}{\out_{j}}{\out^{j-1}}.
\end{equation}
Clearly $\ECX{\mar{V}_{\tin+1}}{\out^{\tin}} =\mar{V}_{\tin}$ and 
$\EX{|\mar{V}_{\tin}|} \leq \ln |\mesS|+\CX \tin < \infty$. Hence $\mar{V}_{\tin}$ is a martingale. 

Furthermore, 
\begin{equation}
  \label{eq:vdefa}
\ECX{|\mar{V}_{\tin+1}-\mar{V}_{\tin}|}{\out^{\tin}} \leq \ln |\mesS|+\CX.
\end{equation}
Recall that we have assumed that \(\PX{\dt_{j}\leq \dt_{j+1}\leq \dt}=1\) and \(\EX{\dt}<\infty\); consequently
\begin{align}
  \label{eq:vdefb}
\EX{\dt_{j}}\leq\EX{\dt_{j+1}}<\infty.
\end{align}
As a result of equations  (\ref{eq:vdefa}) and  (\ref{eq:vdefb})
 we can apply  Doob's optimal stopping theorem, \cite[Theorem 2, p 487]{shiryaev} to \(\mar{V}_{\tin}\) both at stopping time \(\dt_{j}\) and at stopping time \(\dt_{j+1}\),
  i.e.,   \(\EX{\mar{V}_{\dt_{j+1}}}=\EX{\mar{V}_{0}} \) and
 \(\EX{\mar{V}_{\dt_{j}}}=\EX{\mar{V}_{0}}\). Consequently,
\begin{align}
 \label{eq:vdoob}
\EX{ \sum_{\tin\geq 1} \IND{\dt_{j+1}\geq \tin> \dt_{j}} \CMIX{\mes}{\out_{\tin}}{\out^{\tin-1}}} 
= \EX{ \HX(\mes|\out^{\dt_{j}})- \HX(\mes|\out^{\dt_{j+1}})}.
\end{align}
Using equations   (\ref{eq:sdoob}), (\ref{eq:con:JandD}) and (\ref{eq:vdoob})
\begin{equation}
  \label{eq:sit}
\hspace{-.3cm}
  \EX{\ln\tfrac{\PXD(\out_{\dt_{j}+1}^{\dt_{j+1}}|\out^{\dt_{j}})}{\PXAD{\ldsf_{i}}(\out_{\dt_{j}+1}^{\dt_{j+1}}|\out^{\dt_{j}})}}
\leq \EX{\dt_{j+1}-\dt_{j}}
\JX{\tfrac{ \EX{ \HX(\mes|\out^{\dt_{j}})- \HX(\mes|\out^{\dt_{j+1}})}}{\EX{\dt_{j+1}-\dt_{j}}}}
\end{equation}
Hence inequality given in (\ref{3eq:sita}) not only when   \(\PX{\dt_{j+1}=\dt_{j}}=1\) but 
also when  \(\PX{\dt_{j+1}=\dt_{j}}<1\).
\end{proof}

\subsection{Proof of Lemma \ref{lem:consm} for The Case \(\EX{\dt}<\infty\)}
\label{sec:appconsm}
\begin{proof}
 In order to bound \(\Pem{\dmes}\) from below we apply Lemma \ref{lem:con} for \((\dt_1,\ldsf_1)\) and 
\((\dt_2,\ldsf_2)\) given in equations (\ref{eq:smmuep-c0}), (\ref{eq:smmuep-c1}), (\ref{eq:smmuep-c2}) and  
(\ref{eq:smmuep-c3}) and use the  fact that \(\JX{\cdot} \leq \DX\) we get 
\begin{subequations}
\label{eq:n:a2b0a}
  \begin{align}
\label{eq:n:a2b1a}
\ln \Pem{\dmes} 
&\geq 
 \tfrac{-\bent{\Pe+ |\mesS|^{-1}}- 
 \EX{\dt_2}
 \JX{\frac{\EX{\HX(\mes)-\HX(\mes|\out^{\dt_{2}})}}{\EX{\dt_{2}}}}
-\EX{\dt-\dt_2}
\DX
 }{1-\Pe- |\mesS|^{-1}}   \\
 \label{eq:n:a2b2a}
\ln  \PXA{\ldsf_2}{\est \notin \ldsn{2}{\out^{\dt_2}}} 
&\geq 
 \tfrac{-\bent{\Pe+ \PX{\mes \in \ldsn{2}{\out^{\dt_2}}}}- 
\EX{\dt-\dt_2}\DX }{1-\Pe- \PX{\mes \in \ldsn{2}{\out^{\dt_2}}}}   
\end{align}
\end{subequations}
provided that  \( |\mesS|^{-1}+\Pe \leq 1/2\) and \(\PX{\mes \in \ldsn{2}{\out^{\dt_2}}} +\Pe \leq 1/2 \).

We start with bounding  \(\PXA{\ldsf_2}{\est \notin \ldsn{2}{\out^{\dt_2}}} \) from above and 
\(\PX{\mes\notin \ldsn{2}{\out^{\dt_2}}} \) from below.

\begin{itemize}
\item Since \(\min\nolimits_{\dinp \in \inpS, \dout\in \outS} \CT{\dinp}{\dout}=\mtp\)  the posterior probability 
of a message  at time \(\tin+1\)  can  not be smaller than \(\mtp\) times the  posterior probability of the same 
message at time \(\tin\).  Hence   for the  stopping time \(\dt_2\) defined in equation (\ref{eq:smmuep-c2}),
random\footnote{The set \(\ldsf_{2}\) is random in the sense that it depends on previous channel outputs. } 
set \(\ldsf_{2}\) defined in equation (\ref{eq:smmuep-c3}) and \( \delta<\tfrac{1}{2}\) we have
 \begin{align}
\PCX{\mes \in \ldsn{2}{\out^{\dt_2}}}{\out^{\dt_2}=\dout^{\ddt_2}}&>  \mtp \delta 
&&\forall \dout^{\ddt_2} \in \outS^{\dt_2*}.
\end{align}
 As a result of the definition of $\PXAD{\ldsf_2}(\dmes, \dout^{\ddt}) $ given in  equation (\ref{eq:defpxa-a}) we have,
\begin{align}
\label{eq3:intaqq}
\PXAD{\ldsf_2}(\dmes, \dout^{\ddt})
<  \PXD(\dmes, \dout^{\ddt})  \tfrac{\IND{\dmes \in \ldsn{2}{\dout^{\ddt_2}}}}{\mtp \delta}
&&\forall \dmes \in \mesS,  \dout^{\ddt} \in \outS^{\dt*}.
\end{align}
If the decoded message \(\est(\dout^{\ddt})\) is not in   \(\ldsn{2}{\dout^{\ddt_2}}\) and message \(\dmes\) is in \(\ldsn{2}{\dout^{\ddt_2}}\) then 
\(\est(\dout^{\ddt}) \neq \dmes\):
\begin{align}
\label{eq3:intbqq}
\IND{\est(\dout^{\ddt})  \notin \ldsn{2}{\dout^{\ddt_2}}} \IND{\dmes \in \ldsn{2}{\dout^{\ddt_2}}} \leq  \IND{\est(\dout^{\ddt})\neq \dmes}
&&\forall \dmes \in \mesS,  \dout^{\ddt} \in \outS^{\dt}.
\end{align}
Using equations (\ref{eq3:intaqq}) and (\ref{eq3:intbqq}) we get
\begin{align}
\label{eq3:intaqq-n1}
\PXAD{\ldsf_2}(\dmes, \dout^{\ddt}) \IND{\est(\dout^{\ddt})  \notin \ldsn{2}{\dout^{\ddt_2}}}
<  \PXD(\dmes, \dout^{\ddt})  \tfrac{\IND{\est(\dout^{\ddt})\neq \dmes}}{\mtp \delta}
&&\forall \dmes \in \mesS,  \dout^{\ddt} \in \outS^{\dt*}.
\end{align}
If we sum over all \((\dmes,\dout^{\ddt})\)'s in \(\mesS \times {\outS}^{\dt*}\)  and use equations (\ref{eq:defpxd-b}) and (\ref{eq:defpxa-b}) we get,
\begin{align}
  \label{eq:a2b}
\PXA{\ldsf_2}{\est \notin \ldsn{2}{\out^{\dt_2}}}
&<  \tfrac{\PX{\est\neq \mes}}{\mtp \delta} =\tfrac{\Pe}{\mtp \delta}.
\end{align}

\item The probability of an event \(\event_1\)   is lower bounded by the probability of its intersection with any event \(\event_2\), i.e.,   \(\PX{\event_1}\geq \PX{\{\event_1, \event_2\}} \):
\begin{align}
\notag
\Pe
&=\PX{ \est\neq  \mes }\\
\notag
&\geq  \PX{\left\{\est\neq  \mes,  \ldsn{2}{\out^{\dt_2}}={\mesS} \right\}}\\
\label{eq:nn:smuepc1}
&=\PCX{ \est\neq  \mes }{ \ldsn{2}{\out^{\dt_2}}={\mesS} }
\PX{ \ldsn{2}{\out^{\dt_2}}={\mesS} }
\end{align}

Note that if \(\ldsn{2}{\dout^{\ddt_2}}={\mesS}\) then \(\dt\) is reached before any of the messages reach a posterior probability of \(1-\delta\). Thus 
\begin{equation}
\label{eq:nn:smuepc2}
\PCX{ \est\neq  \mes }{ \ldsn{2}{\out^{\dt_2}}={\mesS}}  > \delta
\end{equation}
Thus as a result of equations (\ref{eq:nn:smuepc1}) and  (\ref{eq:nn:smuepc2}) we have
\begin{equation}
  \label{eq:a2b3}
\PX{ \ldsn{2}{\out^{\dt_2}}={\mesS} } 
<  \tfrac{\Pe}{\delta}.
\end{equation}

On the other hand  if \(\ldsn{2}{\dout^{\ddt_2}}\neq {\mesS}\), then  the most likely  message with a probability at least  \((1-\delta)\) is excluded from \(\ldsn{2}{\dout^{\ddt_2}}\). Thus
\begin{equation}
\label{eq:nn:smuepc3}
\PCX{\mes \in \ldsn{2}{\out^{\dt_2}}}{ \ldsn{2}{\out^{\dt_2}}\neq \mesS}\leq \delta
\end{equation}
Using equations  (\ref{eq:a2b3}) and (\ref{eq:nn:smuepc3})  together with  total probability formula  we get
\begin{align}
 \PX{\mes \in \ldsn{2}{\out^{\dt_2}}}
\notag 
&=
\PCX{\mes \in \ldsn{2}{\out^{\dt_2}}}{\ldsn{2}{\out^{\dt_2}}={\mesS}} 
\PX{\ldsn{2}{\out^{\dt_2}}=\mesS}\\
\notag 
&\qquad ~ \quad+ 
\PCX{\mes \in \ldsn{2}{\out^{\dt_2}}}{\ldsn{2}{\out^{\dt_2}}\neq{\mesS}} 
\PX{\ldsn{2}{\out^{\dt_2}}\neq{\mesS}}\\
\notag
&\leq
\PX{\ldsn{2}{\out^{\dt_2}}={\mesS}} 
+
\PCX{\mes \in \ldsn{2}{\out^{\dt_2}}}{\ldsn{1}{\out^{\dt_1}}\neq{\mesS}}
\\
 \label{eq:a2b4}
&< \tfrac{\Pe}{\delta}+\delta.
\end{align}
\end{itemize}
We plug the bounds on  \(\PXA{\ldsf_2}{\est \notin \ldsn{2}{\out^{\dt_2}}} \)  and 
\(\PX{\mes\notin \ldsn{2}{\out^{\dt_2}}} \) given in equations  (\ref{eq:a2b}) and
(\ref{eq:a2b4}) in equation  (\ref{eq:n:a2b0a}) to get
\begin{subequations}
\label{eq:n:a2b0b}
  \begin{align}
\label{eq:n:a2b1b}
\ln \Pem{\dmes} 
&\geq 
 \tfrac{-\bent{\epst{1}}- 
 \EX{\dt_2}
 \JX{\frac{\EX{\HX(\mes)-\HX(\mes|\out^{\dt_{2}})}}{\EX{\dt_{2}}}}
-\EX{\dt-\dt_2}
\DX
 }{1-\epst{1}}   \\
 \label{eq:n:a2b2b}
\ln \tfrac{\Pe}{\mtp \delta }
&\geq 
 \tfrac{-\bent{\epst{1}} - 
\EX{\dt-\dt_2}\DX }{1-\epst{1}}   
\end{align}
\end{subequations}
provided that \(\epst{1}\leq 1/2\) where \(\epst{1}=\Pe+\delta+\tfrac{\Pe}{\delta}+|\mesS|^{-1}\).

Now we bound \(\EX{\HX(\mes|\out^{\dt_{2}})}\) from below. Note that \(\IND{\mes \in \ldsn{2}{\out^{\dt_2}}}\) is a discrete random variable that is either zero or one;  its conditional entropy  given 
\(\out^{\dt_2}\) is given by
\begin{equation}
 \label{eq:a2b4n1a}
\HX(\IND{\mes \in \ldsn{2}{\out^{\dt_2}}}|\out^{\dt_2}) =
\bent{\PCX{\mes \in \ldsn{2}{\out^{\dt_2}}}{\out^{\dt_2}}}.
\end{equation}
Furthermore since \(\IND{\mes \in \ldsn{2}{\out^{\dt_2}}}\) is  a function of \(\out^{\dt_2}\) and \(\mes\),  chain rule entropy implies that 
\begin{align}
 \label{eq:a2b4n1b}
 \HX(\mes|\out^{\dt_2})
&=\HX(\IND{\mes \in \ldsn{2}{\out^{\dt_2}}}|\out^{\dt_2})
+\ECX{\HX(\mes|\out^{\dt_2},\IND{\mes \in \ldsn{2}{\out^{\dt_2}}})}{\out^{\dt_2}}.
\end{align}
Since \(\ldsn{2}{\out^{\dt_2}}\) has  at most  \(|\mesS|\)  elements and its complement,
 \(\mesS \setminus \ldsn{2}{\out^{\dt_2}} \),   has at most  one element, we can bound the conditional  entropy  of the messages as follows
\begin{align}
 \label{eq:a2b4n1}
\HX(\mes|\out^{\dt_2},\IND{\mes \in \ldsn{2}{\out^{\dt_2}}})
&\leq \IND{\mes \in \ldsn{2}{\out^{\dt_2}}} \ln |\mesS|
\end{align}
Thus using equations (\ref{eq:a2b4n1a}), (\ref{eq:a2b4n1b}) and (\ref{eq:a2b4n1}) we get
\begin{align}
 \label{eq:a2b4n2}
\HX(\mes|\out^{\dt_2})
&\leq  \bent{\PCX{\mes \in \ldsn{2}{\out^{\dt_2}}}{\out^{\dt_2}}}
 +\PCX{\mes \!\in \! \ldsn{2}{\out^{\dt_2}}}{\out^{\dt_2}}  \ln |\mesS|.
\end{align}
Then using concavity of the binary entropy function \(\bent{\cdot}\) together with equations (\ref{eq:a2b4}) and (\ref{eq:a2b4n2}) we get
\begin{equation}
  \label{eq:fano}
\EX{  \HX(\mes|\out^{\dt_2})} < \bent{\delta+ \tfrac{\Pe}{\delta}} + (\delta+ \tfrac{\Pe}{\delta}) \ln |\mesS|.
\end{equation}
provided that \(\delta+ \tfrac{\Pe}{\delta}\leq 1/2\).

If we plug in  equation  (\ref{eq:fano}) and the identity \(\HX(\mes)= \ln |\mesS|\) 
in equation (\ref{eq:n:a2b0b}) we get,
\begin{subequations}
\label{eq:n:a2b0c}
  \begin{align}
\label{eq:n:a2b1c}
(1-\epst{1})\tfrac{\ln \Pem{\dmes}}{\EX{\dt}} 
&\geq -\tfrac{\bent{\epst{1}}}{\EX{\dt}}- \fr \JX{\tfrac{(1-\epst{1})\rate -\bent{\epst{1}}/\EX{\dt}}{\fr}} -(1- \fr)\DX\\
 \label{eq:n:a2b2c}
-(1-\epst{1})\ex 
&\geq  \tfrac{-\bent{\epst{1}}+\ln \mtp \delta}{\EX{\dt}}
-(1-\fr)\DX
\end{align}
\end{subequations}
provided that \(\epst{1}\leq 1/2\) where \(\fr=\tfrac{\EX{\dt_2}}{\EX{\dt}}\), 
\(\epst{1}=\Pe+\delta+\tfrac{\Pe}{\delta}+|\mesS|^{-1}\),
\(\rate=\frac{|\mesS|}{\EX{\dt}}\) and
\(\ex=\tfrac{-\ln \Pe}{\EX{\dt}}\).

Note that the inequality given in  equation  (\ref{eq:n:a2b2c}) bounds the value of \(\fr\) from above,
\begin{equation}
 \label{eq:n:a2b2d}
\fr  \leq 1-\tfrac{(1-\epst{1})\ex-\epst{2}}{\DX}
\end{equation}
where \(\epst{2}=\tfrac{\bent{\epst{1}} -\ln \mtp \delta}{\EX{\dt}}\).

Furthermore for any \(\fr_1\leq\fr_2 \leq \tfrac{\widetilde{\rate}}{\CX}\) as a result of concavity of \(\JX{\cdot}\)	we have
\begin{align}
\notag
\fr_1 \JX{\tfrac{\widetilde{\rate}}{\fr_1}}+(1-\fr_1) \DX 
&=\fr_1 \JX{\tfrac{\widetilde{\rate}}{\fr_1}}+(\fr_2-\fr_1)\JX{0}+ (1-\fr_2) \DX \\
 \label{eq:n:a2b1d}
&\leq\fr_2 \JX{\tfrac{\widetilde{\rate}}{\fr_2}}+ (1-\fr_2) \DX .
\end{align}
Using  equations (\ref{eq:n:a2b2d}), (\ref{eq:n:a2b1d}) we see that the bound in equation (\ref{eq:n:a2b1c}) is 
lower bounded by its value at \(\fr = 1-\tfrac{(1-\epst{1})\ex-\epst{2}}{\DX} \)  if 
\(\ex\geq \tfrac{\epst{2}}{1-\epst{1}}\) and by its value at \(\fr=1\) otherwise, i.e.,
\begin{equation*}
\tfrac{\ln \Pem{\dmes}}{\EX{\dt}} 
\geq
\begin{Bmatrix}
  -\ex- \left(1-\tfrac{\ex-\epst{}}{\DX}\right)
 \JX{\tfrac{\rate -\frac{\epst{2}}{1-\epst{1}}}{1-\tfrac{\ex-\epst{}}{\DX}}}
&\qquad &\mbox{if ~} \ex\geq \tfrac{\epst{2}}{1-\epst{1}}\\
-\tfrac{\epst{2}}{1-\epst{1}}-\tfrac{1}{1-\epst{1}}
 \JX{(1-\epst{1})\rate -\epst{2}}
&\qquad &\mbox{if ~} \ex<\tfrac{\epst{2}}{1-\epst{1}}\\
\end{Bmatrix}
\end{equation*}
where  \(\epst{}=\tfrac{\epst{1}\DX+\epst{2}}{1-\epst{1}}\).

Then, for the case \(\ex\geq \tfrac{\epst{2}}{1-\epst{1}}\)   Lemma \ref{lem:consm} follows  from the fact that \(\JX{\cdot}\) is a non-negative decreasing function. For the case \(\ex <  \tfrac{\epst{2}}{1-\epst{1}}\) in Lemma \ref{lem:consm} follows from the fact that   \(\JX{\cdot}\) is a concave   non-negative decreasing function.

\end{proof}

\subsection{Proof of Lemma \ref{lem:conbits} for The Case \(\EX{\dt}<\infty\)}
\label{sec:appconbits}

\begin{proof}
We start with proving the bounds given in equations  (\ref{eq:n:bits-a}) and (\ref{eq:n:bits-b}).
\begin{itemize}
\item    Let us start with the bound on  \(\PXA{\ldsf_i}{\est \notin \ldsn{i}{\out^{\dt_i}}}\)  given in equation  
(\ref{eq:n:bits-a}).  Since \(\min\nolimits_{\dinp \in \inpS, \dout\in \outS} \CT{\dinp}{\dout}=\mtp\), 
 the posterior probability of a ${\dmes}^{i} \in \mesS^{i}$ at time \(\tin+1\) can not be smaller than  
\(\mtp\) times its  value at time \(\tin\).    Hence for \(\delta<1/2\), as a result definitions of \(\dt_i\) and
 \(\ldsn{i}{\out^{\dt_i}}\) given in  equations (\ref{eq:dti}) and (\ref{eq:ati}), we have
\begin{align*}
\PCX{\mes \in \ldsn{i}{\out^{\dt_i}}}{\out^{\dt_i}=\dout^{\ddt_i}}
&> \mtp \delta
&&\forall \dout^{\ddt_i} \in \outS^{\dt_i*},  i\in \left\{1,2,\ldots, \nlay\right\}.
\end{align*}
Then as a result of the definition of \(\PXAD{\ldsf_i}(\dmes, \dout^{\ddt}) \) given in  equation (\ref{eq:defpxa-a}) we have,
\begin{align}
\label{eq:s1ax}
\PXAD{\ldsf_i}(\dmes, \dout^{\ddt})
<  \PXD(\dmes, \dout^{\ddt})  \tfrac{\IND{\dmes \in \ldsn{i}{\dout^{\ddt_i}}}}{\mtp \delta}
&&\forall \dmes \in \mesS,  \dout^{\ddt} \in \outS^{\dt*}, i\in \left\{1,2,\ldots, \nlay\right\}.
\end{align}
For \(\ldsn{i}{\dout^{\ddt_i}}\) given in  equation (\ref{eq:ati}),
if the decoded message \(\est(\dout^{\ddt})\) is not in   \(\ldsn{i}{\dout^{\ddt_i}}\) 
but \(\dmes\) is in \(\ldsn{i}{\dout^{\ddt_i}}\) then 
\(\est^{i}(\dout^{\ddt}) \neq \dmes^{i}\):
\begin{align}
\label{eq:s1axb}
\IND{\est(\dout^{\ddt})  \notin \ldsn{i}{\dout^{\ddt_i}}} 
\IND{\dmes \in \ldsn{i}{\dout^{\ddt_i}}} 
&\leq  \IND{\est^i\neq \dmes^i}
&&\forall \dmes \in \mesS,  
\dout^{\ddt} \in \outS^{\dt},
 i\in \left\{1,2,\ldots, \nlay\right\}.
\end{align} 
Using equations (\ref{eq:s1ax}) and (\ref{eq:s1axb}) we get
\begin{align*}
\PXAD{\ldsf_i}(\dmes, \dout^{\ddt})  \IND{\est(\dout^{\ddt})  \notin \ldsn{i}{\dout^{\ddt_i}}} 
<  \PXD(\dmes, \dout^{\ddt})    \tfrac{\IND{\est^i\neq \dmes^i}}{\mtp \delta}
&&\forall \dmes \in \mesS,  \dout^{\ddt} \in \outS^{\dt*},
 i\in \left\{1,2,\ldots, \nlay\right\}.
\end{align*}
If we sum over all \((\dmes,\dout^{\ddt})\)'s in \(\mesS \times {\outS}^{\dt*}\)  and use equations
 (\ref{eq:defpxd-b}) and (\ref{eq:defpxa-b}) we get,
\begin{align}
\notag
  \PXA{\ldsf_i}{\est \notin \ldsn{i}{\out^{\dt_i}}} 
 &< \tfrac{\PX{\est^{i}\neq \mes^{i}}}{\mtp \delta}
 &&\\
 &= \tfrac{\Peb{i}}{\mtp \delta}
&&\forall  i\in \left\{1,2,\ldots, \nlay\right\}.
\label{eq:s1}
\end{align}

\item Let us now prove the bound on \(\PX{\mes \in \ldsn{i}{\out^{\dt_i}}}\) given in equation  (\ref{eq:n:bits-b}).
\begin{itemize}
\item If \(\ldsn{i}{\out^{\dt_i}}\neq \mesS\), then at \(\dt_i\) there is a \(\dmes^{i}\) with
 posterior probability \((1-\delta)\) and all the messages \(\dmes\) of the 
 form \(\dmes=(\dmes^{i},\dmes_{i+1},\ldots,\dmes_{k})\) are excluded from 
 \(\ldsf_{i}\). Consequently we have
\begin{align}
\label{qeq:n1}
\PCX{\mes\in\ldsn{i}{\out^{\dt_i}}}{\ldsn{i}{\out^{\dt_i}}\neq \mesS} < \delta.
\end{align}
\item If \(\ldsn{i}{\out^{\dt_i}}= \mesS\), then  at \(\dt_i\) there is no  \(\dmes^{i}\) with posterior probability \((1-\delta)\) and  \(\dt_i=\dt\).  Since \(\est^{i}\neq \mes^{i}\) implies that \(\est\neq \mes\) we have
   \begin{align}
\label{qeq:n2}
\PCX{\est \neq \mes}{\ldsn{i}{\out^{\dt_i}}= \mesS} \geq \delta.
\end{align}
As a result of  total probability formula  for \(\PX{\est\neq\mes}\) we have   
\begin{align}
\notag
~\hspace{-1cm}\Pe
&= 
\PCX{\est\neq \mes}{\ldsn{i}{\out^{\dt_i}}= \mesS}\PX{\ldsn{i}{\out^{\dt_i}}= \mesS}
+\PCX{\est\neq \mes}{\ldsn{i}{\out^{\dt_i}}\neq \mesS} \PX{\ldsn{i}{\out^{\dt_i}}\neq\mesS}\\
\label{eq:s2}
&\geq \delta \PX{\ldsn{i}{\out^{\dt_i}}= \mesS}
\end{align}
\end{itemize}
If use  the total probability formula  for \(\PX{\mes \in \ldsn{i}{\out^{\dt_i}}}\) together with  equations 
(\ref{qeq:n1}) and    (\ref{eq:s2}) we get
\begin{align}
\notag
\PX{\mes \in \ldsn{i}{\out^{\dt_i}}}
&
=\PX{\left\{\mes \in \ldsn{i}{\out^{\dt_i}}, \ldsn{i}{\out^{\dt_i}}\neq \mesS\right\}}
+\PX{\left\{\mes \in \ldsn{i}{\out^{\dt_i}}, \ldsn{i}{\out^{\dt_i}}= \mesS\right\}}\\
&
\notag
\leq \PCX{\mes \in \ldsn{i}{\out^{\dt_i}}}{\ldsn{i}{\out^{\dt_i}}\neq \mesS}
+\PX{ \ldsn{i}{\out^{\dt_i}}= \mesS }\\
\notag
&\leq \delta+\tfrac{\Pe}{\delta}.
\end{align}
\end{itemize}

We apply Lemma \ref{lem:con} for  \((\dt_1,\ldsf_1)\),\(\ldots\),\((\dt_k,\ldsf_k)\)
defined in equations (\ref{eq:dti}) and (\ref{eq:ati});  use the bounds 
on \(\PXA{\ldsf_i}{\est \notin \ldsn{i}{\out^{\dt_i}}}\)  and   \(\PX{\mes \in \ldsn{i}{\out^{\dt_i}}}\)
given in (\ref{eq:n:bits-a}) and (\ref{eq:n:bits-b}). Then we can conclude that if 
\(\Pe+ \delta+ \Pe/\delta\leq 1/2\) then
\begin{align}
  \label{eq:s5}
(1-\epst{3}) \ex_i
&\leq \epst{5} + \sum\limits_{j= i+1}^{ \nlay+1} \afr_{j} \JX{\arate{j}}
&i=&1,2,\ldots, \nlay
\end{align}
where \(\rate_i\), \(\ex_i\),  \(\epst{3}\) and  \(\epst{5}\) are defined in Lemma \ref{lem:conbits}, 
 \(\arate{j}\)'s are defined in equation (\ref{eq:defarate}) of Lemma  \ref{lem:con}
and \(\afr_j\)'s  are  defined  as follows\footnote{We use the convention 
\( \dt_0=0 \) and \(\dt_{ \nlay+1}=\dt\).}
\begin{align}
\label{eq:rdti}
\afr_j
&\DEF\frac{\EX{\dt_{j}}-\EX{\dt_{j-1}}}{\EX{\dt}}
&&
\forall j \in \{1,2,\ldots, \nlay+1\}
\end{align}
Depending on the values of \(\afr_j\) and \(\arate{j}\) the bound in equation (\ref{eq:s5}) takes different 
values.  However \(\afr_j\) and \(\arate{j}\) are not changing freely. As a result of equation (\ref{eq:vdoob}) and the fact that $\CMIX{\mes}{\out_{t+1}}{\out^{t}}\leq \CX$ we have
\begin{align}
\label{eq:nb-ent-0}
\arate{j} &\leq \CX
& j&\in \{1,2,\ldots,( \nlay+1)\}.
\end{align}
 In addition   \(\afr_j\)'s and \(\arate{j}\)'s are constrained by the definitions of  \(\dt_j\) and  
 \(\ldsn{j}{\out^{\dt_j}}\) 
given in   equations (\ref{eq:dti}) and (\ref{eq:ati}). At \(\dt_j\) with high probability one element of
 \(\mesS^{j}\) has a posterior probability \((1-\delta)\). Below we use  this fact to bound  
\(\EX{\HX(\mes|\out^{\dt_j})}\) from above.  Then we turn this bound into a constraint  on the
 values of \(\afr_j\)'s and \(\arate{j}\)'s and  use that constraint together with equations  (\ref{eq:s5}), (\ref{eq:nb-ent-0}) to bound 
 \(\ex_i\)'s from above.  
 
For all \(j\) in \(\left\{1,2,\ldots, \nlay \right\}\), \(\IND{\mes \in \ldsn{j}{\out^{\dt_j}}}\) is a discrete random variable that is either zero or one; its conditional entropy  given by
\begin{equation}
 \label{eq:nb-ent-1}
\HX(\IND{\mes \in \ldsn{j}{\out^{\dt_j}}}|\out^{\dt_j}) 
=\bent{\PCX{\mes \in \ldsn{j}{\out^{\dt_j}}}{\out^{\dt_j}}}.
\end{equation}
Furthermore since \(\IND{\mes \in \ldsn{i}{\out^{\dt_i}}}\) is a function of \(\out^{\dt_i}\) and \(\mes\),  
the chain rule entropy implies that 
\begin{align}
 \label{eq:nb-ent-2}
 \HX(\mes|\out^{\dt_i})
&=\HX(\IND{\mes \in \ldsn{i}{\out^{\dt_i}}}|\out^{\dt_i})
+\ECX{\HX(\mes|\out^{\dt_i},\IND{\mes \in \ldsn{i}{\out^{\dt_i}}})}{\out^{\dt_i}}.
\end{align}
Note that  \(\ldsn{i}{\out^{\dt_i}}\) has  at most  \(|\mesS|\)  elements and its complement,
 \(\mesS \setminus \ldsn{i}{\out^{\dt_i}} \),   has at most  \( \tfrac{|\mesS|}{|\mesS^i|}\)
  elements. We can bound the conditional  entropy  of the messages \(\HX(\mes|\out^{\dt_i},\IND{\mes \in \ldsn{i}{\out^{\dt_i}}})\)  as follows
\begin{align}
\notag
  \HX(\mes|\out^{\dt_i},\IND{\mes \in \ldsn{i}{\out^{\dt_i}}})
&\leq \IND{\mes \in \ldsn{i}{\out^{\dt_i}}} \ln |\mesS| +\IND{\mes \notin \ldsn{i}{\out^{\dt_i}}} \ln \tfrac{ |\mesS|}{|\mesS^{i}|}\\
 \label{eq:nb-ent-3}
&=  \ln \tfrac{ |\mesS|}{|\mesS^{i}|}+ \IND{\mes \in \ldsn{i}{\out^{\dt_i}}} \ln |\mesS^{i}|
\end{align}
Thus using equations  (\ref{eq:nb-ent-1}), (\ref{eq:nb-ent-2}) and  (\ref{eq:nb-ent-3}) we get
\begin{align}
 \label{eq:nb-ent-4}
 \HX(\mes|\out^{\dt_i})
&\leq  
\bent{\PCX{\mes \in \ldsn{j}{\out^{\dt_j}}}{\out^{\dt_j}}}
+\ln \tfrac{ |\mesS|}{|\mesS^{i}|}+\PCX{\mes \in \ldsn{i}{\out^{\dt_i}}}{\out^{\dt_i}} \ln |\mesS^{i}|.
\end{align}
If we take the expectation of both sides of the inequality  (\ref{eq:nb-ent-4}) and   
use the concavity of the binary entropy function  we get
\begin{align}
\notag
\EX{ \HX(\mes|\out^{\dt_i})} 
&
\leq  \bent{\PX{\mes \in \ldsn{j}{\out^{\dt_j}}}}
 +\ln \tfrac{ |\mesS|}{|\mesS^{i}|}+\PX{\mes \in \ldsn{i}{\out^{\dt_i}}}   \ln |\mesS^{i}|
\end{align} 
Using the  inequality given  (\ref{eq:n:bits-b}) and the fact that binary entropy function is an increasing function on the interval \([0,1/2]\) we see that
\begin{align}
\EX{ \HX(\mes|\out^{\dt_i})} 
\label{eq:nb-ent-5}
&<  \bent{\Pe+\delta+ \tfrac{\Pe}{\delta}}
 +\ln \tfrac{ |\mesS|}{|\mesS^{i}|}+(\Pe+\delta+ \tfrac{\Pe}{\delta})   \ln |\mesS^{i}|.
\end{align} 
provided that \(\Pe+\delta+ \tfrac{\Pe}{\delta}\leq 1/2\).

Note that as a result of Fano's inequality for \( \EX{\HX(\mes|\out^{\dt})}\)  we have
\begin{align}
\EX{ \HX(\mes|\out^{\dt})} 
\label{eq:nb-ent-5-a}
&<  \bent{\Pe} +\Pe \ln |\mesS|.
\end{align} 

If we divide both sides of the inequalities (\ref{eq:nb-ent-5}) and (\ref{eq:nb-ent-5-a}) to \(\EX{\dt}\), we see that following bounds holds
\begin{subequations}
\label{eq:nb-ent-6}
\begin{align}
\tfrac{\EX{ \HX(\mes|\out^{\dt_i})}}{\EX{\dt}} 
&
\leq  \epst{4} +\rate-\sum\nolimits_{j=1}^{i} \rate_i +\epst{3} \sum\nolimits_{j=1}^{i} \rate_i.
&& i =1,2,\ldots,\nlay\\
\EX{ \HX(\mes|\out^{\dt})} 
&\leq  \bent{\Pe} +\Pe R.
&&
\end{align} 
\end{subequations}
Note that
\begin{align}
\label{eq:nb-ent-7}
\tfrac{\EX{ \HX(\mes|\out^{\dt_i})}}{\EX{\dt}} 
&
=  \rate -\sum\nolimits_{j=1}^{i}  \afr_j \arate{j}
&& i =1,2,\ldots,(\nlay+1)
\end{align} 
Using equations (\ref{eq:nb-ent-6}) and (\ref{eq:nb-ent-7}) we get,
\begin{subequations}
\label{eq:nb-ent-8}
\begin{align}
\sum\nolimits_{j=1}^{i}   \afr_j \arate{j}
&\geq   (1-\epst{3})\sum\nolimits_{j=1}^{i} \rate_j-\epst{4}
&i=&1,2,\ldots,\nlay\\
\sum\nolimits_{j=1}^{\nlay+1}   \afr_j \arate{j}
&\geq   (1-\Pe) \rate-\tfrac{\bent{\Pe}}{\EX{\dt}}
&&
\end{align} 
\end{subequations}
where \(\arate{j}\)'s and \(\afr_j\)'s given in equation (\ref{eq:defarate}) and (\ref{eq:rdti}) respectively.

Thus using  equations  (\ref{eq:defarate}),  (\ref{eq:s5}), (\ref{eq:rdti}),  (\ref{eq:nb-ent-0}) and (\ref{eq:nb-ent-8})
 we reach the  following conclusion. For any  variable length block code satisfying the hypothesis of the
 Lemma \ref{lem:conbits}  and for any positive  \(\delta\) such that \(\Pe+\delta+\tfrac{\Pe}{\delta}\leq\tfrac{1}{2}\)
\begin{subequations}
\label{eq:s8-a}
\begin{align}
\label{eq:s8-a1}
(1-\epst{3}) \ex_i-\epst{5}
&\leq   \sum\nolimits_{j= i+1}^{ \nlay+1} \afr_{j} \JX{\arate{j}}
&i=&1,2,\ldots, \nlay\\
\label{eq:s8-a2}
(1-\epst{3})\sum\nolimits_{j=1}^{i} \rate_j-\epst{4}
&\leq   \sum\nolimits_{j=1}^{i} \afr_j   \arate{j}
&i=&1,2,\ldots,\nlay\\
\label{eq:s8-a3}
(1-\Pe) \rate-\tfrac{\bent{\Pe}}{\EX{\dt}}
&\leq  \sum\nolimits_{j=1}^{\nlay+1}   \afr_j \arate{j}
&&  
\end{align}
\end{subequations}
for some \((\afr_1^{ \nlay+1}\!,\!\arate{1}^{ \nlay+1})\) such that
\begin{subequations}
\label{eq:s8-b}
\begin{align}
\label{eq:s8-b1}
\arate{i}
&\in [0, \CX] 
&i=&1,2,\ldots,( \nlay+1)\\
\label{eq:s8-b2}
\afr_i
&\geq 0
&i=&1,2,\ldots,( \nlay+1)\\
\label{eq:s8-b3}
\sum\nolimits_{i=1}^{ \nlay+1}\afr_i
&=1
&&\end{align}
\end{subequations}

We show below if the constraints given in equation (\ref{eq:s8-a}) is satisfied for some 
\((\afr_1^{ \nlay+1}\!,\!\arate{1}^{ \nlay+1})\) satisfying (\ref{eq:s8-b}), constraints given 
in  (\ref{eq:conbits-a}) is satisfied for some  \((\fr_1^{ \nlay})\) satisfying
  (\ref{eq:conbits-b}).

One can confirm numerically that 
\begin{align*}
(1-\epst{3}) \ln 2 &> \bent{\epst{3}} 
&\forall \epst{3} \in \left[0,\tfrac{1}{5} \right]
\end{align*}
Recall that we have assumed that \(\Pe+\delta+\tfrac{\Pe}{\delta}\leq \tfrac{1}{5}\), i.e.,
 \(\epst{3}\leq \tfrac{1}{5}\). Thus,
 \begin{align}
 \label{eq:r1poss}
(1-\epst{3}) \rate_1-\epst{4}>0.
\end{align}
Let \( \fr_1 \), \(\widetilde{\arate{1}}\) ,  \(\widetilde{\afr}_{2}\)  and \(\widetilde{\arate{2}}\) be
\begin{align*}
\fr_1
&=\tfrac{(1-\epst{3}) \rate_1-\epst{4}}{\arate{1}}\\
\widetilde{\arate{1}}
&= \arate{1}\\
\widetilde{\afr}_{2}
&=\afr_{2}+\afr_{1}-\fr_1\\
\widetilde{\arate{2}}
&=\tfrac{\arate{2} \afr_{2}+ (\afr_{1}-\fr_{1}) \arate{1}}{\widetilde{\afr}_{2}}.
\end{align*}

Note that 
\((\fr_1,\widetilde{\afr}_{2},\afr_{3}^{ \nlay+1},   \widetilde{\arate{1}},\widetilde{\arate{2}},\arate{3}^{ \nlay+1})\) satisfies  (\ref{eq:s8-a2}),  (\ref{eq:s8-a3}) and (\ref{eq:s8-b}) by construction. 
Furthermore as a result of concavity of \(\JX{\cdot}\) we have,
\begin{align*}
\afr_{1} \JX{\arate{1}}+\afr_{2} \JX{\arate{2}}
\leq 
\fr_{1} \JX{\widetilde{\arate{1}}}+\widetilde{\afr}_{2} \JX{\widetilde{\arate{2}}}.
\end{align*}
Thus
\((\fr_1,\widetilde{\afr}_{2},\afr_{3}^{ \nlay+1},   \widetilde{\arate{1}},\widetilde{\arate{2}},\arate{3}^{ \nlay+1})\) 
also satisfies  (\ref{eq:s8-a1}).

 For \(j\geq 2\) we use   \(\widetilde{\afr}_{j}\) and \(\widetilde{\arate{j}}\) to define 
\(\fr_{j}\), \(\widetilde{\afr}_{j+1}\) and \(\widetilde{\arate{}}_{j+1}\) as follows:
\begin{subequations}
\label{eq:itebits}
\begin{align}
\fr_{j}
&=\tfrac{(1-\epst{3})\rate_{j}}{\widetilde{\arate{j}}} \\
\widetilde{\afr}_{j+1}
&=\afr_{j+1}+\widetilde{\afr}_{j}-\fr_{j}\\
\widetilde{\arate{}}_{j+1}
&=\tfrac{\arate{j+1} \afr_{j+1}+ (\widetilde{\afr}_{j}-\fr_{j}) \widetilde{\arate{j}}}{\widetilde{\afr}_{j+1}}.
\end{align}
\end{subequations}
Using the fact that
 \((\fr_1^{j-1},\widetilde{\afr}_{j},\afr_{j+1}^{ \nlay+1},   \widetilde{\arate{1}}^{j},\arate{j+1}^{ \nlay+1})\)  
 satisfies  (\ref{eq:s8-a}) and (\ref{eq:s8-b}) and  the concavity of \(\JX{\cdot}\)  we can show that 
\((\fr_1^{j},\widetilde{\afr}_{j+1},\afr_{j+2}^{ \nlay+1},   \widetilde{\arate{1}}^{j+1},\arate{j+2}^{ \nlay+1})\) 
also satisfies  (\ref{eq:s8-a}) and (\ref{eq:s8-b}).
We repeat the iteration given in equation (\ref{eq:itebits})  until we reach  \(\widetilde{\afr}_{ \nlay+1}\) 
and  \(\widetilde{\arate{}}_{ \nlay+1}\) and we let \(\fr_{ \nlay+1}=\widetilde{\afr}_{ \nlay+1}\).

Then we conclude that  for 
any   variable length block code satisfying the hypothesis of the Lemma \ref{lem:conbits} 
 and for any positive  \(\delta \) such that  \(\Pe+\delta+\tfrac{\Pe}{\delta}\leq \tfrac{1}{5}\)
\begin{subequations}
\label{eq:s9-a}
\begin{align}
\label{eq:s9-a1}
(1-\epst{3}) \ex_i-\epst{5}
&\leq   \sum\nolimits_{j= i+1}^{ \nlay+1} \fr_{j} \JX{\widetilde{\arate{j}}}
&i=&1,2,\ldots, \nlay\\
\label{eq:s9-a2}
(1-\epst{3})\rate_{i} -\epst{4}\IND{i=1}
&=\widetilde{\arate{i}} \fr_i
&i=&1,2,\ldots,\nlay\\
\label{eq:s9-a3}
(\epst{3}-\Pe)\rate+\tfrac{\bent{\epst{3}}-\bent{\Pe}}{\EX{\dt}}
&\leq
\widetilde{\arate{}}_{\nlay+1}\fr_{\nlay+1}
&&
\end{align}
\end{subequations}
for some \((\fr_1,\ldots,\fr_{ \nlay+1},\widetilde{\arate{1}},\ldots,\widetilde{\arate{}}_{ \nlay+1})\) such 
that\footnote{One can replace the inequality in  equation (\ref{eq:s9-a3}) by equality because \(\JX{\cdot}\) 
is a decreasing function.}
\begin{subequations}
\label{eq:s9-b}
\begin{align}
\label{eq:s9-b1}
\widetilde{\arate{i}}
&\in [0, \CX] 
&i=&1,2,\ldots,(\nlay+1)\\
\label{eq:s9-b2}
\fr_i
&\geq 0 
&i=&1,2,\ldots,( \nlay+1)\\
\label{eq:s9-b5}
\sum\nolimits_{i=1}^{ \nlay+1}\fr_i
&=1.
&&\end{align}
\end{subequations}
The Lemma \ref{lem:conbits} follows from  the fact that  \(\JX{\cdot}\leq \DX\).
\end{proof}

\subsection{Codes with Infinite Decoding Time on Channels with Positive Transition Probabilities}\label{app:infdect}

In this section we consider variable length block codes on discrete memoryless channels with 
positive transition probabilities, i.e.,  \(\min_{\dinp\in\inpS, \dout \in \outS}  \CT{\dinp}{\dout}>0\),
and derive lower bounds to the  probabilities of various error events.
These bounds,  i.e., equations (\ref{neq:erb-xac}),  (\ref{neq:erb-xbf}) and (\ref{neq:erb-xcc}),
 enable us to argue that  Lemma \ref{lem:consm} and Lemma \ref{lem:conbits}  hold for variable 
 length block codes with infinite expected decoding time, i.e., \(\EX{\dt}=\infty\).

\subsubsection{\(\Pe>0\)}
\label{app:infdect-a}
On discrete memoryless channel such that 
 \(\min_{\dinp\in\inpS, \dout \in \outS}  \CT{\dinp}{\dout}=\mtp\) 
the posterior probability of any message \(\dmes \in \mesS\) at time \(\tin\) is lower bounded as
\begin{align*}
\PCX{\mes=\dmes}{\out^{\tin}} 
&\geq \left(\tfrac{\mtp}{1-\mtp} \right)^{\tin} \tfrac{1}{|\mesS|}.
\end{align*}
Then conditioned on the event \(\left\{\dt=\tin\right\}\) the  probability of erroneous  decoding 
is  lower bounded as
\begin{align}
\label{neq:erb-xaa}
\PCX{\est\neq \mes}{\dt=\tin}
&\geq \tfrac{|\mesS|-1}{|\mesS|} \left(\tfrac{\mtp}{1-\mtp} \right)^{\tin}.
\end{align} 
Note that since \(\PX{\dt<\infty}=1\), the error probability of any variable length code satisfies
\begin{align}
\label{neq:erb-xab}
\Pe&=\sum_{\tin=1}^{\infty} \PCX{\mes\neq\est}{\dt=\tin} \PX{\dt=\tin}.
\end{align}
Using equation (\ref{neq:erb-xaa}) and (\ref{neq:erb-xab}) we get
\begin{equation}
\label{neq:erb-xac}
\Pe \geq  \tfrac{|\mesS|-1}{|\mesS|}\EX{ \left(\tfrac{\mtp}{1-\mtp} \right)^{\dt}}.
\end{equation}
Note that equation (\ref{neq:erb-xac}) implies that for a variable length code with  infinite expected decoding time not only the rate \(\rate\) but also the error exponent \(\ex\) is zero.

\subsubsection{If  \(\Pe+\tfrac{1}{|\mesS|}<1\) then \(\min_{\dmes}\Pem{\dmes}>0\) }
\label{app:infdect-b}
Note that since \(\PX{\dt<\infty}=1\) and \(|\mesS|<\infty\), 
\begin{align*}
\PCX{\dt<\infty}{\mes=\dmes}& =1
&&\forall \dmes \in\mesS.
\end{align*}
For any variable length block code  such that \(\Pe+\tfrac{1}{|\mesS|}<1\), let \(\tin^*\) be
\begin{align}
\label{neq:erb-xba}
\tin^*=\min \left\{\tin: \max_{\dmes \in \mesS}\PCX{\dt>\tin}{\mes=\dmes} \leq \tfrac{|\mesS|-1}{|\mesS|}-\Pe
\right\}.
\end{align}
Since \(\PCX{\dt<\infty}{\mes=\dmes}=1\) for all \(\dmes \) in \(\mesS\) and \(\mesS\) is finite,  \(\tin^*\) is finite.

Note that for any \(\tin\), \(\dmes \) and \(\widetilde{\dmes}\) we have,
\begin{align}
\label{neq:erb-xbb}
\PCX{\out^{\tin}=\dout^{\tin}}{\mes=\dmes}\geq (\tfrac{\mtp}{1-\mtp})^{\tin} \PCX{\out^{\tin}=\dout^{\tin}}{\mes=\widetilde{\dmes}}
\end{align}
Then using equation  (\ref{neq:erb-xbb}) we get,
\begin{align}
\notag
\Pem{\dmes}
&\geq 
\sum_{\widetilde{\dmes} \neq \dmes} \PCX{\left\{\est=\widetilde{\dmes}, \dt \leq \tin^{*} \right\}}{\mes=\dmes}\\
\notag
&=(\tfrac{\mtp}{1-\mtp})^{\tin^{*}} 
\sum_{\widetilde{\dmes} \neq \dmes} \PCX{\left\{\est=\widetilde{\dmes}, \dt \leq \tin^{*} \right\}}{\mes=\widetilde{\dmes}}\\
\label{neq:erb-xbc}
&\geq(\tfrac{\mtp}{1-\mtp})^{\tin^{*}} 
\sum_{\widetilde{\dmes} \neq \dmes} 
\left(
\PCX{ \est=\widetilde{\dmes} }{\mes=\widetilde{\dmes}}
-
\PCX{ \dt > \tin^{*} }{\mes=\widetilde{\dmes}}
\right)
\end{align}
Note that as a result of equation (\ref{neq:erb-xba}) we have,
\begin{align}
\label{neq:erb-xbd}
\PCX{ \dt > \tin^{*} }{\mes=\widetilde{\dmes}}
&\leq  \left(\tfrac{|\mesS|-1}{|\mesS|}-\Pe \right)
&& \forall \widetilde{\dmes}\in \mesS
\end{align}
Furthermore
\begin{align}
\label{neq:erb-xbe}
\sum_{\widetilde{\dmes} \neq \dmes} 
\PCX{ \est=\widetilde{\dmes} }{\mes=\widetilde{\dmes}}
&\geq |\mesS|(1-\Pe)-1
\end{align}
Thus using equations  (\ref{neq:erb-xbc}), (\ref{neq:erb-xbd})  and (\ref{neq:erb-xbe}) we get
\begin{align}
\label{neq:erb-xbf}
\min_{\dmes\in \mesS}\Pem{\dmes}&\geq(\tfrac{\mtp}{1-\mtp})^{\tin^{*}} 
\left(1-\tfrac{1}{|\mesS|}-\Pe \right)
\end{align}
where \(\tin^*\) is a finite integer defined in equation (\ref{neq:erb-xba}).

\subsubsection{For all \(i \in \left\{1,2,\ldots, \ell\right\}\), \(\Peb{i}>0\)}
\label{app:infdect-c}
For a variable length block code with message set \(\mesS\) of the form
\(\mesS=\mesS_1 \times \mesS_2 \times \ldots \times \mesS_k\)
on a discrete memoryless channel such that 
 \(\min_{\dinp\in\inpS, \dout \in \outS}  \CT{\dinp}{\dout}=\mtp\) 
 the posterior probability of  any element of \(\mesS_{i}\) at time \(\tin\) is lower bounded as
\begin{align*}
\PCX{\mes^{i}=\dmes^{i}}{\out^{\tin}} 
&\geq \left(\tfrac{\mtp}{1-\mtp} \right)^{\tin} \tfrac{1}{|\mesS_i|}
&&\forall\dmes^{i} \in\mesS^{i}, ~ \forall i \left\{1,2,\ldots, \nlay\right\}.
\end{align*}
Then  conditioned on the event \(\left\{\dt=\tin\right\}\) the  probability of decoding the \(i^{th}\) 
sub-message erroneously  is lower bounded as
\begin{align}
\label{neq:erb-xca}
\PCX{\est^{i}\neq \mes^{i}}{\dt=\tin}
&\geq \tfrac{|\mesS^{i}|-1}{|\mesS^{i}|} \left(\tfrac{\mtp}{1-\mtp} \right)^{\tin}.
\end{align} 
Since \(\PX{\dt<\infty}=1\), \(\Peb{i}\)   satisfies
\begin{align}
\label{neq:erb-xcb}
\Pe&=\sum_{\tin=1}^{\infty} \PCX{\mes^{i}\neq\est^{i}}{\dt=\tin} \PX{\dt=\tin}.
\end{align}
Using equation (\ref{neq:erb-xca}) and (\ref{neq:erb-xcb}) we get
\begin{align}
\label{neq:erb-xcc}
\Peb{i}& \geq  \tfrac{|\mesS^{i}|-1}{|\mesS^{i}|}\EX{ \left(\tfrac{\mtp}{1-\mtp} \right)^{\dt}}
&& \forall i \left\{1,2,\ldots, \nlay\right\}.
\end{align}
Equation (\ref{neq:erb-xcc}) implies that for any variable length code with  infinite expected decoding time  
on a DMC without any zero probability transition, not only  the rates but also the error exponents of the 
sub-messages are zero.

\subsection{Proof of Theorem \ref{thm:md}}
\label{sec:appmd}
\begin{proof}
In Section \ref{sec:achsa} it is shown that for any rate \(\rate\in [0,\CX]\), error exponent 
\(\ex\in [0,(1-\tfrac{\rate}{\CX})\DX]\) there exists a reliable sequence \(\SC\) such that \(\rate_{\SC}=\rate\),
  \(\ex_{\SC}=\ex\),  \(\Emd_{,\SC}=\ex+(1-\tfrac{\ex}{\DX}) \JX{\tfrac{\rate}{1-\ex/\DX}}\). Thus as a result of 
  the definition of \(  \Emd(\rate, \ex)\) given in equation (\ref{eq:def:singlemes}) we have
\begin{equation}
\label{eq:mdr-z-fpa}
  \Emd(\rate, \ex) \geq  \ex+\left(1-\tfrac{\ex}{\DX} \right) \JX{\tfrac{\rate}{1-\ex/\DX}}.
\end{equation} 
In Section \ref{sec:con-2}  we have shown that any reliable sequence of codes \(\SC\) with rate \(\rate_\SC\) and error exponent \(\ex_\SC\) satisfies
\begin{align*}
\Emd_{,\SC}&\leq\ex_{\SC}+(1-\tfrac{\ex_{\SC}}{\DX}) \JX{\tfrac{\rate_{\SC}}{1-\ex_{\SC}/\DX}}.
\end{align*}
Thus, using the fact that \(\JX{\cdot}\) is a decreasing concave function we can conclude that
\begin{align*}
\max_{\SC:
\substack{\rate_{\SC}\geq \rate \\ \ex_{\SC}\geq \ex}}
\Emd_{,\SC}&\leq\ex+(1-\tfrac{\ex}{\DX}) \JX{\tfrac{\rate}{1-\ex/\DX}}.
\end{align*}
Consequently  as a result of the definition of \(  \Emd(\rate, \ex)\) given in equation (\ref{eq:def:singlemes}) we have
\begin{equation}
\label{eq:mdr-z-fpc}
  \Emd(\rate, \ex) \leq  \ex+\left(1-\tfrac{\ex}{\DX} \right) \JX{\tfrac{\rate}{1-\ex/\DX}}.
\end{equation} 
Thus using equations (\ref{eq:mdr-z-fpa}) and (\ref{eq:mdr-z-fpc}) we can conclude that
\begin{equation}
\label{eq:mdr-z-fp}
  \Emd(\rate, \ex) =  \ex+\left(1-\tfrac{\ex}{\DX} \right) \JX{\tfrac{\rate}{1-\ex/\DX}}.
\end{equation} 
In order to prove the concavity of \(  \Emd(\rate, \ex)\) in \((\rate, \ex)\) pair,
let \((\rate_a,\ex_a)\) and \((\rate_b,\ex_b)\) be two pairs such that 
\begin{subequations}
\label{eq:smfeas}
\begin{align}
\rate_a&\in [0,\CX] &
\ex_a&\leq (1-\tfrac{\rate_a}{\CX})\DX\\
\rate_b&\in [0,\CX] &
\ex_b&\leq (1-\tfrac{\rate_b}{\CX})\DX.
\end{align}
\end{subequations}
Then for any \(\tsc\in [0,1]\) let \(\rate_{\tsc}\) and \(\ex_{\tsc}\) be
\begin{subequations}
\label{eq:smconc}
\begin{align}
\rate_\tsc&=  \tsc \rate_a+(1-\tsc) \rate_b\\
\ex_\tsc    &=  \tsc \ex_a   +(1-\tsc) \ex_b.
\end{align}
\end{subequations}
From equations (\ref{eq:smfeas}) and  (\ref{eq:smconc}) we have
\begin{align}
\label{eq:smfeas-tsc}
\rate_\tsc&\in [0,\CX] &
\ex_\tsc&\leq (1-\tfrac{\rate_\tsc}{\CX}).
\end{align}
Furthermore using the concavity of \(\JX{\cdot}\) we get,
\begin{align}
\notag
\tsc  \Emd(\rate_a,& \ex_a) +  (1-\tsc)  \Emd(\rate_b, \ex_b) \\
\notag
&=\tsc
\left(
\ex_a+\left(1-\tfrac{\ex_a}{\DX} \right) \JX{\tfrac{\rate_a}{1-\ex_a/\DX}}
\right)	
+  (1-\tsc)
\left(
\ex_b+\left(1-\tfrac{\ex_b}{\DX} \right) \JX{\tfrac{\rate_b}{1-\ex_b/\DX}}
\right)\\
\notag
&=\ex_\tsc+ 
\tsc \left(1-\tfrac{\ex_a}{\DX} \right) \JX{\tfrac{\rate_a}{1-\ex_a/\DX}}
+  (1-\tsc)\left(1-\tfrac{\ex_b}{\DX} \right) \JX{\tfrac{\rate_b}{1-\ex_b/\DX}}
\\
\notag
&\leq \ex_\tsc+ \left(1-\tfrac{\ex_\tsc}{\DX} \right) \JX{\tfrac{\tsc\rate_a+(1-\tsc)\rate_b}{1-\ex_\tsc/\DX}}\\
\label{eq:smcon-tsc}
&= \Emd(\rate_\tsc, \ex_\tsc). 
\end{align}
Thus \(\Emd(\rate, \ex)\) is jointly concave in rate exponent pairs. 
\end{proof}

\subsection{Proof of Theorem \ref{thm:bits}}
\label{sec:appbits}
\begin{proof}
In Section \ref{sec:achsb} it is shown that for any positive integer \( \nlay\) a \rev~  \((\vec{\rate},\vec{\ex})\) is achievable if there exists  a time sharing vector \(\vec{\fr}\) such that,
\begin{subequations}
\label{eq:bitsr-sc}
   \begin{align}
\label{eq:bitsa-sc}
\ex_{i} &\leq (1-\sum\nolimits_{j=1}^{ \nlay}\fr_j) \DX + \sum\nolimits_{j=i+1}^{ \nlay} \fr_j \JX{\tfrac{\rate_j}
{\fr_j}} && \forall i\in \{1,2,\ldots, \nlay\}\\
\label{eq:bitsb-sc}
\rate_i &\leq \CX \fr_i&& \forall i\in \{1,2,\ldots, \nlay\}\\
\label{eq:bitsbc-sc}
\fr_i &\geq 0 && \forall i\in \{1,2,\ldots, \nlay\}\\
\label{eq:bitsc-sc}
\sum\nolimits_{j=1}^{ \nlay}\fr_j&\leq 1     &&
   \end{align}
\end{subequations}
Thus the existence of a time sharing vector \(\vec{\fr}\)  satisfying (\ref{eq:bitsr-sc}) is
 a sufficient condition for the achievablity of a \rev~   \((\vec{\rate},\vec{\ex})\).

For any reliable code sequence \(\SC\) whose message sets are of the form 
\(  \mesS^{(\inx)}=\mesS_{1}^{(\inx)} \times \mesS_{2}^{(\inx)} \times \ldots \times \mesS_{ \nlay}^{(\inx)}\),
Lemma \ref{lem:conbits} with \(\delta=\tfrac{-1}{\ln \Pe}\) implies that 
there exists a sequence  \(\vec{\fr}_{\inx}\)  such that 
\begin{subequations}
\label{eq:conbits-aproof}
  \begin{align}
\label{eq:conbits1proof}
(1-\epst{3,\inx}) \ex_{i,\inx} - \epst{5,\inx}
&\leq  (1- \sum\nolimits_{j=1}^{\nlay}\fr_{j,\inx}) \DX+ 
\sum\nolimits_{j=i+1}^{\nlay}\fr_{j,\inx} \JX{\tfrac{(1-\epst{3,\inx})\rate_{j,\inx}}{\fr_j,\inx}}&&i=1,2,\ldots, \nlay\\
\label{eq:conbits2proof}
(1-\epst{3,\inx})\rate_{i,\inx}-\epst{4,\inx}\IND{i=1}
&\leq \CX \fr_{i,\inx} &&i=1,2,\ldots, \nlay\\
\label{eq:conbits3aproof}
\fr_{i,\inx}
&\geq 0 &&i=1,2,\ldots, \nlay\\
\label{eq:conbits3proof}
\sum\nolimits_{j=1}^{ \nlay} \fr_{,\inx}
&\leq1 &&
\end{align}
\end{subequations}
where 
\(\rate_{i,\inx}=\tfrac{|\ln \mesS_{i}^{(\inx)}|}{\EXS{\inx}{\dt^{(\inx)}}}\), 
\(\ex_{i,\inx}=\tfrac{-\ln \Peb{i}^{(\inx)}}{\EXS{\inx}{\dt^{(\inx)}}}\),  
\(\epst{3,\inx}=\tfrac{\Pe^{(\inx)}+1-\Pe^{(\inx)} \ln \Pe^{(\inx)}}{-\ln \Pe^{(\inx)}}\), 
\(\epst{4,\inx}=\tfrac{\bent{\epst{3,\inx}}}{\EXS{\inx}{\dt^{(\inx)}}}\)
\(\epst{5,\inx}=\tfrac{\bent{\epst{3,\inx}}-\ln \mtp \delta}{\EXS{\inx}{\dt^{(\inx)}}}\).

Note that as a result of equation  (\ref{eq:conbits-aproof}) all members of the sequence \( \vec{\fr}_{\inx}\) 
are from a compact metric space.\footnote{Let the metric be 
\(\lVert \vec{\fr}-\vec{\afr} \rVert=\max_{j} |\fr_j-\afr_j|\).}  Thus there exists  a convergent subsequence, 
converging to a \(\vec{\fr}\). Using equation (\ref{eq:conbits-aproof}),  definitions of  \( \rate_{\SC,i}\) and 
\(\ex_{\SC,i}\) given in Definition \ref{def:bits}  we can 
conclude that \(\vec{\fr}\) satisfies 
  \begin{subequations}
\label{eq:bitsr-nc}
   \begin{align}
\label{eq:bitsa-nc}
\ex_{\SC,i} &\leq (1-\sum\nolimits_{j=1}^{ \nlay}\fr_j) \DX +
 \sum\nolimits_{j=i+1}^{ \nlay} \fr_j \JX{\tfrac{\rate_{\SC,j}}
{\fr_j}} && \forall i\in \{1,2,\ldots, \nlay\}\\
\label{eq:bitsb-nc}
\rate_{\SC,i}&\leq \CX \fr_{i} && \forall i\in \{1,2,\ldots, \nlay\}\\
\label{eq:bitsbc-nc}
\fr_i &\geq 0 && \forall i\in \{1,2,\ldots, \nlay\}\\
\label{eq:bitsc-nc}
\sum\nolimits_{j=1}^{ \nlay}\fr_j&\leq 1.     &&
   \end{align}
\end{subequations}
According to Definition \ref{def:bits} describing the \emph{bit-wise} \uep~problem 
a \rev~ \((\vec{\rate},\vec{\ex})\) is achievable 
only if there exists a reliable code sequence \(\SC\) such that 
\((\vec{\rate}_{\SC},\vec{\ex}_{\SC})=(\vec{\rate},\vec{\ex})\).
Consequently the existence of a time sharing vector satisfying (\ref{eq:bitsr-sc})
is also a necessary condition for the achievablity of 
a \rev~ \((\vec{\rate},\vec{\ex})\) 

Thus we can conclude that  a \rev~ \((\vec{\rate},\vec{\ex})\) is achievable if and only if there exists a \(\vec{\fr}\) satisfying (\ref{eq:bitsr-sc}).

In order to prove the convexity of region of achievable  \revs, let \((\vec{\rate}_a,\vec{\ex}_a)\) and 
\((\vec{\rate}_b,\vec{\ex}_b)\) be two achievable \revs. Then there exist triples  
\((\vec{\rate}_a,\vec{\ex}_a,\vec{\fr}_a)\)
 and  \((\vec{\rate}_b,\vec{\ex}_b,\vec{\fr}_b)\) satisfying (\ref{eq:bitsr-sc}). 
 
 For any \(\tsc\in [0,1]\) let   \(\vec{\rate}_\tsc\), \(\vec{\ex}_\tsc\) and \(\vec{\fr}_\tsc\) be
\begin{align*}
  \vec{\rate}_\tsc
  &=\tsc   \vec{\rate}_a + (1-\tsc)   \vec{\rate}_b\\
  \vec{\ex}_\tsc
  &=\tsc    \vec{\ex}_a + (1-\tsc)  \vec{\ex}_b\\
  \vec{\fr}_\tsc
  &=\tsc   \vec{\fr}_a + (1-\tsc)   \vec{\fr}_b.
\end{align*}
As \(\JX{\cdot}\) is concave and   the  triples
 \((\vec{\rate}_a,\vec{\ex}_a,\vec{\fr}_a)\) and  \((\vec{\rate}_b,\vec{\ex}_b,\vec{\fr}_b)\) satisfy
 the constraints given in (\ref{eq:bitsr-sc}), the triple \((\vec{\rate}_\tsc,\vec{\ex}_\tsc,\vec{\fr}_\tsc)\)
also  satisfies the constraints given in (\ref{eq:bitsr-sc}). 
 Consequently  the \rev~ \((\vec{\rate}_\tsc,\vec{\ex}_\tsc)\) is
 achievable and the region of achievable \revs~ is convex. 
 \end{proof}

%\bibliographystyle{plain}
%\bibliography{main}
%\end{document}

\newcommand{\noopsort}[1]{} \newcommand{\printfirst}[2]{#1}
  \newcommand{\singleletter}[1]{#1} \newcommand{\switchargs}[2]{#2#1}

 \end{document}